\pdfoutput=1
\documentclass[11pt]{article}
\usepackage{graphicx}
\usepackage{latexsym}
\usepackage{mathrsfs}
\usepackage[overload]{textcase}

.5 scaled \magstep4
\font\medio=cmr9.5 scaled \magstep2
\outer\def\beginsection#1\par{\medbreak\bigskip
      \message{#1}\leftline{\bf#1}\nobreak\medskip
\vskip-\parskip
      \noindent}
\begin{document}
\bibliographystyle{unsrt}

\titlepage
\vspace{1cm}
\begin{center}
{\Large {\bf Relic gravitons at intermediate frequencies}}\\
\vspace{1cm}
{\Large {\bf and the expansion history of the Universe}}\\
\vspace{1.5 cm}
Massimo Giovannini \footnote{e-mail address: massimo.giovannini@cern.ch}\\
\vspace{1cm}
{{\sl Department of Physics, CERN, 1211 Geneva 23, Switzerland }}\\
\vspace{0.5cm}
{{\sl INFN, Section of Milan-Bicocca, 20126 Milan, Italy}}
\vspace*{1cm}
\end{center}
\vskip 0.3cm
\centerline{\medio  Abstract}
\vskip 0.5cm
The early expansion history of the Universe is constrained by combining the most recent limits on the cosmic gravitons in the audio band and the claimed evidences  of the nHz domain. The simplest scenario stipulates that between the end of inflation and the formation of light nuclei the evolution consists of a single phase expanding at a rate that is either faster or slower than the one of radiation. If there are instead multiple post-inflationary stages evolving at different rates, the spectral energy density always undershoots the signals potentially attributed to relic gravitons by the pulsar timing arrays at intermediate frequencies but ultimately develops a local maximum. After examining further complementary possibilities (like the presence of a secondary stage of inflation at low-scales) we analyze the early modifications of the effective expansion rate and argue that if the refractive index of the relic gravitons increases during a conventional inflationary epoch the spectral energy density is blue above the fHz and then flattens out in the $\mu$Hz region. In this instance the signal is compatible with the unconfirmed nHz observations, with the most recent limits of the wide-band  interferometers and with the further constraints customarily imposed on the backgrounds of relic gravitons produced during inflation. 
\noindent
\vspace{5mm}
\vfill
\newpage

\renewcommand{\theequation}{1.\arabic{equation}}
\setcounter{equation}{0}
\section{Introduction}
\label{sec1}
The early evolution of the space-time curvature is the primary source of cosmic gravitons \cite{AA1,AA1a} but when a conventional stage of inflationary expansion is followed by a radiation-dominated epoch, the spectral energy density in critical units\footnote{We recall that $h_{0}$ is the Hubble rate expressed in units of $100\,\mathrm{Hz}\, \mathrm{km}/\mathrm{Mpc}$. Since $h_{0}^2$ appears 
in the denominator of $\Omega_{gw}(\nu, \tau_{0})$ it is common practice to phrase the discussions directly in terms of  $h_{0}^2 \Omega_{gw}(\nu, \tau_{0})$ which does not depend on the specific value of $h_{0}$.}
at the present conformal time $\tau_{0}$ (denoted hereunder by $h_{0}^2\,\Omega_{gw}(\nu,\tau_{0})$) is quasi-flat for $\nu > 100 \, \mathrm{aHz}$ \cite{AA2} where $\nu$ denotes the comoving frequency. Between few aHz and $100$ aHz  $h_{0}^2\,\Omega_{gw}(\nu,\tau_{0})$ scales as $\nu^{-2}$ \cite{AA3} and this regime encompasses the wavelengths that reentered the Hubble radius after matter-radiation equality. From the nHz domain to the audio band (between few Hz and $10$ kHz)  the spectral energy density of inflationary origin is, at most, ${\mathcal O}(10^{-16.5})$ and  the deviations from scale-invariance in the direction of blue spectral indices are excluded at least in the conventional situation where the possible corrections to $h_{0}^2\,\Omega_{gw}(\nu,\tau_{0})$ always lead to decreasing spectral slope. This happens, for instance, in the single-field case where, thanks to the consistency relations, the tensor spectral index $n_{T}$ is notoriously related to the tensor to scalar ratio $r_{T}$ as $n_{T} \simeq - r_{T}/8$. Since $r_{T}$ is currently assessed from the analysis of the temperature and polarization anisotropies of the Cosmic Microwave Background (CMB) \cite{RT1,RT2,RT3} $n_{T}$ cannot be positive. There are finally known sources of damping that further reduce the inflationary result (see \cite{CC0} for a recent review) and, most notably, the free-streaming of neutrinos \cite{STRESSNU1,STRESSNU2,STRESSNU3,STRESSNU4,STRESSNU5} 
for frequencies below the nHz.
\begin{figure}[!ht]
\centering
\includegraphics[height=6cm]{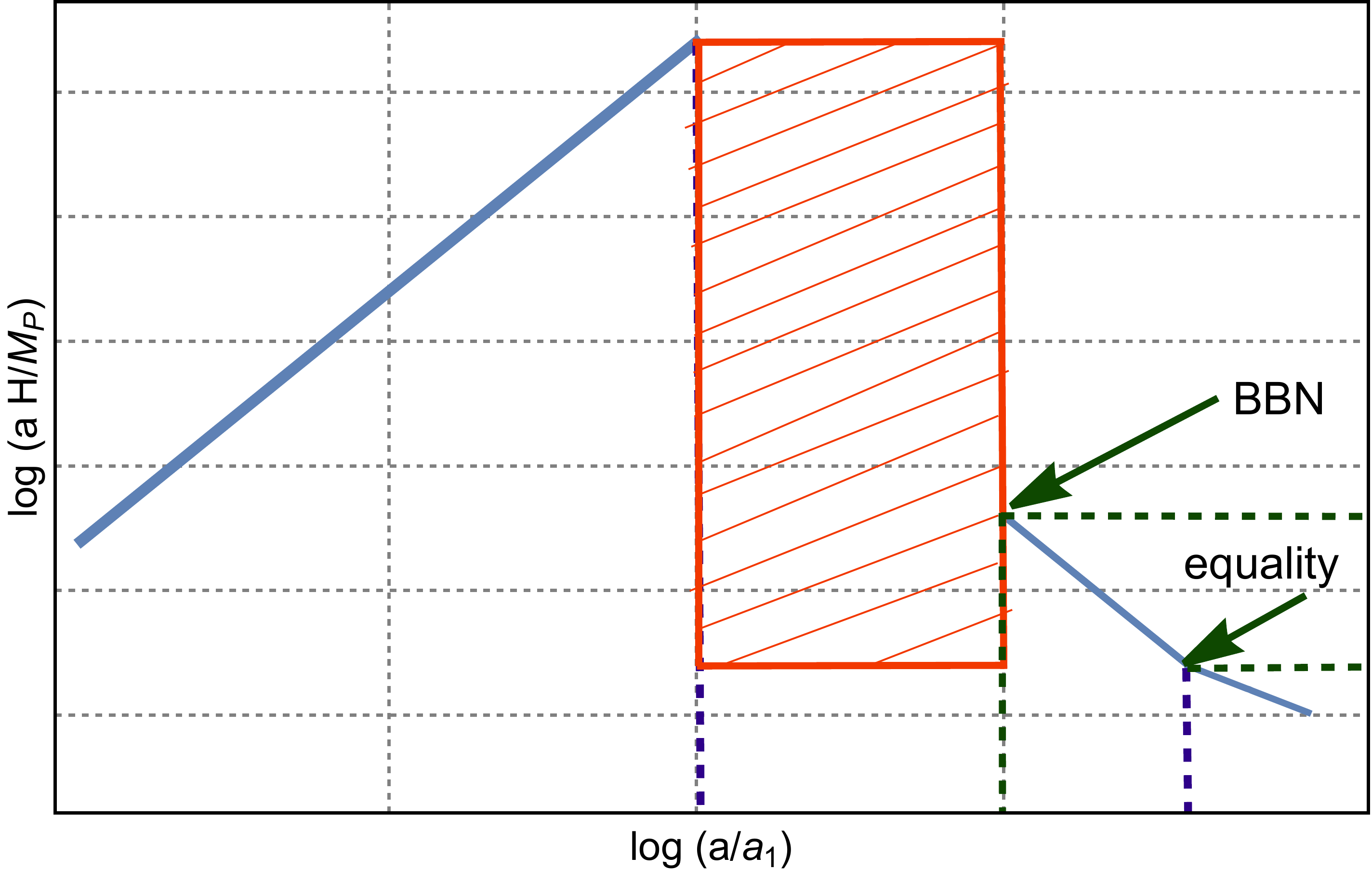}
\caption[a]{The effective expansion rate is qualitatively illustrated in terms 
of the scale factor; common logarithm are employed on both axes. The shaded region extends from the end of the inflationary epoch down to the nucleosynthesis curvature scale which is here considered as a strict lower bound on the duration of a post-inflationary phase potentially different from radiation.}
\label{FF1}      
\end{figure}

\begin{figure}[!ht]
\centering
\includegraphics[height=5.3cm]{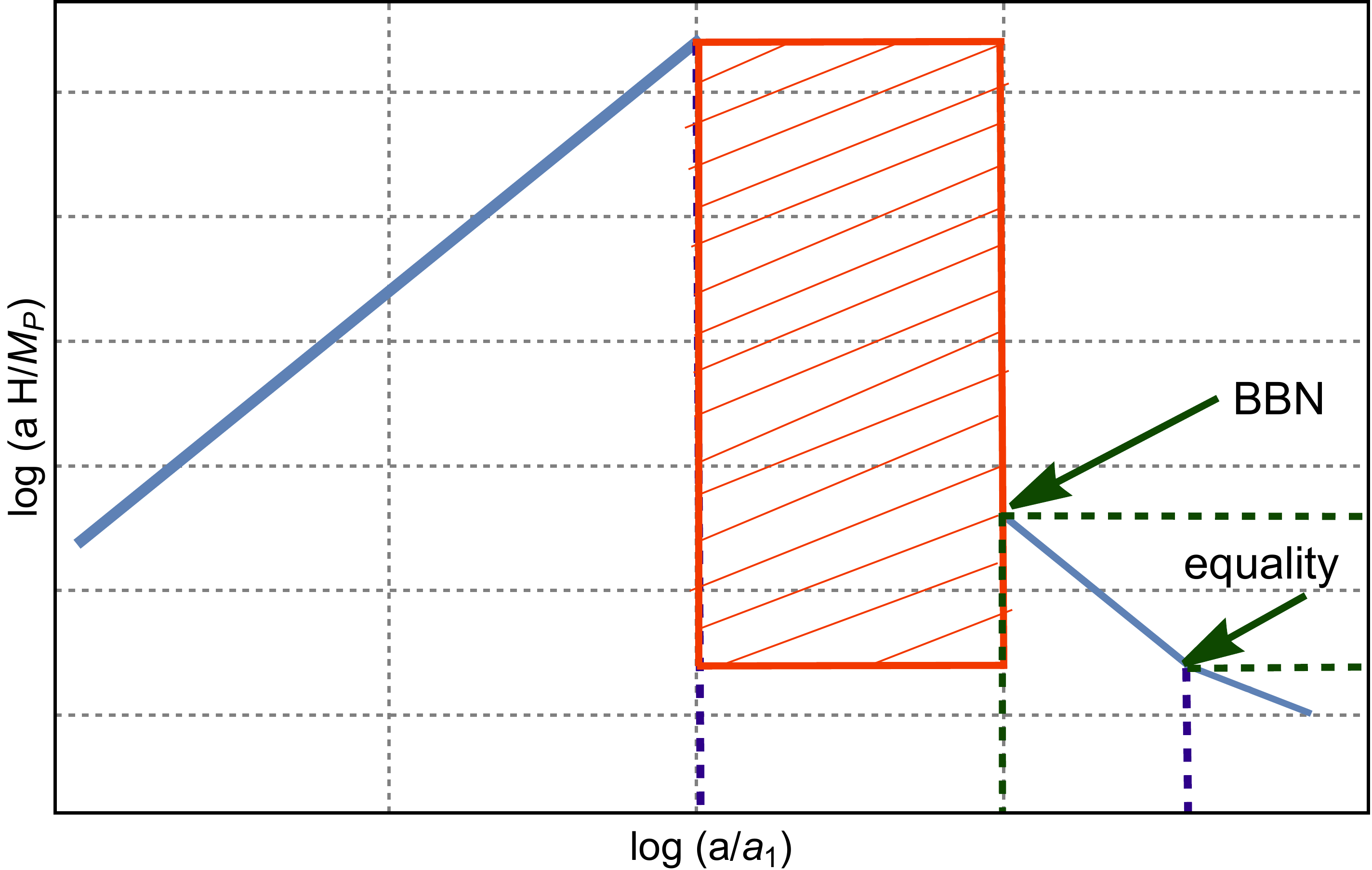}
\includegraphics[height=5.3cm]{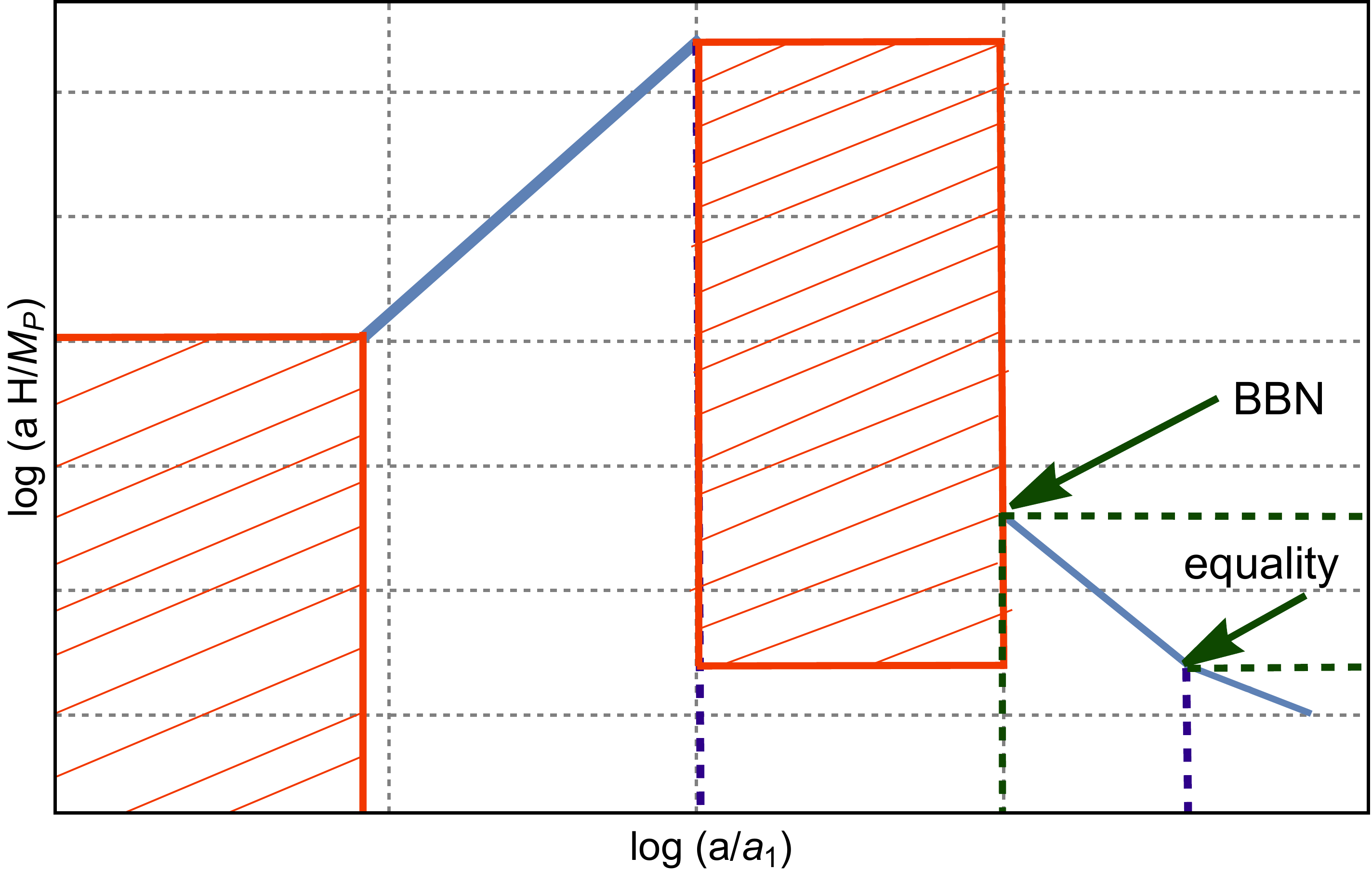}
\caption[a]{In the plot at the left the effective horizon is illustrated when the inflationary 
evolution is modified. In the plot at the right we instead illustrate the most general situation where 
the evolution of the effective horizon is modified during and after inflation; the plot at the right is, in practice, the combination of Fig. \ref{FF1} with the plot at the left. Common logarithms are employed on both axes and in both cartoons.}
\label{FF2}      
\end{figure}

The approximate scale-invariance of the spectral energy density is not only determined by the presence of an early inflationary stage but also by the expansion rate of the post-inflationary evolution. As a consequence the estimate $h_{0}^2\,\Omega_{gw}(\nu,\tau_{0}) = {\mathcal O}(10^{-16.5})$ for $\nu\geq \mathrm{nHz}$ holds when the inflationary epoch is followed by a radiation-dominated stage approximately lasting down to the curvature scale of matter-radiation equality, as originally estimated in \cite{AA2} and subsequently confirmed by various analyses with complementary approaches \cite{BB1,BB2} (see also \cite{CC0} and references therein).  If the assumption of post-inflationary radiation dominance is dropped, the spectral energy density gets modified both in the nHz range and in the audio band. The  perspective of this investigation is therefore to establish a correspondence between the early evolution of the space-time curvature and the frequency dependence of $h_{0}^2 \, \Omega_{gw}(\nu,\tau_{0})$ at intermediate and high-frequencies. Figures \ref{FF1} and \ref{FF2} illustrate, for the sake of concreteness, the effective expansion rate $a H/M_{P}$ as a function of the scale factor in three different classes of profiles that will be closely scrutinized in the present investigation. In Fig. \ref{FF1} the inflationary epoch is followed by a generalized stage of expansion (illustrated by the shaded rectangle) that subsequently turns into the standard radiation-dominated epoch after big-bang nucleosynthesis (BBN in what follows).
As we shall see $a\,H/M_{P}$ determines the evolution of the mode functions and ultimately governs the shape of the spectral energy density. This is is why it is particularly useful to discuss its global evolution during and after inflation.  The shaded box appearing between the end of the inflationary stage and the onset of the radiation-dominated evolution defines the extension of the post-inflationary stage where the evolution can be in principle different from radiation. At the maximum the standard Hubble rate in Planck units is ${\mathcal O}(10^{-6})$ while that the onset of the nucleosynthesis stage the same quantity is of the order of $10^{-44}$ or slightly larger\footnote{This figure is purely illustrative and it approximately correspond to a temperature of the order of the MeV. Depending on the expansion rate,  the stage preceding the nucleosynthesis epoch in Fig. \ref{FF1} is covered in different time-scales.}. 

The inflationary evolution takes place in the leftmost part of 
Fig. \ref{FF1} where the effective horizon evolves linearly with the scale factor (i.e. $a H/M_{P} \propto a$). In the rightmost part of the same cartoon the the expansion is first dominated by radiation (i.e. $a \, H\propto 1/a$) and then by dust (i.e. $a \, H \propto 1/\sqrt{a}$). Since there are no compelling 
reasons why  the expansion rate must coincide with radiation during the whole post-inflationary evolution,
we first assume that the shaded box is filled by a single stage expanding at a rate that is either faster or slower than the one of radiation. It can also happen that various successive phases (all characterized by different expansion rates) are present one after the other. 
The logic described by Fig. \ref{FF1} has an inflationary counterpart that is illustrated in Fig. \ref{FF2} where the initial $e$-folds of the inflationary expansion
are replaced by a shaded region that is not necessarily associated with a different evolution 
of the space-time curvature but rather with a modification of the evolution of the tensor modes of the 
geometry. As we shall see the simplest possibility along this direction is provided by the presence of a refractive index. If an inflationary modification is combined with a generalized post-inflationary evolution 
we shall eventually end-up in the situation described in the right panel of Fig. \ref{FF2}. 

The purpose of the present paper is to outline a model-independent approach to the 
spectra of the relic gravitons by adopting a minimal set of assumptions 
that delicately improve on the concordance paradigm. The aim is not to endorse 
a particular scenario but rather to scrutinize the frequency dependence of the 
spectral energy density of the relic gravitons in the light of the most recent bounds and, along this perspective, the shaded areas of Figs. \ref{FF1} and \ref{FF2} will be 
replaced by various profiles that ultimately lead to different templates of $h_{0}^2 \Omega_{gw}(\nu,\tau_{0})$. The obtained results are then confronted with the current phenomenological bounds in all the available ranges of frequency with the aim of constraining the rate and the duration post-inflationary expansion Universe. In particular, in the nHz region, the pulsar timing arrays (PTA) recently reported a potential signal that could be attributed to the relic gravitons \cite{CCPP1,CCPP2,CCPP3,NANO1}. These analyses follow the upper limits analyzed through the years since the early 1990s (see e.g. \cite{PUL1,PUL4}). The PTA measurements at intermediate frequencies must be complemented with current bounds coming from the operating interferometers in the audio band (i.e. 
between few Hz and $10$ kHz). In particular the joint analysis of the Kagra, Ligo and Virgo (KLV) collaborations has been recently published \cite{LIGO1} and it follows a succession of limits on the spectral energy density of relic gravitons (see also \cite{LIGO2} and Ref. \cite{CC0} for a review of the previous bounds obtained through the years). 

The layout of the paper is therefore the following. In section \ref{sec2} the main notations 
are introduced together with a swift account of the various bounds associated 
with the relic gravitons both at intermediate and high-frequencies. In section \ref{sec3} 
we consider the case of a single post-inflationary stage expanding at a rate that is either 
faster or slower than radiation. The limits 
imposed on the expansion rate and on its duration by the PTA measurements and by the KLV bounds are addressed. The discussion of section \ref{sec3} is further amplified in section \ref{sec4} where the post-inflationary epoch does not consist of a single stage of expansion but rather of a succession of different phases; also in this situation, the spectral energy density is studied in conjunction with the current bounds on the relic graviton backgrounds. While the analysis of sections \ref{sec3} and \ref{sec4} refers, in practice, to the class of profiles generically illustrated in Fig. \ref{FF1}, the results of \ref{sec5} concern the 
early modifications of the effective expansion rate suggested in Fig. \ref{FF2}. In particular 
the spectral energy density produced by the evolution of the refractive index during a conventional inflationary 
phase is confronted with the nHz measurements and with the most recent limits coming from the audio band. The concluding considerations are collected in section \ref{sec6}.

\renewcommand{\theequation}{2.\arabic{equation}}
\setcounter{equation}{0}
\section{Relic gravitons and the evolution of the space-time curvature}
\label{sec2}
Since the tensor modes of the geometry are directly coupled to the evolution of $a\, H/M_{P}$, in what follows, after introducing the main notations,  we analyze the connection between the expansion rate and the spectral energy density in critical units. The relevant bounds associated with backgrounds of relic gravitons are then presented by focussing, in particular, on the results of the pulsar timing arrays (in the nHz range) and on the most recent limits coming from the wide-band interferometers (in the audio band).

\subsection{Effective action and energy density}
The action for relic gravitons written in its covariant form solely depends on the curvature scale through the Riemann tensor \cite{AC1}. In the case of a spatially flat background geometry of Friedmann-Robertson-Walker type the effect of the space-time curvature is therefore related to $a \, H$ that also determines the specific form of the spectral energy density\footnote{In the concordance paradigm the background geometry is spatially flat and can be expressed as $\overline{g}_{\mu\nu} = a^2(\tau) \, \eta_{\mu\nu}$ where $\tau$ is the conformal time coordinate, 
$a(\tau)$ is the scale factor and $\eta_{\mu\nu}$ is the Minkowski metric with signature $(+, \,-\, -\, -)$. Standard notations will be used and, in particular, the prime denotes a derivation with 
respect to $\tau$ so that ${\mathcal H}= a^{\prime}/a$.}.
When the full metric is decomposed as the sum of a background value and of its corresponding fluctuation as $g_{\mu\nu} = \overline{g}_{\mu\nu} + f_{\mu\nu}$, the action of the relic gravitons in its covariant form becomes:
\begin{equation}
S_{g} = \frac{1}{8 \ell_{P}^2} \int d^{4} x\, \sqrt{-\overline{g}} \biggl[ \overline{\nabla}_{\rho} f_{\mu\nu} \,
 \overline{\nabla}^{\rho} \,f^{\,\,\mu\nu} + 2 \, \overline{R}^{\mu\,\,\,\,\,\,\,\,\,\,\,\nu}_{\,\,\,\,\,\rho\sigma}\, f_{\mu\nu} \, f^{\rho \sigma}
\biggr],
\label{SEC1one}
\end{equation}
where the background covariant derivatives are computed with respect to the barred 
metric and $ \ell_{P} = \sqrt{ 8 \pi G}$ is the inverse of the reduced Planck mass $\overline{M}_{P}$:
\begin{equation}
\overline{M}_{P} = \frac{1}{\ell_{P}}, \qquad\qquad \overline{M}_{P} = \frac{M_{P}}{\sqrt{8 \pi}} , \qquad \qquad M_{P} = 1.22\times 10^{19} \, \mathrm{GeV}  = 1.85 \times 10^{43}\,\, \mathrm{Hz}.
\label{notations}
\end{equation}
The fluctuation $f_{\mu\nu}$ described by the action (\ref{SEC1one}) is traceless (i.e. $\overline{g}^{\mu\nu} \, f_{\mu\nu} =0$) divergenceless (i.e. $\overline{\nabla}_{\mu} \, f^{\mu\, \nu} =0$) and orthogonal to the field of the fundamental observer (i.e. 
$f_{\mu\nu} u^{\mu}=0$). Recalling the explicit form of the Riemann tensor in the case of a spatially flat cosmological background geometry, Eq. (\ref{SEC1one}) can also be written as \cite{AC2,AC3}:
\begin{equation}
S_{g} = \frac{1}{8 \ell_{P}^2} \int d^{4} x\, \, \sqrt{- \overline{g}}\,\, 
\overline{g}^{\mu\nu} \partial_{\mu} \, h_{i\, j}  \,\,\partial_{\nu} h^{i \, j},
\label{SEC1two}
\end{equation}
where $h_{i\,j}$ is related to $f_{\mu\nu}$ as $f_{i \, j} = - a^2(\tau) \, h_{i\, j}$. If the background metric is parametrized as 
$\overline{g}_{\mu\nu} = a^2(\tau) \eta_{\mu\nu}$, Eq. (\ref{SEC1two}) becomes:
\begin{equation}
S_{g} = \frac{1}{8 \ell_{P}^2} \int d^{3} x\, \, \int d\tau a^2 \biggl[\partial_{\tau} \, h_{i\, j}  \,\,\partial_{\tau} h^{i \, j} - \partial_{k} h_{i\, j} \, \partial^{k} h^{i \, j}\biggr].
\label{SEC1three}
\end{equation}
 The canonical Hamiltonian associated with 
the action of Eqs. (\ref{SEC1two})--(\ref{SEC1three}) is:
\begin{equation}
H_{g}(\tau) = \int d^{3} x \biggl[ \frac{ 8 \ell_{P}^2}{a^2} \pi_{i\, j} \pi^{i\, j} + 
\frac{a^2}{8 \ell_{P}^2} \partial_{k} h_{i \, j} \partial^{k} h^{i\, j} \biggr], \qquad
\pi_{i\, j} = \frac{a^2}{ 8 \ell_{P}^2 } \, \partial_{\tau} h_{i\, j},
\label{SEC1four}
\end{equation}
where $\pi_{i\,j}$ are the canonical momenta. Thanks to Eq. (\ref{SEC1four}) we have that the evolution of $h_{i\,j}$ and $\pi_{i\,j}$ can be determined from the corresponding Hamilton's equations: 
\begin{equation}
\partial_{\tau} h_{i\, j} = \frac{8 \ell_{P}^2}{a^2} \pi_{i\, j} , \qquad \qquad \partial_{\tau} \pi_{i \, j} = \frac{a^2}{8 \ell_{P}^2} \nabla^2 h_{i\, j}.
\label{SEC1fourb}
\end{equation}
The quantum field operators associated with $h_{i\, j}(\vec{x},\tau)$ and $\pi_{i\,j}(\vec{x},\tau)$ 
can be represented in Fourier space as:
\begin{equation}
\widehat{h}_{i\, j}(\vec{x},\tau) = \frac{1}{(2 \pi)^{3/2}} \int d^{3} q \, \, \widehat{h}_{i \, j} (\vec{q},\tau) \, e^{- i \vec{q}\cdot\vec{x}}, \qquad 
\widehat{\pi}_{i\, j}(\vec{x},\tau) = \frac{1}{(2 \pi)^{3/2}} \int d^{3} p \, \, \widehat{\pi}_{i \, j} (\vec{p},\tau) \, e^{- i \vec{p}\cdot\vec{x}},
\label{FOURa}
\end{equation}
where the $\widehat{h}_{i \, j}^{\dagger}(\vec{q},\tau) = \widehat{h}_{i \, j}(-\vec{q},\tau)$ and $\widehat{\pi}_{i \, j}^{\dagger}(\vec{q},\tau) = \widehat{\pi}_{i \, j}(-\vec{q},\tau)$ since 
$\widehat{h}_{i\, j}(\vec{x},\tau)$ and $\widehat{\pi}_{i\, j}(\vec{x},\tau)$ 
are both Hermitian. The explicit expression of Eq. (\ref{SEC1fourb}) 
in Fourier space is therefore given by:
\begin{equation}
\partial_{\tau} \widehat{h}_{i\, j} = \frac{8 \ell_{P}^2}{a^2} \widehat{\pi}_{i\, j} , \qquad \qquad \partial_{\tau} \widehat{\pi}_{i \, j} = - k^2 \frac{a^2}{8 \ell_{P}^2}  \widehat{h}_{i\, j}.
\label{HPI}
\end{equation}
In the form (\ref{HPI}) the equations are invariant under the inversion of the scale factor; more precisely 
for $a\to 1/a$ we have that the system is invariant provided $\widehat{h}_{i\, j} \to 8 \ell_{P}^2 \,\widehat{\pi}_{i\, j}/k$ and $\widehat{\pi}_{i\, j} \to - \, k \,\widehat{h}_{i\, j}/(8 \ell_{P}^2)$.  

\subsection{The evolution of the mode functions and the effective horizon}
As usual, the explicit form of the operators in Fourier space must involve a sum over the polarizations:
\begin{eqnarray}
\widehat{h}_{i\, j}(\vec{q}, \tau) &=& \sqrt{2} \,\ell_{P} \sum_{\lambda} \biggl[ e_{i\, j}^{(\lambda)}(\hat{q}) \,\, F_{q,\lambda}(\tau)\, \widehat{a}_{\vec{q}, \lambda}
+ e_{i\, j}^{(\lambda)}(-\hat{q}) \,\, F_{q,\lambda}^{\ast}(\tau)\, \widehat{a}^{\dagger}_{- \vec{q}, \lambda} \biggr],
\label{FOURb}\\
\widehat{\pi}_{i\, j}(\vec{p}, \tau) &=& \frac{1}{4 \sqrt{2}\, \ell_{P}} 
\sum_{\lambda} \biggl[ e_{i\, j}^{(\lambda)}(\hat{p}) \,\, G_{p,\lambda}(\tau)\, \widehat{a}_{\vec{p}, \lambda}
+ e_{i\, j}^{(\lambda)}(-\hat{p}) \,\, G_{p,\lambda}^{\ast}(\tau)\, \widehat{a}^{\dagger}_{- \vec{p}, \lambda} \biggr],
\label{FOURc}
\end{eqnarray}
where $e_{i\,j}^{(\lambda)}(\hat{k})$ (with $\lambda =\oplus, \,\,\otimes $) accounts for the two tensor polarizations\footnote{The two tensor polarizations are 
defined as $e_{ij}^{\oplus}(\hat{k}) = (\hat{m}_{i} \, \hat{m}_{j} - \hat{n}_{i} \, \hat{n}_{j})$ and as 
$e_{ij}^{\otimes}(\hat{k}) =( \hat{m}_{i} \, \hat{n}_{j} + \hat{n}_{i} \, \hat{m}_{j}$); $\hat{m}$, $\hat{n}$ and $\hat{k}$ are three mutually orthogonal unit vectors obeying $\hat{m} \times \hat{n} = \hat{k}$.}. Furthermore, for each tensor polarization the creation and the destruction operators obey the standard commutation relations $[ \widehat{a}_{\vec{q},\, \lambda},  \widehat{a}^{\dagger}_{\vec{p},\, \lambda^{\prime}}] = \delta_{\lambda,\, \lambda^{\prime}} 
\delta^{(3)}(\vec{q} - \vec{p})$.  The mode functions $F_{k,\lambda}(\tau)$ and $G_{k,\lambda}(\tau)$ 
introduced in Eqs. (\ref{FOURb})--(\ref{FOURc})  obey the following set of equations that follows 
from Eq. (\ref{SEC1fourb}):
\begin{equation}
\partial_{\tau} F_{k, \lambda} = \frac{G_{k,\,\lambda}}{a^2}, \qquad 
\partial_{\tau} G_{k,\lambda}  = - k^2 \, a^2 F_{k,\lambda}.
\label{SEC1seven}
\end{equation}
The commutator between the canonically conjugate operators in Fourier 
space given by Eqs. (\ref{FOURb})--(\ref{FOURc}) is given by:
\begin{equation}
[ \widehat{h}_{i\, j}(\vec{q}, \tau), \, \widehat{\pi}_{m\, n}(\vec{p}, \tau)] = i \, {\mathcal S}_{i\, j\, m \, n}(\hat{q}) \,\,\,\delta^{(3)}(\vec{q} + \vec{p}).
\label{CANnorm}
\end{equation}
In Eq. (\ref{CANnorm}) ${\mathcal S}_{i\, j\, m \, n}(\hat{q})$ follows from the sum over the two 
tensor polarizations and it is explicitly given by:  
\begin{equation}
{\mathcal S}_{i\,j\,m\,n}(\hat{k}) = [ p_{i\,m}(\hat{k})\, p_{j\,n}(\hat{k}) + p_{j\,m}(\hat{k})\, p_{i\,n}(\hat{k}) - p_{i\,j}(\hat{k})\, p_{m\,n}(\hat{k})]/4, \qquad \qquad p_{i \, j}(\hat{k}) = (\delta_{i\, j} - \hat{k}_{i}\, \hat{k}_{j}),
\label{POLT1}
\end{equation} 
where $\hat{k}^{i} = k^{i}/|\vec{k}|$ is the usual unit vector. The expression of Eq. (\ref{CANnorm}) holds as long as the two mode functions $F_{k, \lambda}$ and $G_{k,\,\lambda}$ are subjected to the Wronskian normalization condition:
\begin{equation}
F_{k,\lambda}(\tau) \,G^{\ast}_{k,\lambda}(\tau) - F_{k,\lambda}^{\ast}(\tau) \,G_{k,\lambda}(\tau) = \,i.
\label{SEC1eight}
\end{equation}
Equation (\ref{SEC1eight}) is therefore essential to guarantee the canonical form of the commutation relations 
(\ref{CANnorm}). If Eqs. (\ref{FOURb})--(\ref{FOURc}) are inserted into Eq. (\ref{FOURa}) the full expression of the canonically conjugate operators is readily obtained:
\begin{eqnarray}
\widehat{\,h\,}_{i\, j}(\vec{x},\tau) &=& \frac{ \sqrt{2} \, \ell_{P}}{(2\pi)^{3/2}} \int\,d^{3} k \, 
\sum_{\lambda} \, e^{(\lambda)}_{i\, j}(\hat{k}) \biggl[ F_{k,\lambda}(\tau)\, \widehat{a}_{\vec{k},\,\lambda}  \, e^{- i \vec{k}\cdot\vec{x}} + F_{k,\lambda}^{\ast}(\tau)\, \widehat{a}_{\vec{k},\,\lambda}^{\dagger}  \, e^{i \vec{k}\cdot\vec{x}}\biggr],
\label{SEC1five}\\
\widehat{\,\pi\,}_{i\, j}(\vec{x},\tau) &=& \frac{1}{4 \sqrt{2} \,(2\pi)^{3/2} \ell_{P}} \int\,d^{3} k \, 
\sum_{\lambda} \, e^{(\lambda)}_{i\, j}(\hat{k}) \biggl[ G_{k,\lambda}(\tau)\, \widehat{a}_{\vec{k},\,\lambda}  \, e^{- i \vec{k}\cdot\vec{x}} + G_{k,\lambda}^{\ast}(\tau)\, \widehat{a}_{\vec{k},\,\lambda}^{\dagger}  \, e^{i \vec{k}\cdot\vec{x}}\biggr].
\label{SEC1six}
\end{eqnarray}
For the present purposes it is quite useful to express the problem in terms of the 
evolution of the two rescaled mode functions 
$f_{k,\,\lambda} = a F_{k, \,\lambda}$ and $g_{k, \, \lambda} =G_{k,\lambda}/a$;
thanks to Eq. (\ref{SEC1seven}) we have:
\begin{equation}
f_{k,\, \lambda}^{\prime} = g_{k,\, \lambda} + {\mathcal H} \, f_{k,\lambda},\qquad g_{k,\, \lambda}^{\prime} = - k^2 f_{k,\,\lambda} - {\mathcal H} \, g_{k,\,\lambda},
\label{TWOFIRST}
\end{equation}
where ${\mathcal H} = a^{\prime}/a$ and the prime denotes a derivation with respect to the conformal 
time coordinate $\tau$. The same duality symmetry of Eq. (\ref{HPI}) relates the  two expressions 
appearing in Eq. (\ref{TWOFIRST}):  for $a\to 1/a$ we have that when $f_{k,\, \lambda} \to g_{k,\,\lambda}/k$ and  $g_{k,\,\lambda} \to - k \, f_{k,\, \lambda}$ the two equations appearing in Eq. (\ref{TWOFIRST}) are transformed one into the other.
The two first-order (coupled) differential equations appearing in Eq. (\ref{TWOFIRST}) can be decoupled by deriving once (with respect to $\tau$) either the first or the second equation. The decoupled evolution for $f_{k,\,\lambda}$ is:
\begin{equation} 
f_{k,\,\lambda}^{\prime\prime} + \biggl[ k^2 - \frac{a^{\prime\prime}}{a} \biggr] f_{k,\,\lambda} =0, \qquad\qquad  g_{k,\,\lambda} = f_{k,\,\lambda}^{\prime} - {\mathcal H} f_{k,\,\lambda}.
\label{PAM}
\end{equation}
Once the evolution of $f_{k,\,\lambda}$ is determined from Eq. (\ref{PAM}), 
$g_{k,\,\lambda}$ follows from the second relation which is now just a definition 
and not  a dynamical equation. The mode functions $f_{k,\,\lambda}$ and $g_{k,\,\lambda}$ must obey the Wronskian normalization condition $W_{\lambda}(k, \tau) = f_{k,\,\lambda}(\tau) \, g_{k,\,\lambda}^{\ast}(\tau) -  f_{k,\,\lambda}^{\ast}(\tau) \, g_{k,\,\lambda}(\tau)= \,i$ that follows directly from Eq. (\ref{SEC1eight}). 

From  Eq. (\ref{PAM}) it follows that the evolution of the mode functions is chiefly determined, in practice, by $a\,H$ that appears in the profiles of Figs. \ref{FF1} and \ref{FF2}. The first observation, in this 
respect, is that  $a^{\prime\prime}/a = - a^2 \overline{R}/6$ where $\overline{R}$ Ricci scalar of the background so that we can also write: 
\begin{equation}
\frac{a^{\prime\prime}}{a} =  a^2 H^2 \biggl( 2  + \frac{\dot{H}}{H^2} \biggr),
\label{RGC1}
\end{equation}
where the overdot now denotes now a derivation with respect to the cosmic time coordinate. The result of Eq. (\ref{RGC1}) follows from the observation that ${\mathcal H} = a \, H$ and from the relation between cosmic and conformal times (i.e. $a(\tau) d\tau = dt$). During an inflationary stage of expansion $| \dot{H}/H^2 |\ll 1$ and the right hand side of Eq. (\ref{RGC1}) 
can be approximated by $a^2 H^2$. Conversely, during a stage of decelerated expansion $\dot{H}/H^2 = {\mathcal O}(1)$ and, up to a numerical factor, the right hand side of Eq. (\ref{RGC1}) is given by $a^2 H^2$. Equation (\ref{RGC1}) clarifies that $a\, H$ offers a fair estimate of the evolution of the space-time curvature and this is ultimately the rationale behind the Figs. \ref{FF1} and \ref{FF2}. In the following sections the different classes of evolutions will be distinguished by considering the profiles 
of $a\, H/M_{P}$ as a function of the scale factor\footnote{This is in fact 
just an illustrative strategy and not as a computational tool. For instance it would be inaccurate 
to determine the normalization and the evolution of the mode functions by just neglecting 
the terms ${\mathcal O}(\dot{H}/H^2)$ in Eq. (\ref{RGC1}). However, from the physical viewpoint, the profile $a\, H/M_{P}$
is exactly the quantity we would like to infer, eventually,  from the analysis of $h_{0}^2 \Omega_{gw}(\nu,\tau_{0})$.}.

\subsection{The spectral energy density}
The energy density, the pressure and the anisotropic stress of the relic gravitons do not have a unique gauge-invariant and frame-invariant expression since the equivalence principle ultimately forbids the localization of the energy-momentum of the gravitational field. The classic proposals for the energy density of the gravitational field discussed 
through the years  \cite{AC3a,AC3b,AC3c} (see also \cite{AC3d,AC3e})
have been recently compared in the context of the relic gravitons \cite{AC4} 
and it has been argued that the energy-momentum pseudotensor of the relic gravitons 
should not violate the weak energy condition and it should be derived in general terms, i.e. 
without explicitly demanding that the rate of variation of the background 
geometry is either faster or slower than the frequencies of the corresponding gravitons.
An energy-momentum pseudo-tensor with these features
follows from the effective action of the relic gravitons 
by considering the tensor fluctuations and the background metric as independent 
variables. In its simplest realization the effective action coincides with 
the result of Eqs. (\ref{SEC1two})--(\ref{SEC1three}) and the derived energy density 
is given by \cite{AC4}:
\begin{equation}
\widehat{\rho}_{gw}(\vec{x},\tau) = \frac{1}{8 \ell_{P}^2 a^2} \biggl[\partial_{\tau} \widehat{h}_{i\, j} \partial_{\tau} \widehat{h}^{i\, j} +\partial_{k} \widehat{h}_{i\, j} \partial_{k} \widehat{h}^{i\, j}\biggr].
\label{ENDENS}
\end{equation} 
The quantum average of the energy density follows by inserting 
Eqs. (\ref{SEC1five})--(\ref{SEC1six}) into Eq. (\ref{ENDENS}) 
and the final result can be expressed as:
\begin{equation}
\langle \widehat{\rho}_{gw}(\vec{x},\tau)  \rangle = \frac{1}{8 \ell_{P}^2 a^2} \int \frac{d \, k}{k} \bigg[ k^2 \, P_{T}(k,\tau) + Q_{T}(k,\tau)\biggr],
\label{AVENDENS}
\end{equation}
where the power spectra $P_{T}(k,\tau)$ and $Q_{T}(k,\tau)$ are: 
\begin{eqnarray}
P_{T}(k,\tau) &=& \frac{4 \ell_{P}^2}{\pi^2} \, k^3\, \bigl| F_{k}(\tau)\bigr|^2 = \frac{4 \ell_{P}^2}{\pi^2\, a^2} \, k^3\, \bigl| f_{k}(\tau)\bigr|^2,
\label{PSPT1}\\
Q_{T}(k,\tau) &=& \frac{4 \ell_{P}^2}{\pi^2 a^4} \, k^3\, \bigl| G_{k}(\tau)\bigr|^2 = \frac{4 \ell_{P}^2}{\pi^2\, a^2} \, k^3\, \bigl| g_{k}(\tau)\bigr|^2.
\label{PSQT1}
\end{eqnarray}
In Eqs. (\ref{PSPT1})--(\ref{PSQT1}) the index $\lambda$ appearing in the mode functions has been suppressed since the two polarizations obey the same equation. There are differences and the analogies between $P_{T}(k,\tau)$ and $Q_{T}(k,\tau)$; in particular,  if $F_{k}(\tau)$ and $G_{k}(\tau)$ 
obey Eqs. (\ref{SEC1seven}) the power spectrum $Q_{T}(k,\tau)$ inherits a 
term $|G_{k}(\tau)|^2/a^4$. This is just a matter of definition since 
what appears in Eq. (\ref{ENDENS}) are not the canonical momenta 
but rather the time derivatives of the field operators. From Eqs. (\ref{SEC1five})--(\ref{SEC1six}) the expectation values of the quadratic combinations appearing in Eq. (\ref{ENDENS}) are associated with $P_{T}(k,\tau)$ and 
$Q_{T}(k,\tau)$ respectively:
\begin{eqnarray}
\langle \widehat{h}_{i\, j}(\vec{k}, \tau) \, \widehat{h}_{m\, n}(\vec{p}, \tau) \rangle &=& \frac{2 \pi^2}{k^3} \, {\mathcal S}_{i\, j\, m\,n}(\widehat{k}) \,\,P_{T}(k,\tau) \, \, \delta^{(3)}(\vec{k} + \vec{p}),
\label{PSPT}\\
\langle \partial_{\tau} \widehat{h}_{i\, j}(\vec{k}, \tau) \, \partial_{\tau} \widehat{h}_{m\, n}(\vec{p}, \tau) \rangle &=& \frac{2 \pi^2}{k^3} \, {\mathcal S}_{i\, j\, m\,n}(\widehat{k}) \,\,Q_{T}(k,\tau) \, \, \delta^{(3)}(\vec{k} + \vec{p}).
\label{PSQT}
\end{eqnarray}
It is useful to recall that the Brill-Hartle averaging 
scheme \cite{AC3b} reproduces the results of the quantum averaging of Eq. (\ref{AVENDENS}) when the wavelengths are shorter than the Hubble radius but it is not defined in the opposite limit. Despite claims to the contrary the issue of the averaging  
is not secondary \cite{AC4}. From Eq. (\ref{AVENDENS}) we finally introduce the spectral energy density in critical units, namely: 
\begin{equation}
\Omega_{gw}(k,\tau) = \frac{1}{\rho_{crit}}\frac{d \, \langle \widehat{\rho}_{gw}  \rangle}{d \ln{k}} = \frac{ k^2 \, P_{T}(k,\tau) + Q_{T}(k,\tau)}{24 \, H^2 \, a^2},\qquad\qquad \rho_{crit} = 3 \, H^2\, \overline{M}_{P}^2,
\label{ENDENS2}
\end{equation}
where, according to the notations established in Eq. (\ref{notations}), $\ell_{P} = 1/\overline{M}_{P}$. There are some 
misleading notations in the current literature suggesting that what we defined 
as $\Omega_{gw}(k,\tau)$ is in fact given by the quotient between the energy density 
of the relic gravitons and the critical energy density. These definitions are grossly 
incorrect since $\rho_{gw}(\vec{x},\tau)/\rho_{crit}$ does not coincide with 
Eq. (\ref{ENDENS2}). We can finally insert Eqs. (\ref{PSPT1})--(\ref{PSQT1}) into Eq. (\ref{ENDENS2}) and obtain
the following expression for the spectral energy density in critical units:
\begin{equation}
\Omega_{gw}(k,\tau) = \frac{k^3}{6 \pi^2 \, H^2 \, \overline{M}_{P}^2 \, a^4}\biggl[ k^2 \, \bigl| f_{k}(\tau)\bigr|^2 + \bigl| g_{k}(\tau)\bigr|^2 \biggr].
\label{ENDENS3}
\end{equation}
If we refer to the case of Fig. \ref{FF1} the early initial conditions of the mode functions 
are assigned  during the inflationary stage and this will be the initial conditions 
assumed in the following two sections. The slow-roll parameters affect the definition of the conformal time 
coordinate $\tau$. In fact, by definition
\begin{equation}
\tau = \int \frac{dt}{a(t)} = - \frac{1}{a H} + \epsilon \int \frac{d a}{a^2 H} \qquad \Rightarrow\qquad a H = - \frac{1}{\tau ( 1 - \epsilon)},
\label{tauHa}
\end{equation}
where the second equality follows after integration by parts assuming that $\epsilon = - \dot{H}/H^2$ is slowly-varying during the inflationary stage. The solution of Eq. (\ref{PAM}) with the appropriate boundary conditions for $\tau \to - \infty$ is then given by:
\begin{eqnarray}
f_{k}(\tau) &=& \frac{{\mathcal N}_{\mu}}{\sqrt{2 k}} \, \sqrt{- k\tau} \, H_{\mu}^{(1)}(-k\tau), \qquad {\mathcal N}_{\mu} = \sqrt{\pi/2}\,\,e^{i \pi (\mu +1/2)},
\nonumber\\
g_{k}(\tau) &=& - {\mathcal N}_{\mu} \,\sqrt{\frac{k}{2}} \,  \sqrt{- k\tau} \, H_{\mu-1}^{(1)}(-k\tau), \qquad \mu= \frac{3 - \epsilon}{2(1 - \epsilon)},
\label{INCOND1}
\end{eqnarray}
where $H_{\mu}^{(1)}(-k\tau)$ denotes the Hankel function of the first kind \cite{abr1,abr2}.  With the initial conditions of Eq. (\ref{INCOND1}) the spectral energy density can be computed for different classes of post-inflationary evolutions that are analyzed in the forthcoming sections. Before discussing this point it is however necessary to recall the most recent phenomenological constraints on the spectral energy density of the relic gravitons.

\subsection{Pulsar timing arrays}
The pulsar timing arrays (PTA in what follows) recently reported evidence 
of a potential signal in the nHz band. Using the spectral energy density in critical units as a pivotal 
variable the features of this purported signal would imply, in the present notations, that:
\begin{equation}
q_{0}^2\times 10^{-8.86}  < \, h_{0}^2\,\Omega_{gw}(\nu,\tau_{0}) < \,q_{0}^2 \times10^{-9.88} , \qquad\qquad 3\,\mathrm{nHz} < \nu< 100 \, \mathrm{nHz}.
\label{PTAb1}
\end{equation}
In Eq. (\ref{PTAb1}) we introduced  the numerical factor $q_{0}$ that depends on the 
specific experimental determination. For instance the the Parkes Pulsar Timing Array collaboration (PPTA in what follows) \cite{CCPP1} (see also \cite{PPTA1,PPTA2}) suggests $q_{0}= 2.2$. Similarly the International Pulsar Timing Array collaboration (IPTA in what follows) estimates $q_{0}= 2.8$ \cite{CCPP2} while the European Pulsar Timing Array collaboration (EPTA in what follows) \cite{CCPP3} gives $q_{0} = 2.95$ (see also \cite{EPTA1,EPTA2}).  The results  of PPTA, IPTA and EPTA seem, at the moment, to be broadly compatble with the NANOgrav 12.5 yrs data \cite{NANO1} (see also \cite{NANO2,NANO3}) implying $q_{0} =1.92$. 

It is relevant to point out that neither the observations of Refs. \cite{CCPP1,CCPP2,CCPP3} nor the ones of Ref. \cite{NANO1} can be interpreted yet as an evidence of relic gravitons. The property of a PTA is that the signal 
from relic gravitons will be correlated across the baselines, while that from the other noise will not. 
Since these correlation have not been observed so far, the interpretation suggested in 
Eq. (\ref{PTAb1}) is still preliminary, to say the least.  To be fair the pragmatic strategy followed here 
will be to interpret Eq. (\ref{PTAb1}) as an upper limit whenever the 
corresponding theoretical signal is too low in the nHz region. Conversely  
if $h_{0}^2\,\Omega_{gw}(\nu,\tau_{0})$ happens to be grossly compatible 
with the range of Eq. (\ref{PTAb1}) it will be interesting to see if the associated spectral 
energy density fits within the PTA window.

The PTA collaborations express their results in terms of the chirp amplitude so that the parametrization of Eq. (\ref{PTAb1}) is not the one directly employed by the observers. In this respect we just note that, up to a numerical factor, the square of the chirp amplitude coincides with the power spectrum so that its relation with the spectral energy density may be easily determined:
\begin{equation} 
P_{T}(\nu, \tau_{0}) = 2 \, h_{c}^2(\nu,\tau_{0}), \qquad \qquad
\Omega_{gw}(\nu, \tau_{0}) = \frac{2 \pi^2}{3 H_{0}^2} \, \nu^2 \, h_{c}^2(\nu, \tau_{0}).
\label{NOTT0}
\end{equation}
The various PTA collaborations \cite{CCPP1,CCPP2,CCPP3,NANO1} normalize the chirp amplitude 
at a pivot frequency $\nu_{ref} = \mathrm{yr}^{-1}$:
\begin{equation}
h_{c}(\nu,\tau_{0}) = {\mathcal Q} \biggl(\frac{\nu}{\nu_{ref}}\biggr)^{\beta}, \qquad \qquad \nu_{ref} = \frac{1}{\mathrm{yr}}=   31.68\,\, \mathrm{nHz}.
\label{NOTT1}
\end{equation}
For the different estimates of Refs. \cite{CCPP1,CCPP2,CCPP3,NANO1} the value of ${\mathcal Q}$ 
is always ${\mathcal O}(10^{-15})$ and, for this reason, we parametrize it 
as ${\mathcal Q} = q_{0} \times 10^{-15}$ where $q_{0}$ is exactly the constant appearing 
in Eq. (\ref{PTAb1}). From Eq. (\ref{NOTT0}) the spectral energy density in the nHz band can be finally expressed as:
\begin{equation}
h_{0}^2\,\Omega_{gw}(\nu,\tau_{0}) = 6.290\times 10^{-10} \,\, \, q_{0}^2\, \biggl(\frac{\nu}{\nu_{ref}}\biggr)^{ 2 + 2 \beta}.
\label{NOTT8}
\end{equation}
The different values of $q_{0}$ discussed above (see after Eq. (\ref{PTAb1})) are associated with 
a specific value of $\beta$ which has now become conventional. For instance 
the EPTA finds that the most favoured model to be the common uncorrelated red noise 
described by ${\mathcal Q}= 5.13^{+4.20}_{-2.73}\times 10^{-15}$ with $\gamma = 3.78^{+0.69}_{-0.59}$ \cite{CCPP2}. Within our notation we recall that $\beta = (3-\gamma)/2$. If the spectral index is instead fixed as $\gamma = 13/3$ (i.e. $\beta = -2/3$) we have that the EPTA collaboration finds \cite{CCPP2} ${\mathcal Q}= 2.95^{+0.89}_{-0.72}\times 10^{-15}$. The other values of $q_{0}$ quoted after Eq. (\ref{PTAb1}) always refer to the case $\beta = -2/3$. For the sake of simplicity in what follows we shall consider the case $\beta = -2/3$ implying that the slope of the spectral energy density is actually $+2/3$, as it follows from Eq. (\ref{NOTT8}).

\subsection{Big-bang nucleosynthesis limits}
While the PTA measurements constrain the spectral energy density at intermediate 
frequencies, the bounds coming from big-bang nucleosynthesis \cite{bbn1,bbn2,bbn3} imply a constraint on the integral $h_{0}^2\,\Omega_{gw}(\nu,\tau_{0})$:
\begin{equation}
h_{0}^2  \int_{\nu_{bbn}}^{\nu_{max}}
  \Omega_{gw}(\nu,\tau_{0}) d\ln{\nu} = 5.61 \times 10^{-6} \Delta N_{\nu} 
  \biggl(\frac{h_{0}^2\,\Omega_{\gamma0}}{2.47 \times 10^{-5}}\biggr),
\label{CC2}
\end{equation}
where $\Omega_{\gamma0}$ is the (present) critical fraction of CMB photons. The limit  of Eq. (\ref{CC2}) sets an indirect constraint  on the extra-relativistic species possibly present at the time of nucleosynthesis. Since Eq. (\ref{CC2}) is relevant in the context of neutrino physics, the limit is often expressed for practical reasons  in terms of $\Delta N_{\nu}$ representing the contribution of supplementary neutrino species. The actual bounds on $\Delta N_{\nu}$ range from $\Delta N_{\nu} \leq 0.2$ 
to $\Delta N_{\nu} \leq 1$;  the integrated spectral density in Eq. (\ref{CC2}) is thus between $10^{-6}$ and $10^{-5}$. It is relevant to point out, as 
we shall see, that the upper limit of integration (labeled by $\nu_{max}$) depends on the specific 
post-inflationary evolutions\footnote{In the forthcoming discussion an important 
element is the determination of $\nu_{max}$ that depends on the duration of the post-inflationary 
evolution and on the corresponding expansion rates. For $\nu> \nu_{max}$ the 
spectra of relic gravitons are exponentially suppressed since these wavelengths 
never cross the Hubble radius and are not amplified (see e.g. \cite{transfer1,transfer2}.}. Conversely, the lower limit of integration in Eq. (\ref{CC2}) is given by  the frequency corresponding to the Hubble rate at the nucleosynthesis epoch: 
\begin{equation}
\nu_{bbn}= 2.252\times 10^{-11} \biggl(\frac{N_{eff}}{10.75}\biggr)^{1/4} \biggl(\frac{T_{bbn}}{\,\,\mathrm{MeV}}\biggr) 
\biggl(\frac{h_{0}^2\,\Omega_{R0}}{4.15 \times 10^{-5}}\biggr)^{1/4}\,\,\mathrm{Hz} \simeq 0.01\, \mathrm{nHz},
\label{CC3}
\end{equation}
where  $N_{eff}$ denotes the effective number of relativistic degrees of freedom entering the total energy density of the plasma and $T_{bbn}$ is the temperature of big-bang nucleosynthesis.  We finally remark that the bound of Eq. (\ref{CC2}) 
could be relaxed if the nucleosynthesis takes place in the presence 
of matter-antimatter domains \cite{bbn2}. This possibility 
will not be specifically considered hereunder and we shall instead enforce 
the bound of Eqs. (\ref{CC2})--(\ref{CC3}) in its conservative version.

\subsection{The Kagra-Ligo-Virgo bound}
The Kagra, Ligo and Virgo collaborations, in their attempt to constraint the stochastic backgrounds of relic gravitons, 
 reported a constraint \cite{LIGO1} implying, in the case of a flat spectral energy density,
\begin{equation}
\Omega_{gw}(\nu, \tau_{0}) < 5.8 \times 10^{-9}, \qquad\qquad 20 \,\, \mathrm{Hz} < \nu_{KLV} < 76.6 \,\, \mathrm{Hz},
\label{CONS2}
\end{equation}
where $\nu_{KLV}$ denotes the Kagra-Ligo-Virgo frequency; as already pointed out in the introduction, for the sake of conciseness we shall commonly refer to this limit as the KLV bound.
The class of limits associated with Eq. (\ref{CONS2}) improves on a series of bounds previously deduced by the Ligo and Virgo collaborations (see \cite{CC0} for a recent review). In particular 
in Ref. \cite{LIGO2} the analog of Eq. (\ref{CONS2}) implied $ \Omega_{gw}(\nu, \tau_{0}) < 6 \times 10^{-8}$ for a comparable frequency interval and always in the case of a flat spectral energy density. While the bound of Eq. (\ref{CONS2}) could be immediately 
used also in our case since at high-frequency the spectral energy density is 
nearly scale-invariant, it is useful to elaborate on the result of Ref. \cite{LIGO1}. 
Even if in Eq. (\ref{CONS2}) we just quoted the most constraining limit, the Kagra-Ligo-Virgo  collaboration actually reports a threefold bound which could be parametrized as 
\begin{equation}
\Omega_{gw}(\nu, \tau_{0}) = \overline{\Omega}(\zeta) \biggl(\frac{\nu}{\nu_{ref}}\biggr)^{\zeta}, \qquad \qquad \nu_{ref} = 25 \,\, \mathrm{Hz}.
\label{NOT1}
\end{equation}
In terms of Eq. (\ref{NOT1}) the results of Ref. \cite{LIGO1} read 
$\overline{\Omega}(0) < 5.8 \times 10^{-9}$ (valid in the case $\zeta =0$),
$\overline{\Omega}(2/3) < 3.4 \times 10^{-9}$ (when $\zeta= 2/3$) 
and $\overline{\Omega}(3) < 3.9 \times 10^{-10}$ (when $\zeta= 3$).
When the value of $\zeta$ increases the bound becomes more restrictive 
once the reference frequency has been kept fixed. The three results are unified in the following interpolating formula
\begin{equation}
\log{\overline{\Omega}}(\zeta) < -\,8.236 -\, 0.335\, \zeta- 0.018\, \zeta^2.
\label{NOT2}
\end{equation}
The quadratic fit is slightly more accurate the linear one, 
the essence of the arguments does not change in the two cases since the different points fall approximately on the same straight line (i.e. $ - 8.223 - 0.393\, \zeta$).  
In principle the expression could be applied for arbitrary values of $\zeta$. However, as we shall see, one of the most relevant cases is the one where $\zeta \ll 1$ (i.e. quasi-flat spectral energy density at high-frequency). 

\renewcommand{\theequation}{3.\arabic{equation}}
\setcounter{equation}{0}
\section{The case of a single post-inflationary phase}
\label{sec3}
In the simplest class of profiles the shaded rectangle of Fig. \ref{FF1} is 
replaced by a single decelerated stage of expansion where the rate 
is either faster or slower than radiation. After considering the typical scales 
of the problem, the constraints on the spectral energy density 
of the produced gravitons are specifically analyzed.

\subsection{The profiles of $a\, H/M_{P}$ for a single post-inflationary stage}
Two scales are particularly relevant for the forthcoming considerations: the expansion rates at the end of inflation (i.e. $H_{1}$) and at the onset of big-bang nucleosynthesis (i.e. $H_{bbn}$).
If we assume the validity of the consistency relations $H_{1}$ follows from the amplitude of the scalar power spectrum ${\mathcal A}_{{\mathcal R}}$ and  from the 
tensor to scalar ratio $r_{T}$.  In particular, taking into account the limits on the 
$r_{T}$  \cite{RT1,RT2,RT3} we can estimate\footnote{The current limits on $r_{T}$ would imply $r_{T} <0.064$ \cite{RT1,RT2,RT3}. In what follows we shall therefore demand $r_{T}\leq 0.06$. }:
\begin{equation}
\frac{H_{1}}{M_{P}} = 5.32 \times 10^{-6} \biggl(\frac{{\mathcal A}_{{\mathcal R}}}{2.41\times 10^{-9}}\biggr)^{1/2} \biggl(\frac{r_{T}}{0.06}\biggr)^{1/2}.
\label{H1}
\end{equation}
It is understood that we shall always use $k_{p} =0.002\, \mathrm{Mpc}^{-1}$ as the pivot scale for assigning the scalar and tensor power spectra. The frequency corresponding 
to $k_{p}$ falls in the aHz range and will be used for the low-frequency normalization 
of the spectral energy density:
\begin{equation}
\nu_{p}= k_{p}/(2\pi)= 3.09 \, \biggl( \frac{k_{p}}{0.002\, \, \mathrm{Mpc}^{-1}}\biggr) \,\,\mathrm{aHz}.
\label{nup}
\end{equation} 
 If the lowest scale coincides with the BBN epoch we have that, at least,  $H_{bbn} \simeq 10^{-44} M_{P}$ where, for the sake of concreteness, we considered a putative BBN temperature $T_{bbn} = {\mathcal O}(\mathrm{MeV})$. With these two numerical estimates we have that the expansion rate  
encompasses the following range: 
\begin{equation}
- 44 \leq \log{\biggl(\frac{H}{M_{P}}\biggr)} \leq -5.27 + \frac{1}{2}\log{\bigg(\frac{{\mathcal A}_{{\mathcal R}}}{2.41\times 10^{-9}}\biggr)} + \frac{1}{2}\log{\biggl(\frac{r_{T}}{0.06}\biggr)}.
\label{FF01}
\end{equation}
From Eq. (\ref{FF01}) it follows that radiation may dominate 
{\em before} the BBN epoch so that the minimal $H$, in a given model, might be larger than $H_{bbn}$; however, in spite of the specific scenario, it cannot be smaller. This is, in a nutshell, the logic  
adopted in Eq. (\ref{FF01}) and in the forthcoming discussions.

While  the range deduced in Eq. (\ref{FF01}) does not depend on the intermediate expansion rate, 
$a H/M_{P}$ is sensitive to the different expansion histories. To clarify this statement 
it is sufficient  to consider the ratio
\begin{equation}
\frac{a_{bbn} \,H_{bbn}}{a_{1} \, H_{1}} \leq 10^{-38} \biggl(\frac{a_{bbn}}{a_{1}}\biggr).
\label{FF02}
\end{equation}
Even if the right hand side of Eq. (\ref{FF02}) has been approximately estimated from the 
typical values discussed in Eq. (\ref{FF01}),  still the ratio $(a_{bbn}/a_{1})$ 
 depends on the expansion rate and cannot be deduced in absolute terms. For instance, if the curvature scale evolves like radiation then  $(a_{bbn}/a_{1}) = \sqrt{H_{1}/H_{bbn}}$ which means that, overall, the ratio appearing in Eq. (\ref{FF02}) is of the order of $10^{-19}$.
\begin{figure}[!ht]
\centering
\includegraphics[height=5.5cm]{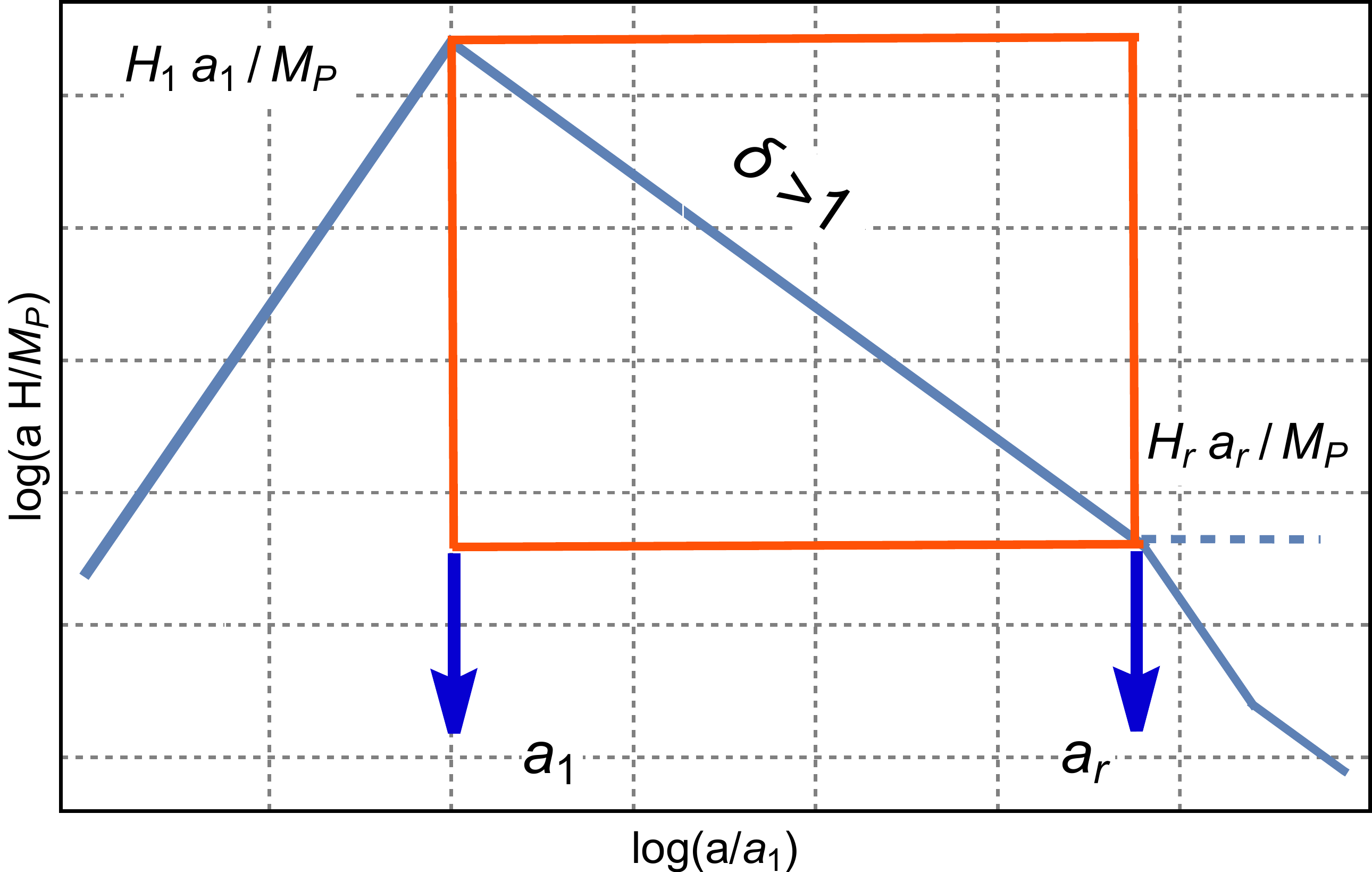}
\includegraphics[height=5.5cm]{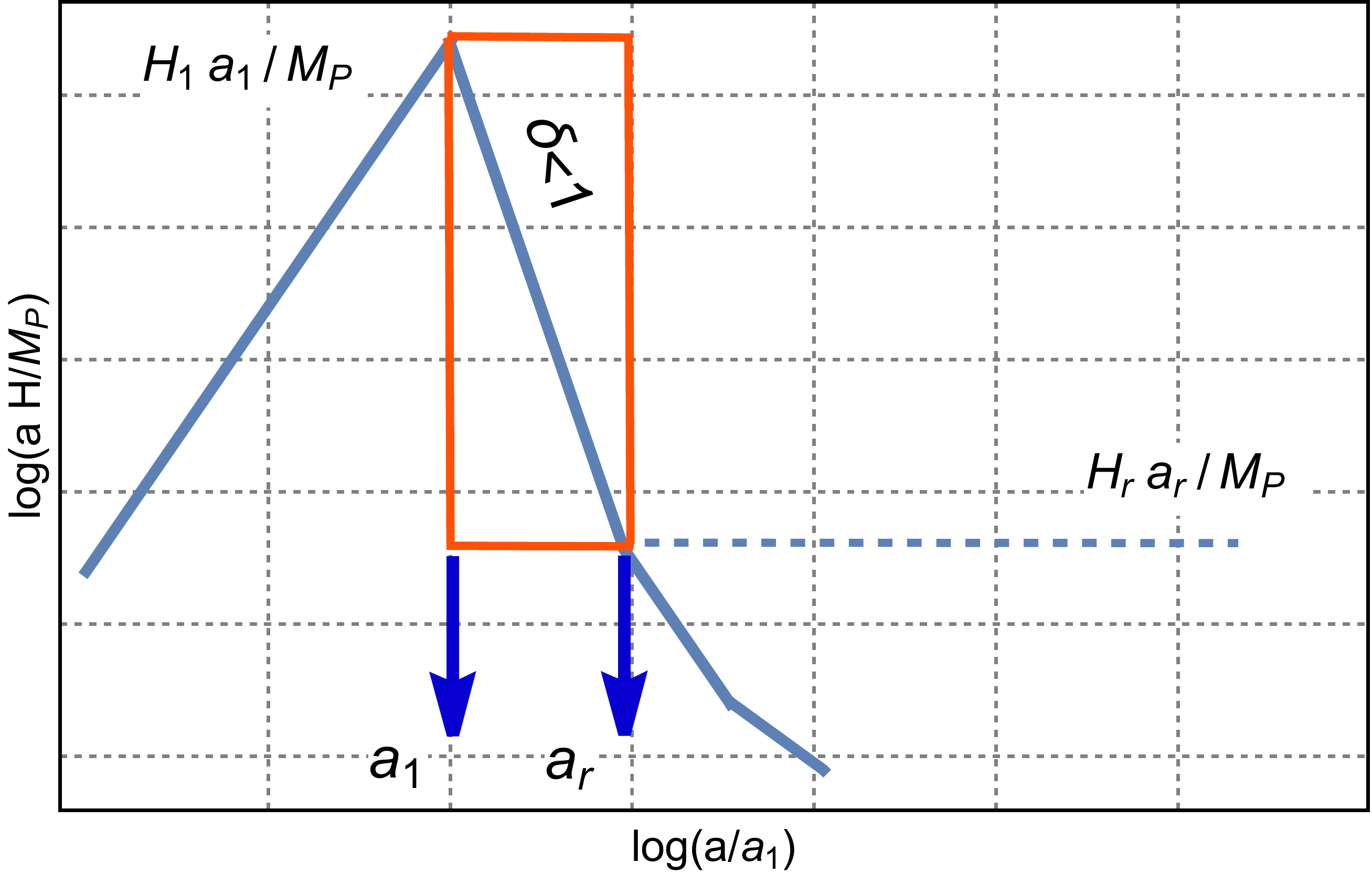}
\caption[a]{The profiles of $a\,H/M_{P}$ are  illustrated in terms 
of the scale factor; common logarithms are employed on both axes. After an inflationary 
stage where $a\,H/M_{P}$ evolves linearly with the scale factor the background decelerates and, in the left panel, the expansion rate is faster than radiation while in the plot at the right it is slower than radiation. The two rectangles approximately define the region where the expansion rate may differ from radiation. A swift comparison  suggests that the same gap in $a\, H$ 
is covered in different redshifts depending on the expansion rate. 
The two profiles also define the pivotal frequencies of the spectrum 
(i.e. $\nu_{max}$ and $\nu_{r}$) that are associated, respectively, with $a_{1}\, H_{1}$ and 
with $a_{r}\, H_{r}$ (see, in this respect, Eqs.(\ref{SING3})--(\ref{SING3a}) and (\ref{SING4})).}
\label{FF4a}      
\end{figure}
If the the expansion rate does not coincide with radiation, as it is suggested in the profiles of Fig. \ref{FF4a},  the ratio  $(a_{bbn}/a_{1})$ can be either larger or smaller 
than $10^{19}$. In Fig. \ref{FF4a} 
the conventional radiation-dominated stage has been replaced by an intermediate stage 
where the background expands at a rate that is either faster or slower 
than radiation. For the sake of concreteness we assume that during  the intermediate 
stage the scale factor evolves, in conformal time, as $a(\tau) \propto \tau^{\delta}$ with $\delta >0$. Within this parametrization the background expands faster than radiation 
when $\delta > 1$ (see the plot at the left in Fig. \ref{FF4a}) while for $0 < \delta < 1$ the expansion is slower than radiation (see the  plot at the right in Fig. \ref{FF4a}). 

For different values of $\delta$ the same gap in the $a\, H$  is covered by different amounts of redshift. For the same reason the maximal number of $e$-folds accessible to the large-scale observations (denoted here by $N_{max}$) is larger when $\delta < 1$ and it is smaller when $\delta > 1$.  The value of $N_{max}$ can be computed by fitting the redshifted inflationary event horizon inside the current Hubble patch, namely by requiring $H_{1}^{-1} (a_{0}/a_{1}) \simeq H_{0}^{-1}$ \cite{HOR}. This condition 
can be made explicit by considering the timeline illustrated in Fig. \ref{FF4a}
and the result, for different values of $\delta$, is:
\begin{equation}
e^{N_{max}} = \biggl(\frac{\pi \, r_{T}}{8}\, {\mathcal A}_{{\mathcal R}}\, \Omega_{R0} \biggr)^{1/4} \, 
\sqrt{\frac{M_{P}}{H_{0}}}\, \xi^{\frac{\delta -1}{2(\delta +1)}},\qquad\qquad \xi= \frac{H_{r}}{H_{1}},
\label{SING1}
\end{equation}
where $\Omega_{R0}$ is the critical fraction of massless species at the present time; we shall assume 
the same value of the concordance paradigm where the only massless species are the photons 
and the neutrinos. Since $\xi$ gives $H_{r}$ in units of $H_{1}$ we can easily argue, by definition, that  $\xi < 1$; consequently  
Eq. (\ref{SING1}) implies that $N_{max}$ increases in the range $0<\delta <1$ and decreases for $\delta >1$. In even more explicit terms Eq. (\ref{SING1}) becomes:
\begin{eqnarray}
N_{max} &=& 61.88 - \ln{\biggl(\frac{h_{0}}{0.7}\biggr)} +  \frac{\delta -1}{2 (\delta + 1)} \ln{\xi} 
\nonumber\\
&+& \frac{1}{4} \ln{\biggl(\frac{r_{T}}{0.06}\biggr)}
+ \frac{1}{4} \ln{\biggl(\frac{{\mathcal A}_{{\mathcal R}}}{2.41\times 10^{-9}}\biggr)} + 
\frac{1}{4} \ln{\biggl(\frac{h_{0}^2 \, \Omega_{R0}}{4.15 \times 10^{-5}}\biggr)}.
\label{SING2}
\end{eqnarray} 
When $0<\delta < 1$  the second term at the right-hand side of Eq. (\ref{SING2}) increases the 
value of $N_{max}$. Conversely when $\delta > 1$ the values of $N_{max}$ are 
smaller than in the radiation case (i.e. for $\delta \to 1$). Since, by definition, $ 2 \pi\,\nu_{max} = a_{1} H_{1}$ and $2 \pi \nu_{r} = a_{r} \, H_{r}$, the value of $N_{max}$ affects the maximal frequency of the spectrum of the relic gravitons (see, e.g. \cite{CC0,BB1,BB2}).
Consequently the value of $\nu_{max}$ can be written as:
\begin{equation}
\nu_{max}= \overline{\nu}_{max}\,\, \, \xi^{ \frac{\delta-1}{2 (\delta + 1)}}, \qquad \delta >0, \qquad \xi< 1,
\label{SING3}
\end{equation}
where $\overline{\nu}_{max}$ corresponds to the maximal frequency of the spectrum 
evaluated in the case of $\delta\to 1$ and it is given by:
\begin{equation}
\overline{\nu}_{max}= 269.33 \,\biggl(\frac{r_{T}}{0.06}\biggr)^{1/4} \,\biggl(\frac{{\mathcal A}_{{\mathcal R}}}{2.41\times 10^{-9}}\biggr)^{1/4} \, \biggl(\frac{h_{0}^2 \, \Omega_{R0}}{4.15 \times 10^{-5}}\biggr)^{1/4} \,\,\,\mathrm{MHz}.
\label{SING3a}
\end{equation}
According to Eqs. (\ref{SING3})--(\ref{SING3a}), for a fixed value of $\xi < 1$, the maximal frequency is therefore modified depending on the value of $\delta$. In particular, when
the Universe expands faster than radiation (i.e. $\delta > 1$) we have  $\nu_{max} < \overline{\nu}_{max}$; conversely, if the Universe expands at a rate which is slower than radiation (i.e. $\delta < 1$) Eq. (\ref{SING3}) requires that $\nu> \overline{\nu}_{max}$. While $\nu_{max}$ depends both on the 
expansion rate and on the overall duration of the intermediate phase, $\nu_{r}$ only depends 
upon $\xi$ (and {\em not} on $\delta$):
\begin{equation}
\nu_{r}= 269.33 \,\biggl(\frac{r_{T}}{0.06}\biggr)^{1/4} \,\biggl(\frac{{\mathcal A}_{{\mathcal R}}}{2.41\times 10^{-9}}\biggr)^{1/4} \, \biggl(\frac{h_{0}^2 \, \Omega_{R0}}{4.15 \times 10^{-5}}\biggr)^{1/4} \,\sqrt{\xi} \,\,\,\mathrm{MHz}.
\label{SING4}
\end{equation}
By taking the ratio between $\nu_{r}$ and $\nu_{max}$ we can note that $ (\nu_{r}/\nu_{max} ) = \xi^{1/(\delta +1)}$ which implies, as expected, that $ \nu_{r} < \nu_{max}$ for any value of $\xi < 1$ and $\delta > 0$.

\subsection{The high-frequency and the low-frequency slopes}
While the overall normalization of the spectral energy density must take into account various sources of suppression that are customarily included numerically \cite{STRESSNU1,STRESSNU2,STRESSNU3} (see also \cite{CC0,transfer1,transfer2}), the frequency slopes of the spectral energy density have definite analytic expressions that follow from the profiles of Fig. \ref{FF4a} and from 
the corresponding evolution of the mode functions. The same approach, with minor differences, leads to the spectral slopes associated with the profiles that are analyzed in the forthcoming sections. Since in the leftmost part of the plots of Fig. \ref{FF4a} the background inflates
the mode functions must be determined for $a<a_{1}$ (i.e. $\tau< -\tau_{1}$)
and they coincide with the result of Eq. (\ref{INCOND1}). For $\tau \geq - \tau_{1}$ $f_{k}(\tau)$ and $g_{k}(\tau)$ are fixed, by continuity, in terms of their values for $\tau = -\tau_{1}$:
\begin{equation}
\left(\matrix{ f_{k}(\tau) &\cr
g_{k}(\tau)/k&\cr}\right) = \left(\matrix{ A_{f\, f}(k, \tau, \tau_{1})
& A_{f\,g}(k,\tau, \tau_{1})&\cr
A_{g\,f}(k,\tau, \tau_{1}) &A_{g\,g}(k,\tau, \tau_{1})&\cr}\right) \left(\matrix{ \overline{f}_{k} &\cr
\overline{g}_{k}/k&\cr}\right),
\label{SING4a}
\end{equation}
where, for the sake of conciseness, we defined $\overline{f}_{k}= f_{k}(-\tau_{1})$ and $\overline{g}_{k}= g_{k}(-\tau_{1})$ as the values of the mode functions for $\tau = - \tau_{1}$.
The determinant of the matrix appearing in Eq. (\ref{SING4a}) 
must be equal to $1$ and this follows from the Wronskian normalization obeyed 
by the mode functions.  Even if $(\overline{f}_{k},\,\, \overline{g}_{k})$ are complex quantities 
the entries of the matrix appearing at the right hand side of Eq. (\ref{SING4a}) are all real and they are 
given as products of Bessel functions of first and second kind:
\begin{eqnarray}
A_{f\, f}(k, \tau, \tau_{1}) &=& \frac{\pi}{2} \sqrt{q x_{1}} \sqrt{ k y} \biggl[ J_{\nu+1}( q x_{1}) Y_{\nu}(k y) - Y_{\nu+1}(q x_{1}) J_{\nu}(k y) \biggr],
\nonumber\\
A_{f\, g}(k, \tau, \tau_{1}) &=& \frac{\pi}{2} \sqrt{q x_{1}} \sqrt{ k y} \biggl[ J_{\nu}( q x_{1}) Y_{\nu}(k y) - Y_{\nu}(q x_{1}) J_{\nu}(k y) \biggr],
\nonumber\\
A_{g\, f}(k, \tau, \tau_{1}) &=& \frac{\pi}{2} \sqrt{q x_{1}} \sqrt{ k y} \biggl[ Y_{\nu+1}( q x_{1}) J_{\nu +1}(k y) - J_{\nu+1}(q x_{1}) Y_{\nu+1}(k y) \biggr],
\nonumber\\
A_{f\, g}(k, \tau, \tau_{1}) &=& \frac{\pi}{2} \sqrt{q x_{1}} \sqrt{ k y} \biggl[ Y_{\nu}( q x_{1}) J_{\nu+1}(k y) - Y_{\nu+1}(k y) J_{\nu}(q x_1) \biggr].
\label{SING4b}
\end{eqnarray}
In Eq. (\ref{SING4b}), using the standard notations 
$J_{\alpha}(z)$ and $Y_{\alpha}(z)$ are the Bessel functions of  index $\alpha$
and argument $z$ \cite{abr1,abr2}. It is relevant to appreciate that arguments of the Bessel's functions 
appearing in each product of Eq. (\ref{SING4b}) are different when $\tau\neq -\tau_{1}$ but they coincide 
when $ \tau \to -\tau_{1}$.  For this reason the variables $y=y(\tau, q)$, $q= q(\epsilon,\delta)$ and $\nu= \nu(\delta)$ that appear in Eq. (\ref{SING4b}) are defined, respectively, as:
\begin{equation}
y = y(\tau,q) = \tau + \tau_{1} \biggl( 1 + q\biggr), \qquad
q =q(\epsilon,\delta)= \delta\, (1 -\epsilon), \qquad  \nu = \nu(\delta) = \delta - 1/2,
\label{SING4c}
\end{equation}
where we assumed, for simplicity\footnote{The dependence on the various arguments has been explicitly indicated even if it will be dropped hereunder to maintain a concise notation. In the case $\delta<1/2$ Eq. (\ref{SING4b}) has a slightly different analytic expression leading, at the very end, 
to the same overall expression of the high-frequency slope. }, $\delta \geq 1/2$. The important point 
for the following discussions is that different contributions appearing in 
Eq. (\ref{SING4b}) are not of the same order. In the expression of $f_{k}(\tau)$ 
the term containing $A_{f\,f}(k,\tau,\tau_{1})$ dominates against the other in the physical limit $x_{1} = k \tau_{1} \ll 1$ which is always verified when the spectral energy density is estimated after the various wavelengths reenter the Hubble radius; in practice this limit  only amounts to consider all the frequencies smaller than $\nu_{max}$. If the result of Eq. (\ref{SING4}) is inserted into 
Eqs. (\ref{ENDENS2})--(\ref{ENDENS3}) we obtain a general expression 
of $\Omega_{gw}(k,\tau)$ that can be further simplified in the limit 
$k \tau_{1} \ll 1$. The 
 spectral energy density that corresponds to the wavelengths reentering during the $\delta$-phase is approximately given by:
\begin{eqnarray}
\Omega_{gw}(k,\tau) &=& {\mathcal B}(\delta, \epsilon) \biggl(\frac{H_{1}}{M_{P}}\biggr)^2 \biggl(\frac{a_{1}^2 H_{1}}{a^2\, H}\biggr)^2 \biggl( \frac{k}{a_{1} \, H_{1}}\biggr)^{n_{T}}, 
\label{SING4d}
 \end{eqnarray}
 where ${\mathcal B}(\delta, \epsilon)$ is a numerical factor which can be accurately 
computed but it is not essential for the determination of the slope. Note that in Eq. (\ref{SING4d}) we restored $M_{P}$ by recalling its relation with $\overline{M}_{P}$ given in Eq. (\ref{notations}). The spectral index $n_{T}$ that appears in Eq. (\ref{SING4d}) determines the high-frequency slope of the spectral energy density.  If the consistency relations are enforced the slow-roll parameter can be traded for the tensor-to-scalar ratio $r_{T}$ so that the high-frequency slope is ultimately given by:
\begin{equation}
n_{T}(\delta,\,r_{T}) = \frac{32 - 4 r_{T}}{16 - r_{T}} - 2 \delta.
\label{SING5}
\end{equation}
Equation (\ref{SING5}) implies that the high-frequency spectral slope is increasing 
when the post-inflationary expansion rate is slower than radiation and it is decreasing when 
the expansion rate is faster than radiation:
\begin{eqnarray}
&& n_{T}>0 \qquad \mathrm{for}\qquad \delta < 1 - \frac{r_{T}}{16} + {\mathcal O}(r_{T}^2),
\nonumber\\
&& n_{T}<0 \qquad \mathrm{for}\qquad \delta > 1 - \frac{r_{T}}{16} + {\mathcal O}(r_{T}^2).
\label{SING5aa}
\end{eqnarray}
The same analysis leading to Eq. (\ref{SING5}) determines the conventional low-frequency slope which is applicable for the frequencies $\nu< \nu_{r} \simeq a_{r} \, H_{r} $ and which is given by 
Eq. (\ref{SING5}) evaluated in the limit $\delta \to 1$: 
\begin{equation}
\overline{n}_{T} = \lim_{\delta\to 1} \, n_{T}(\delta,\,r_{T}) = - \frac{2 \, r_{T}}{16 - r_{T}} = - \frac{r_{T}}{8}  + {\mathcal O}(r_{T}^2).
\label{SING5a}
\end{equation}
Equation (\ref{SING5a}) corresponds, as expected, to the slope of the
spectral energy density obtained for the transition between a conventional 
inflationary stage of expansion and a radiation phase \cite{AA2,AA3}.
Note finally that, thanks to the consistency relations, $r_{T} \simeq 16 \, \epsilon$ 
so that the result of Eq. (\ref{SING5a}) also implies that $\overline{n}_{T} = - \,2 \, \epsilon$. 

\subsection{The shapes of the spectra and their phenomenological signatures}
\begin{figure}[!ht]
\centering
\includegraphics[height=5.5cm]{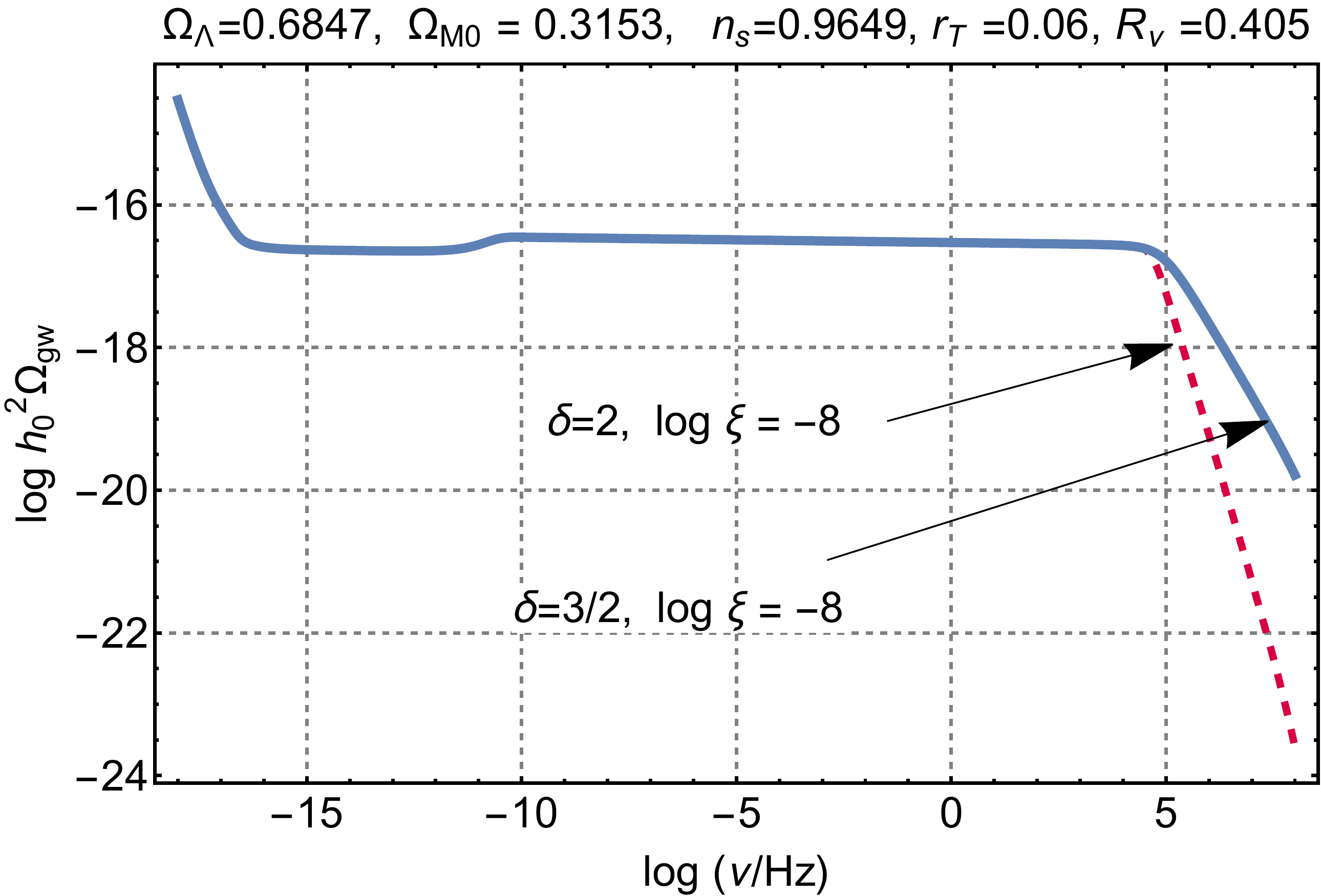}
\includegraphics[height=5.5cm]{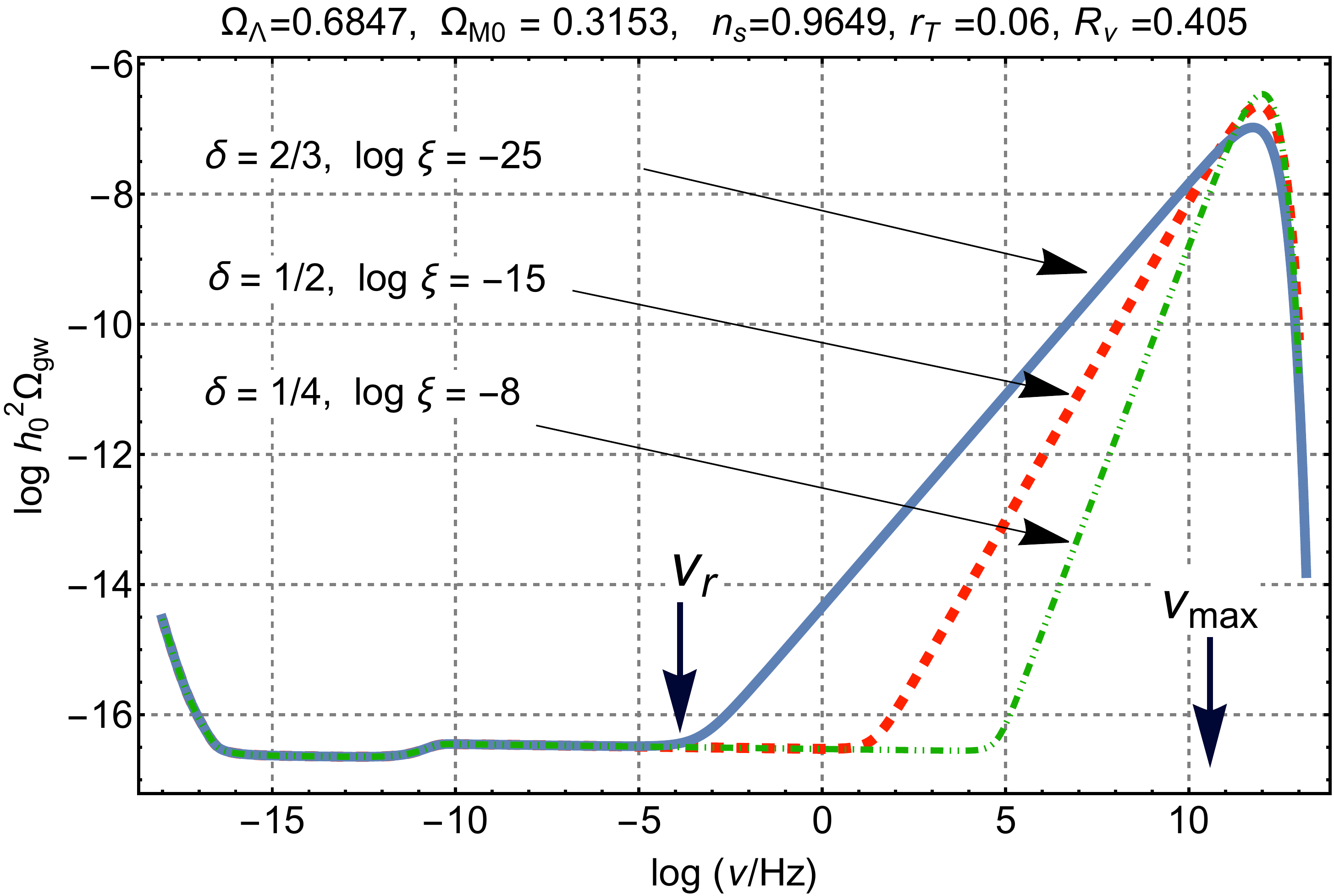}
\caption[a]{The common logarithm of $h_{0}^2 \Omega_{gw}(\nu,\tau_{0})$ is illustrated as a function 
of the common logarithm of the frequency in the cases $\delta >1$ (panel at the left) and $\delta < 1$ (panel at the right). The selected parameters correspond to the last Planck release supplemented by the 
more constraining bounds on $r_{T}$ obtained later on \cite{RT1,RT2,RT3}. }
\label{FF6a}      
\end{figure}

From the simultaneous analysis of Fig. \ref{FF4a} and of the high-frequency slope 
given in Eq. (\ref{SING5})--(\ref{SING5aa}) we see that for  $\nu> \nu_{r}$ 
the spectral energy density decreases when the expansion is faster than radiation 
(i.e. in the case $\delta > 1$). The low-frequency slope $\overline{n}_{T}$ of Eq. (\ref{SING5a}) is always 
decreasing since it corresponds to wavelengths leaving the Hubble radius during inflation and reentering 
when the background is already dominated by radiation. Consequently the spectral 
energy density is maximal in the aHz region and it has been illustrated in the left panel of Fig. \ref{FF6a}.
For the parameters of the plot at the left in Fig. \ref{FF6a} the frequency $\nu_{r}$ is of the order of $20$ kHz as it follows from Eq. (\ref{SING4}) 
when $\xi= 10^{-8}$. Smaller values of $\xi$ reduce both $\nu_{r}$ and $\nu_{max}$ as we can 
appreciate from Eqs. (\ref{SING3})--(\ref{SING3a}) and (\ref{SING4}) in the case $\delta > 1$. 
What really matters, however, is that  below $\nu_{r}$ the different curves coincide 
and the spectral energy density is the one deduced in the concordance 
paradigm for the wavelengths exiting the Hubble radius during 
inflation and reentering in the radiation stage\footnote{Given a specific value of $\xi$, it is important 
to appreciate that $\nu_{r}$ is the same, in spite of the range of $\delta$.}. 

 Recalling the typical frequency of Eq. (\ref{CC3}), the  neutrino free-streaming suppresses $h_{0}^2\,\Omega_{gw}(\nu,\tau_{0})$  for   $\nu < \nu_{bbn}$ \cite{STRESSNU1,STRESSNU2} (see also \cite{STRESSNU3,STRESSNU4,STRESSNU5}). Other sources of suppression taken into account in Fig. \ref{FF6a} and in the remaining plots include the late-time dominance of dark energy and the evolution of relativistic species (see e.g. \cite{CC0} for a review). The spectra of Fig. \ref{FF6a} have been deduced by using 
for the fiducial parameters the last Planck data release and the simplest possibility 
has been considered namely the case of three massless neutrinos where 
$R_{\nu} = \rho_{\nu}/(\rho_{\gamma} + \rho_{\nu}) =0.405$, as indicated on top of each plots.  All in all, when $\delta > 1$ there are in practice no further constraints besides the low-frequency limits that translate into the upper bound on $r_{T}$ \cite{RT1,RT2,RT3}.  
 
So far we discussed the case where the post-inflationary expansion rate is faster than radiation.
In the complementary interval $0< \delta < 1$, the expansion rate is slower than radiation and Eq. (\ref{SING5a}) implies that the high-frequency slope is increasing while the low-frequency slope decreases. In the right plot of Fig. \ref{FF6a} the corresponding spectral energy density is illustrated and since its slope increases for  $\nu > \nu_{r}$ the BBN of Eqs. (\ref{CC2})--(\ref{CC3}) are not always satisfied. Conversely, when the BBN constraint is enforced (as it happens for the parameters 
selected in Fig. \ref{FF6a}) $h_{0}^2 \, \Omega_{gw}(\nu,\tau_{0})$  is always much smaller than the potential evidences of the PTA collaborations and of the KLV bounds discussed, respectively, in Eqs. (\ref{PTAb1}) and (\ref{CONS2})--(\ref{NOT1}). To clarify this aspect it is useful to note  that $h_{0}^2 \, \Omega_{gw}(\nu,\tau_{0})$ can be written, with compact notations, as:
\begin{equation}
h_{0}^2 \, \Omega_{gw}(\nu, \tau_{0}) = {\mathcal N}_{\rho} \, r_{T}(\nu_{p}) \biggl(\frac{\nu}{\nu_{p}}\biggr)^{\overline{n}_{T}} \, \, {\mathcal T}^2_{low}(\nu/\nu_{eq}) \,  {\mathcal T}^2_{high}(\nu/\nu_{r}, \delta), \qquad {\mathcal N}_{\rho} = 4.165 \times 10^{-15} \biggl(\frac{h_{0}^2\,\Omega_{R0}}{4.15\times 10^{-5}}\biggr),
\label{SING6}
\end{equation}
where $r_{T}(\nu_{p})$  depends on $\nu_{p}$ that has been already introduced in Eq. (\ref{nup}); the equality frequency $\nu_{eq}$ is instead defined by:
\begin{equation}
\nu_{eq} =  \frac{k_{\mathrm{eq}}}{2 \pi} = 1.597\times 10^{-17} \biggl(\frac{h_{0}^2\,\Omega_{M0}}{0.1411}\biggr) \biggl(\frac{h_{0}^2\,\Omega_{R0}}{4.15 \times 10^{-5}}\biggr)^{-1/2}\,\, \mathrm{Hz},
\label{SING8}
\end{equation}
and $k_{eq} = 0.0732\, h_{0}^2\,\Omega_{M0}\, \mathrm{Mpc}^{-1}$ (as usual, $\Omega_{M0}$ is the present fraction in dusty matter). In Eq. (\ref{SING6}) ${\mathcal T}^2_{low}(\nu/\nu_{eq})$ and ${\mathcal T}^2_{high}(\nu/\nu_{r}, \delta)$ are the transfer functions directly computed for the spectral energy 
density \cite{transfer1,transfer2}. The low-frequency transfer 
function ${\mathcal T}_{low}(\nu/\nu_{eq})$ has a definite form that can be written as:
\begin{equation}
{\mathcal T}_{low}(\nu, \nu_{eq}) = \sqrt{1 + c_{2}\biggl(\frac{\nu_{eq}}{\nu}\biggr) + b_{2}\biggl(\frac{\nu_{eq}}{\nu}\biggr)^2},\qquad c_{eq}= 0.5238, \qquad b_{eq}=0.3537.
\label{SING7}
\end{equation}
The transfer function for the spectral 
energy density does not coincide with the transfer function computed for the 
spectral amplitude \cite{transfer1,transfer2};  it is obtained by integrating numerically the mode functions across the radiation-matter transition for each $k$-mode
and by computing $\Omega_{gw}(\nu,\tau)$ for different frequencies. The advantage of the transfer function for the energy density is that while $\Omega_{gw}(\nu,\tau)$ is a mildly oscillating function of $k \tau$, the spectral amplitude exhibits much larger oscillations that need to be averaged, as originally suggested in \cite{transfer3a,transfer3b}.  Unlike 
${\mathcal T}_{low}(\nu/\nu_{eq})$, the high-frequency transfer function ${\mathcal T}_{high}(\nu/\nu_{r}, \delta)$ depends on the value of $\delta$ so that it does not have a general form. 

Since for $\nu> \nu_{r}$ the high-energy transfer function has the slope $n_{T}$ (i.e. ${\mathcal T}_{high}^{2} \to (\nu/\nu_{r})^{n_{T}}$) for the analytic estimates of the limits imposed on the spectral energy 
density we can express $h_{0}^2\,\Omega_{gw}(\nu,\tau_{0})$ 
in the following approximate form: 
\begin{equation}
h_{0}^2 \, \Omega_{gw}(\nu, \tau_{0})  = {\mathcal N}_{\rho} \, r_{T} \biggl(\frac{\nu}{\nu_{p}}\biggr)^{\overline{n}_{T}} \, \, {\mathcal T}^2_{low}(\nu_{r}/\nu_{eq}) \, \biggl(\frac{\nu}{\nu_{r}}\biggr)^{n_{T}}, \qquad \qquad \nu_{r} \leq \nu \leq \nu_{max}. 
\label{SING9}
\end{equation}
Equation (\ref{SING9}) rests on the observation that ${\mathcal T}_{low}(\nu_{r}/\nu_{eq}) \to 1$ for $\nu \geq \nu_{r}$;
 in the same limit it is also true that $\overline{n}_{T} \ll 1$. In this situation the prefactor  is practically frequency-independent so that we can write:
\begin{equation}
h_{0}^2 \, \Omega_{gw}(\nu, \tau_{0})  = \overline{{\mathcal N}}_{\rho}(r_{T}, \nu)  \biggl(\frac{\nu}{\nu_{r}}\biggr)^{n_{T}}, \qquad\qquad \nu > \nu_{r},
\label{SING9a}
\end{equation}
where $\overline{{\mathcal N}}_{\rho}(r_{T}, \nu)$ is defined as
\begin{equation}
 \overline{{\mathcal N}}_{\rho}(r_{T}, \nu) = {\mathcal N}_{\rho} \, r_{T} \biggl(\frac{\nu}{\nu_{p}}\biggr)^{\overline{n}_{T}} \, \, {\mathcal T}^2_{low}(\nu_{r}/\nu_{eq}), \qquad \qquad \frac{ d \ln{\overline{{\mathcal N}}_{\rho}}}{ d \ln{\nu}} = 
 - \frac{r_{T}}{8} \ll 1.
 \label{SING9b}
 \end{equation}
Even though the prefactor $\overline{{\mathcal N}}_{\rho}(r_{T}, \nu)$ has a mild frequency dependence coming from neutrino free-streaming, for simplified analytic estimates this 
dependence can be however ignored, at least in the first approximation. 
Along this perspective we can estimate $\overline{{\mathcal N}}_{\rho} = {\mathcal O}(10^{-16.5})$.
Since the prefactor is mildly sensitive to $r_{T}$, it is understood that its numerical value 
corresponds to $r_{T} =0.06$. 

 If $0 < \delta < 1$   the spectral energy density develops a maximum\footnote{This is, incidentally,  the situation of the stiff models where a post-inflationary phase expanding slower than radiation naturally arises \cite{ST1,ST2,ST3,ST3a} and implies a maximum for ${\mathcal O}(100)$ GHz (see also \cite{ST4,ST5,ST6}). In this respect it is appropriate to remark that when the rate
 is slower than radiation (i.e. $0< \delta < 1$) there can be various model-dependent constraints that introduce further limits on $\xi$ and $\nu_{r}$ that have been thoroughly 
 examined in the past \cite{ST1,ST2,ST3,ST3a,ST4,ST5,ST6} and have ben originally introduced by Ford \cite{FORD1} (see also \cite{transfer1,transfer2} for a numerical analysis of the constraints and of the related transfer functions at high-frequencies). } that may fall even above the GHz (see also the right plot in Fig. \ref{FF6a}).  In terms of Eqs. (\ref{SING9a})--(\ref{SING9b}) the BBN constraint discussed in Eqs. (\ref{CC2})--(\ref{CC3}) assumes a particularly simple analytical 
 form. Indeed, since the largest contribution to the integral of 
Eq. (\ref{CC2}) comes from typical frequencies ${\mathcal O}(\nu_{max})$, Eqs. (\ref{SING9a})--(\ref{SING9b}) can be used to set a limit on the integral by requiring $h_{0}^2\,\Omega_{gw}(\nu_{max}, \tau_{0}) < 10^{-6}$; this requirement implies:
\begin{equation}
\log{\xi} > \frac{(1+ \delta) (16 - r_{T})}{2[ 16 ( 1 - \delta) - r_{T} ( 2 - \delta)]} \bigl[ 6 + \log{\overline{{\mathcal N}}_{\rho}(r_{T})}\bigr],
\label{SING12a}
\end{equation}
where we used that $(\nu_{max}/\nu_{r}) \propto \xi^{- 1/(\delta +1)}$ (see Eqs. (\ref{SING3})--(\ref{SING3a}) and (\ref{SING4})). In Eq. (\ref{SING12a}) it is always true that $r_{T} \ll \delta$ 
so that Eq. (\ref{SING12a}) translates into
$\log{\xi} > -5.25 (1 + \delta)/(1 -\delta)$ (where we took $r_{T}=0.06$ and consequently estimated $\log{\overline{{\mathcal N}}_{\rho}} = - 16.5$). Equations (\ref{CC2}) and (\ref{SING12a}) imply then a lower bound on $\xi$. For instance 
if $\delta= 1/2$ we would have that $\log{\xi} > -15.75$. As $\delta$ decreases below $1/2$ the lower bound on $\xi$ gets larger; so for instance for $\delta=1/3$ we will have that the bound will be $\xi > 10^{-10.5}$, and so on\footnote{ Note that a further lower bound on $\xi$ is obtained by requiring that $\nu_{r} > \nu_{bbn}$; but in this case the bound is much less restrictive and it only demands $\xi > 10^{-38}$.}.
\begin{figure}[!ht]
\centering
\includegraphics[height=7cm]{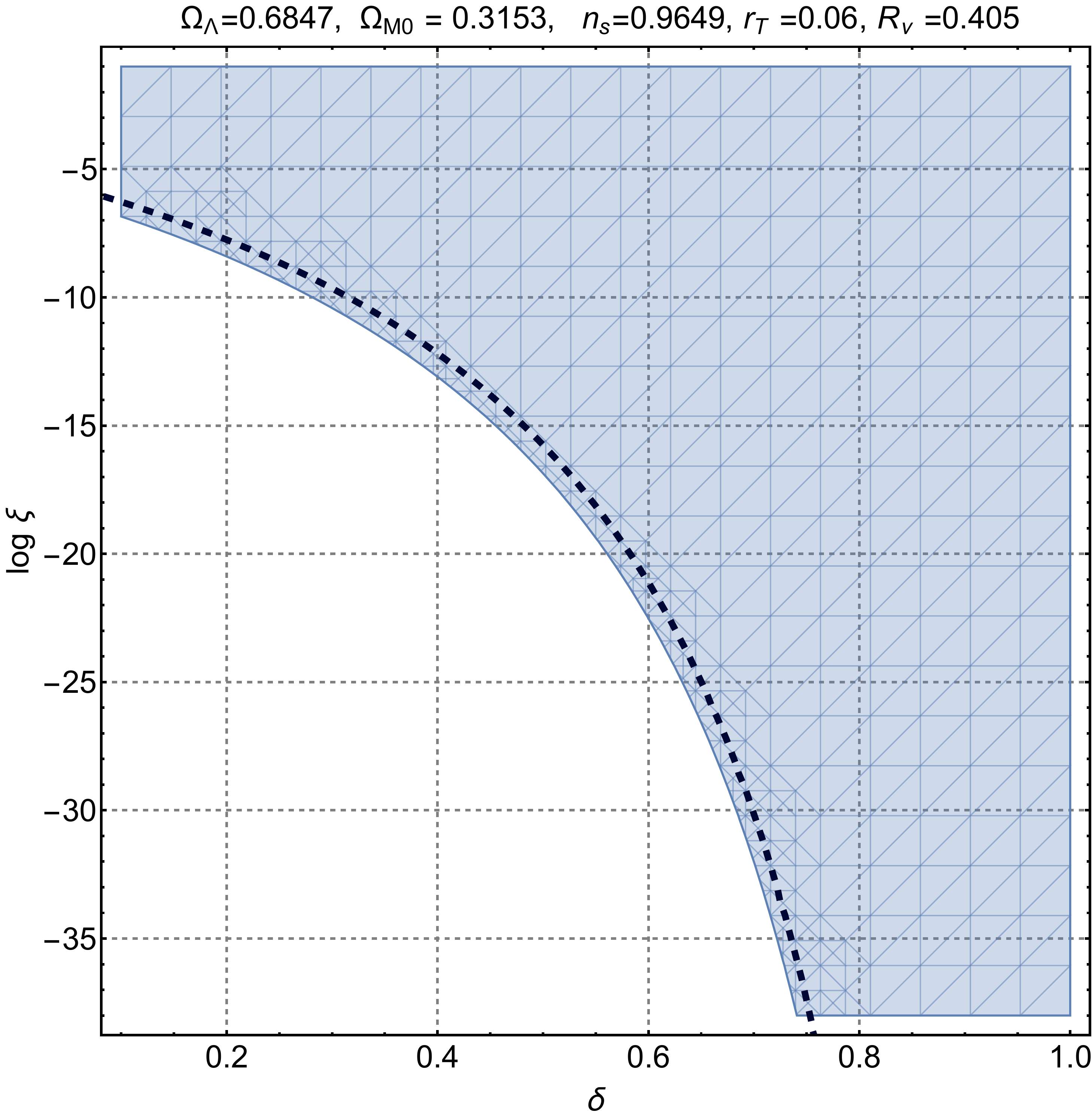}
\includegraphics[height=7cm]{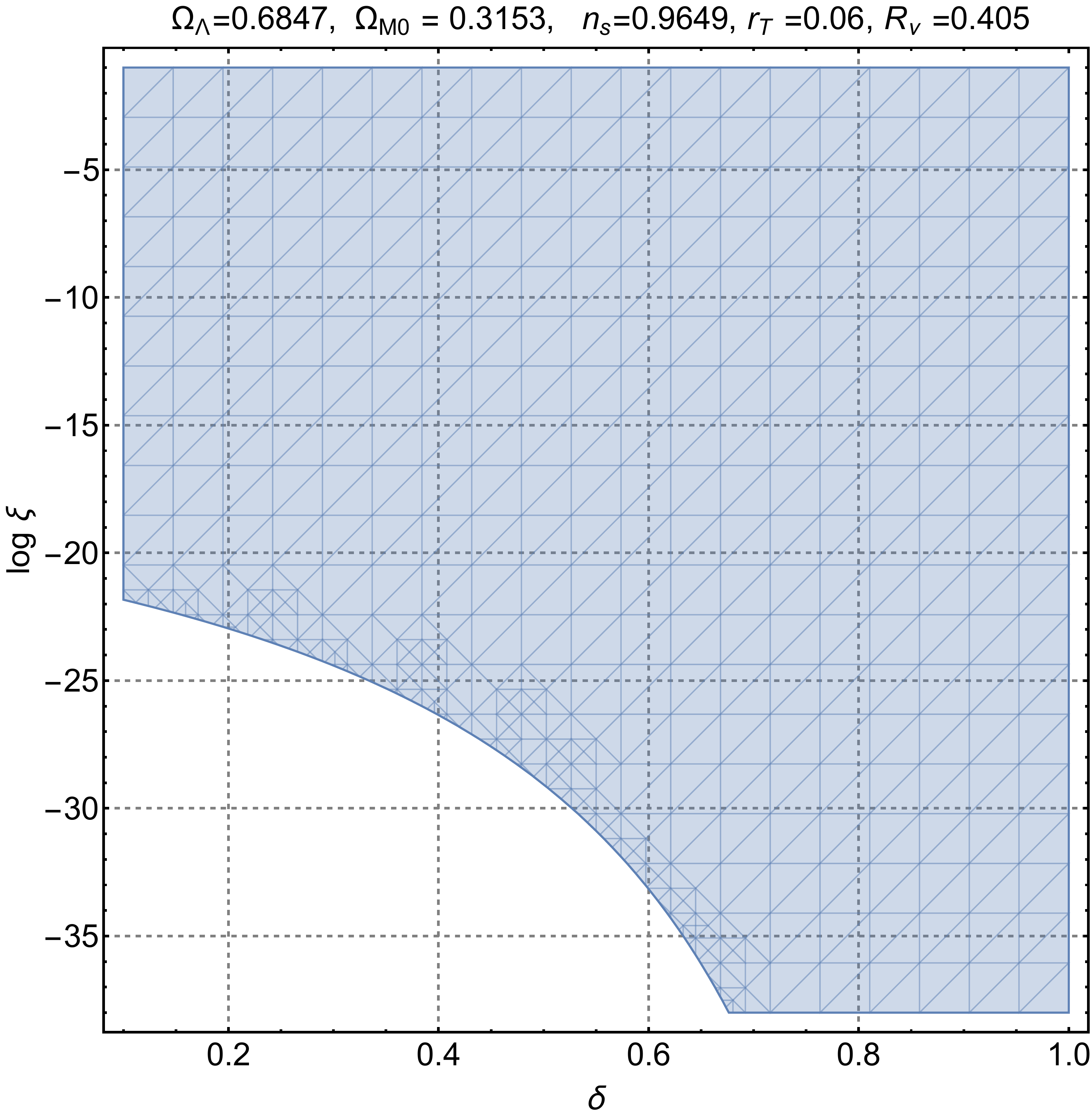}
\caption[a]{In the plot at the left the BBN bound of Eqs. (\ref{CC2})--(\ref{CC3}) 
is illustrated in the $(\xi, \delta)$ plane and for $0< \delta <1$. The dashed line corresponds to the analytic approximation 
of Eq. (\ref{SING12a}). The shaded area corresponds to the allowed region of the $(\delta, \, \xi)$ plane where the BBN bound is enforced. In the plot at the right we illustrate the KLV bound of Eqs.(\ref{CONS2}) and (\ref{NOT1})--(\ref{NOT2}). By comparing the two allowed regions 
the BBN bound turns out to be the most constraining.}
\label{FF6ab}      
\end{figure}
The bound obtained from Eqs. (\ref{SING9a})--(\ref{SING9b}) and (\ref{SING12a}) can be checked by evaluating the BBN limit 
in numerical terms. For this purpose in the left plot of Fig. \ref{FF6ab} 
we illustrated, with the shaded region, the BBN constraint directly computed from 
Eq. (\ref{CC2}) without appealing to the approximations of Eqs. (\ref{SING9a})--(\ref{SING9b}).
In the same plot the dashed curve illustrates the analytic bound obtained in Eq. (\ref{SING12a})
for the case $r_{T} \to 0.06$. The two determinations compare quite well and 
corroborate the approximation schemes of Eqs. (\ref{SING9a})--(\ref{SING9b}). 
In the right plot of Fig. \ref{FF6ab} we plotted the KLV bound of Eqs.(\ref{CONS2}) and (\ref{NOT1})--(\ref{NOT2}). As in the left plot the shaded area represents the allowed region. 
By comparing the left and the right plots we therefore conclude, as anticipated 
that the BBN bound is still more constraining than the KLV limit.

All in all if the post-inflationary evolution is dominated by a single phase expanding faster than radiation (i.e. $\delta > 1$) the spectral energy density is suppressed at high-frequencies and the only physical constraints come from the aHz  region, as in the conventional situation where $\delta \to 1$. Conversely if the expansion is slower than radiation (i.e. $0< \delta < 1$) $h_{0}^2\,\Omega_{gw}(\nu,\tau_{0})$ has a maximum in the high-frequency regime and, in this case,  the relevant limit comes from the abundance of massless species at the BBN epoch. While this limit forbids an arbitrary long duration of the post-inflationary evolution, in both cases the PTA constraints and the Kagra-Ligo-Virgo bounds are always well satisfied. If we however interpret the evidence suggested by the PTA collaborations as a genuine signal potentially coming from the relic gravitons we must conclude that a single post-inflationary stage does not lead to a sufficiently large spectral energy density in the PTA window.

\renewcommand{\theequation}{4.\arabic{equation}}
\setcounter{equation}{0}
\section{Multiple post-inflationary phases and their spectra}
\label{sec4}
In the case of a single post inflationary phase the spectral energy density only depends upon he expansion rate 
and upon the duration of the intermediate stage. Even if, for multiple phases, the number of parameters 
increases the spectral energy density at intermediate frequencies is still constrained.
\subsection{Evolution of $a\, H/M_{P}$ in the case of multiple post-inflationary phases}
 \begin{figure}[!ht]
\centering
\includegraphics[height=5.6cm]{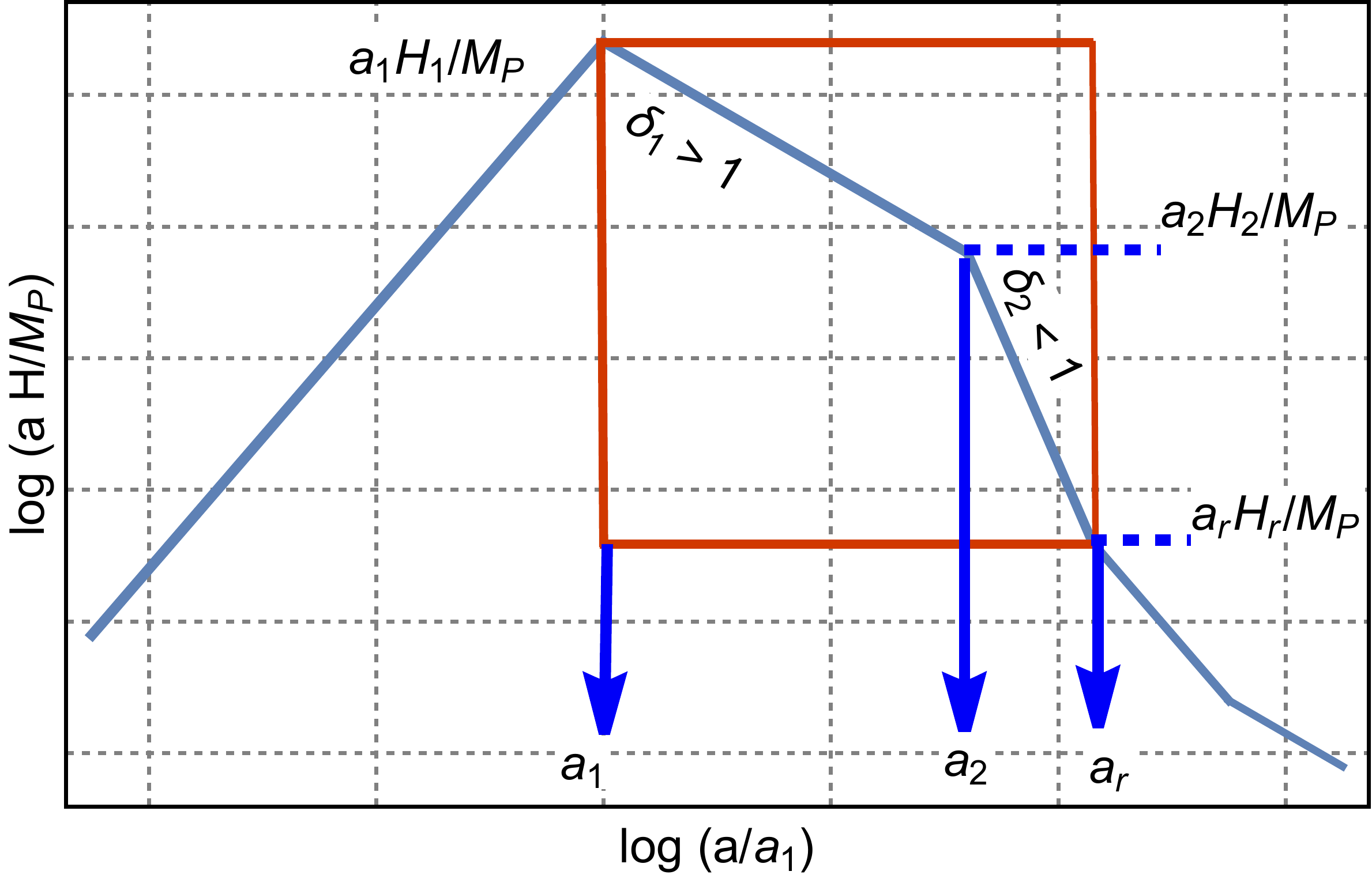}
\includegraphics[height=5.6cm]{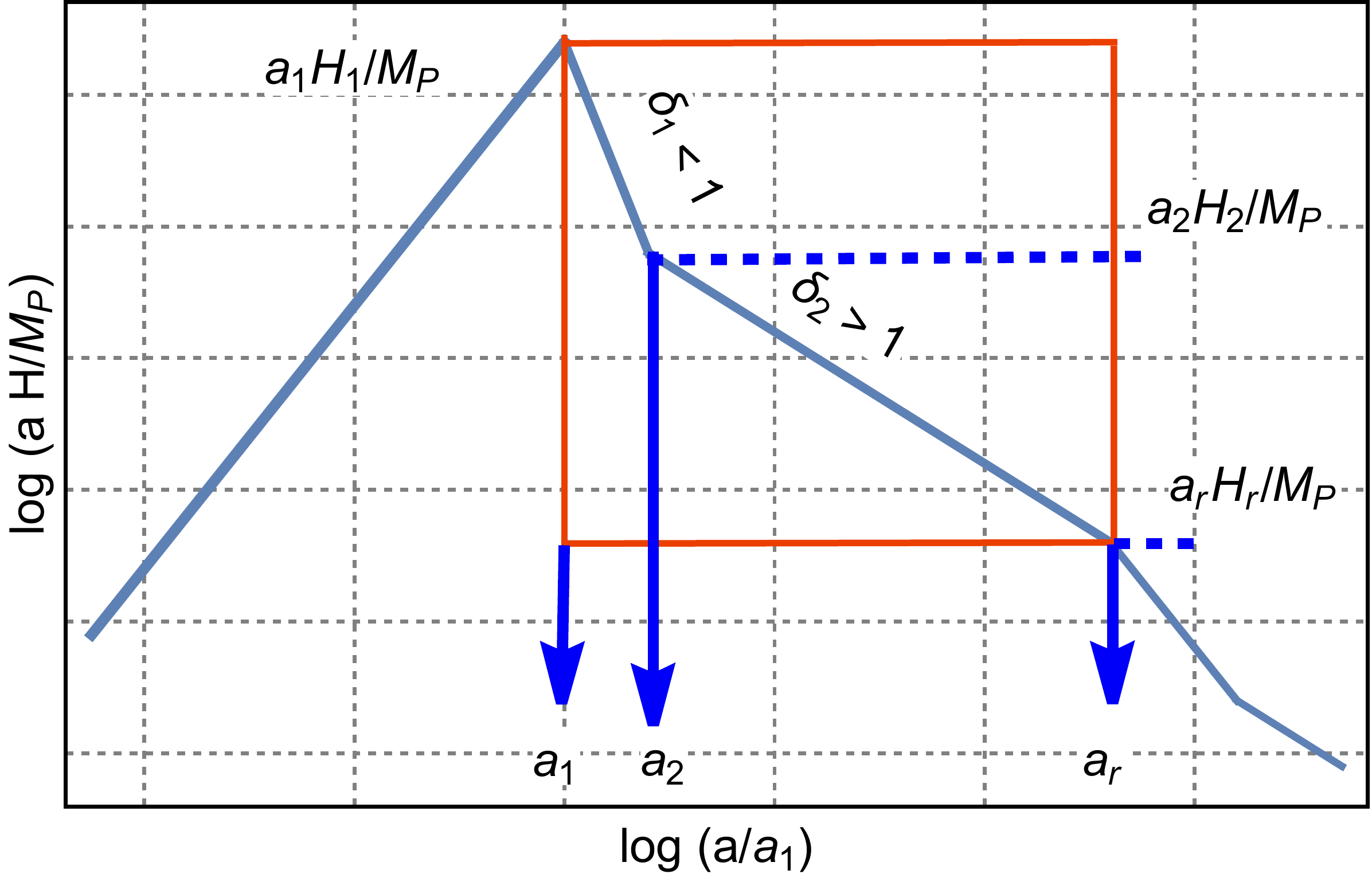}
\caption[a]{ When $\delta_{1} >1 $ and $\delta_{2} < 1$ the two successive phases expand, respectively, faster and slower than radiation (see the left panel). In the plot at the right the hierarchy 
is inverted and $\delta_{1} < 1$ while $\delta_{2} > 1$. In both panels $\delta_{1}>0$ and $\delta_{2}>0$ have 
exactly the same meaning of $\delta$ appearing in Fig. \ref{FF4a} with the difference that there are now two 
post-inflationary stages.}
\label{FF7a}      
\end{figure}
\begin{figure}[!ht]
\centering
\includegraphics[height=5.5cm]{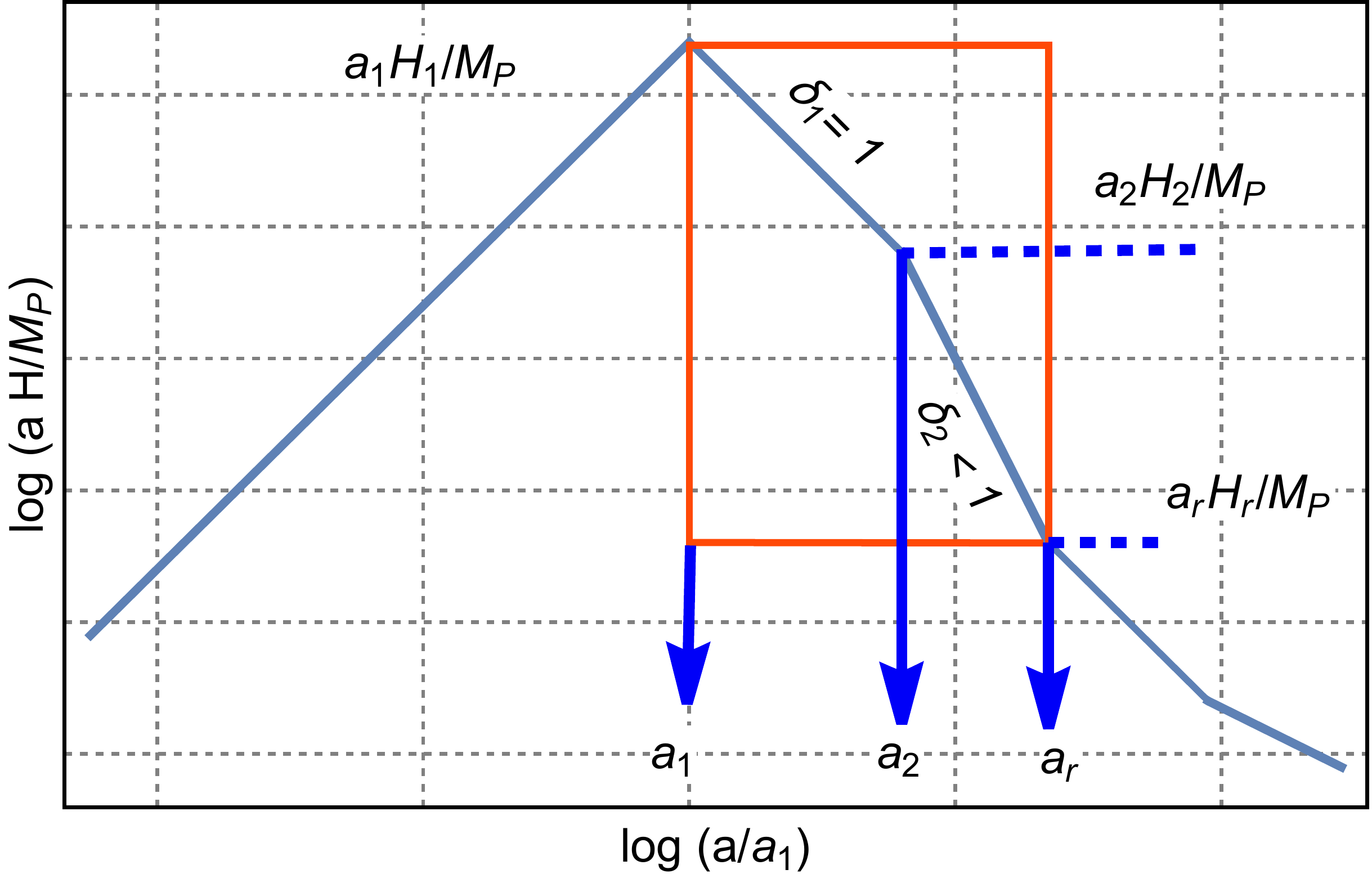}
\includegraphics[height=5.5cm]{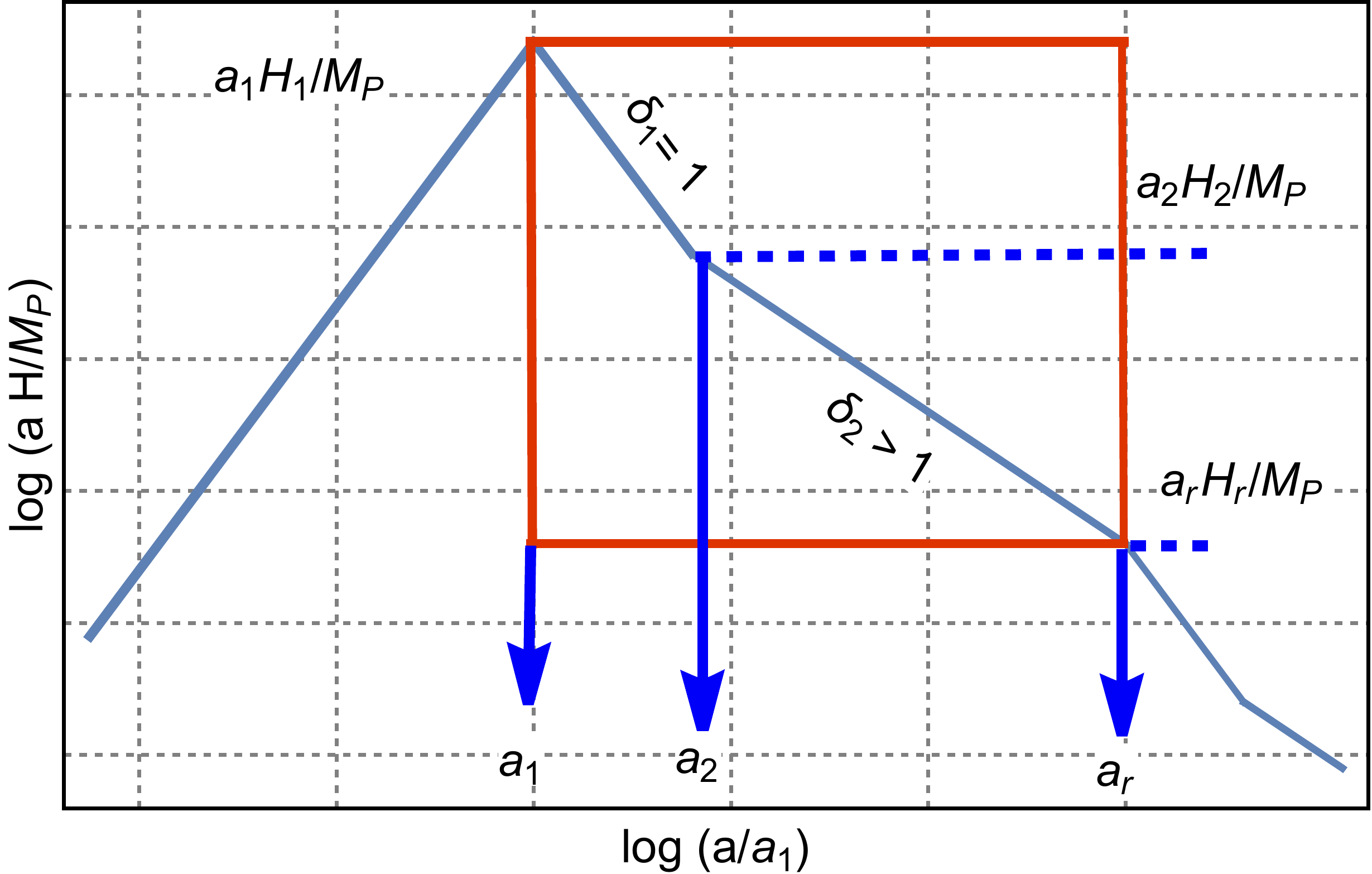}
\caption[a]{The two plots are obtained from Fig. \ref{FF7a} by setting $\delta_{1} \to 1$. In the 
left panel $\delta_{2} >1$ while in the right panel $\delta_{2} <1$. 
The case $\delta_{1} \to 1$ and $\delta_{2} \neq 1$ is dynamically 
different from the one already examined in section \ref{sec3}. On the contrary 
the choice $\delta_{1} \neq 1$ and $\delta_{2} \to 1$ gives again the timeline of Fig. \ref{FF4a} 
with the only difference that radiation dominates earlier.}
\label{FF8a}      
\end{figure}
\begin{figure}[!ht]
\centering
\includegraphics[height=5.5cm]{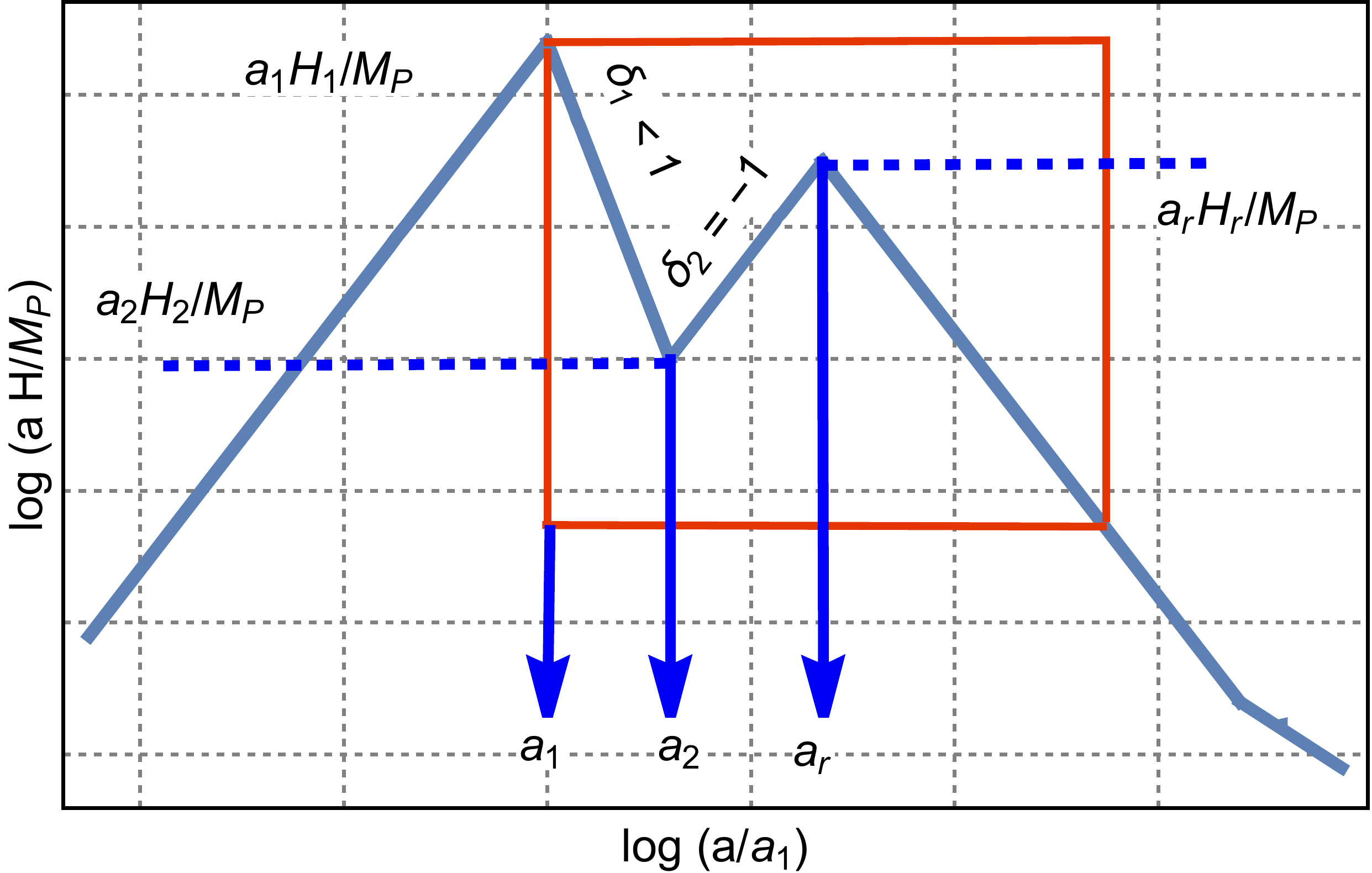}
\includegraphics[height=5.5cm]{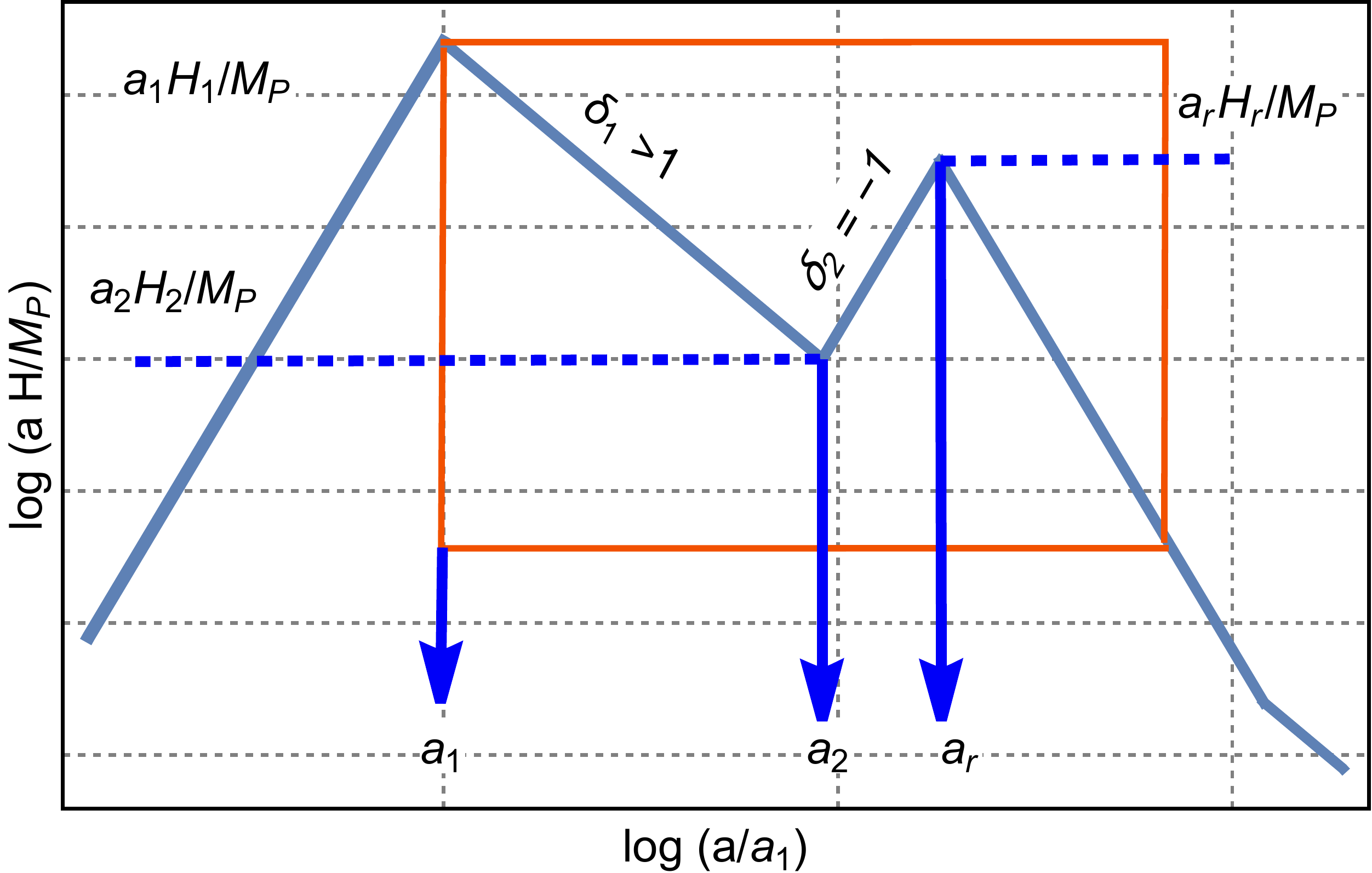}
\caption[a]{The evolution of $a\, H/M_{P}$ is approximately illustrated when a 
second stage of inflation takes place after the radiation-dominated phase. In the 
intermediate (decelerated) epoch of expansion we may have that  $\delta_{1}\to1$ (as it happens in the case of radiation) even if, in these two panels, we illustrate the more general situation.}
\label{FF10a}      
\end{figure}

Assuming that between $H_{1}$ and $H_{r}$ there are two intermediate phases there are three qualitatively different timelines that encompass the plausible dynamical evolutions:
\begin{itemize}
\item{{\it (i)}} the two successive phases expand either faster or slower than radiation; 
\item{{\it (ii)}} the first stage coincides with radiation while the second expands with an arbitrary rate; 
\item{{\it (iii)}} the intermediate phase includes a second epoch of inflationary expansion at a lower scale.
\end{itemize}
The profiles of $a\, H/M_{P}$ corresponding {\it (i)},  {\it (ii)} and  {\it (iii)} are illustrated, respectively, in Figs. \ref{FF7a}, \ref{FF8a} and \ref{FF10a} where the two successive phases are characterized by the scale factors  $a(\tau) \propto \tau^{\delta_{1}}$ and $a(\tau) \propto \tau^{\delta_{2}}$ with $\delta_{1} > 0$ and $\delta_{2}>0$. Within this parametrization that generalizes the one of the previous section if the two stages of expansion are both different from radiation we have, as in Fig. \ref{FF7a}, that $\delta_{1} \neq 1$ and $\delta_{2} \neq 1$. In analogy with $\xi$ introduced in Eq. (\ref{SING1}) we also introduce  $\xi_{1}$ and $\xi_{2}$ which now account for the duration of the two phases (associated, respectively, with $\delta_{1}$ and $\delta_{2}$):
\begin{equation}
\xi_{1}= \frac{H_{2}}{H_{1}}< 1, \qquad\qquad \xi_{2}= \frac{H_{r}}{H_{2}}< 1.
\label{MULTI1}
\end{equation}
From the profiles of Figs. \ref{FF7a} and \ref{FF8a} the  maximal number of $e$-folds
now depends on $(\delta_{1},\,\delta_{2})$ and $(\xi_{1},\, \xi_{2})$: 
\begin{eqnarray}
N_{max} &=& 61.88 - \ln{\biggl(\frac{h_{0}}{0.7}\biggr)} +  \frac{\delta_{1} -1}{2 (\delta_{1} + 1)} \ln{\xi_{1}}  +  \frac{\delta_{2} -1}{2 (\delta_{2} + 1)} \ln{\xi_{2}} 
\nonumber\\
&+& \frac{1}{4} \ln{\biggl(\frac{r_{T}}{0.06}\biggr)}
+ \frac{1}{4} \ln{\biggl(\frac{{\mathcal A}_{{\mathcal R}}}{2.41\times 10^{-9}}\biggr)} + 
\frac{1}{4} \ln{\biggl(\frac{h_{0}^2 \, \Omega_{R0}}{4.15 \times 10^{-5}}\biggr)}.
\label{MULTI2}
\end{eqnarray} 
As in the case Eq. (\ref{SING2}) $N_{max}$ is estimated by requiring that the inflationary event 
horizon redshifted at the present time coincides exactly with the current value 
of the Hubble radius. For a fixed value of $\xi_{1}$ and $\xi_{2}$ the value of $N_{max}$ gets comparatively larger when 
when the Universe expands at a rate slower than radiation. This happens for both profiles of Figs. \ref{FF7a} and 
for the left plot of Fig.  \ref{FF8a} where $\delta_{1}\to 1$ and $\delta_{2} <1$. 
The expression of $\nu_{max}$  coincides with $\nu_{1}$ and it  
depends on ($\xi_{1}$, $\xi_{2}$) and on ($\delta_{1}$, $\delta_{2}$):
\begin{equation}
\nu_{1}= \, \nu_{max} =\xi_{1}^{ \frac{\delta_{1}-1}{2 (\delta_{1} + 1)}}\,\,
 \xi_{2}^{ \frac{\delta_{2}-1}{2 (\delta_{2} + 1)}}\,\,\, \overline{\nu}_{max},\qquad \xi_{i} < 1, \qquad \delta_{i} >0,
\label{MULTI3}
\end{equation}
where $i = 1, \, 2$ and $\overline{\nu}_{max}={\mathcal O}(270) \,\,\mathrm{MHz}$ is the maximal frequency determined when the two intermediate stages reduce to a single radiation-dominated phase (i.e. $\delta_{1} \to 1$ and $\delta_{2}\to 1$); note 
that, by definition, the expression of $\overline{\nu}_{max}$ coincides 
with Eq. (\ref{SING3a}).  Equation (\ref{MULTI3}) 
extends the results of Eqs. (\ref{SING3})--(\ref{SING3a}). Moreover, 
as in the case of a single phase, $\nu_{2}$ and $\nu_{r}$ are related to  $a_{2} H_{2}$ and $a_{r} \, H_{r}$ which are explicitly illustrated in Fig. \ref{FF7a}:
\begin{equation}
\frac{\nu_{2}}{\nu_{max}} = \xi_{1}^{1/(\delta_{1}+1)} = \biggl(\frac{H_{2}}{H_{1}}\biggr)^{1/(\delta_{1}+1)}, \qquad\qquad \frac{\nu_{r}}{\nu_{2}} = \xi_{2}^{1/(\delta_{2}+1)} = \biggl(\frac{H_{r}}{H_{2}}\biggr)^{1/(\delta_{2}+1)}.
\label{MULTI4}
\end{equation}
A particular realisation of the profile reported Fig. \ref{FF7a} 
is obtained by setting $\delta_{1} \to 1$: this choice is described in Fig. \ref{FF8a}
where, as we shall see, the high-frequency spectrum is quasi-flat while $N_{max}$ and $\nu_{max}$ only depend on $\delta_{2}$ and $\xi_{2}$. The hierarchy of Fig. \ref{FF8a} can also be reversed by positing $\delta_{2} = 1$ and $\delta_{1} \neq 1$. This choice coincides however with the case 
already explored in the previous section: indeed to assume $\delta_{2} =1$ just implies a longer radiation-dominated phase preceded by a stage where the expansion rate differs from radiation. In other words, if $\delta_{2} \to 1$  Eqs. (\ref{MULTI3})--(\ref{MULTI4}) have the same content of Eqs. (\ref{SING3})--(\ref{SING4}); in particular we have  $\nu_{2} \to \nu_{r}$, $\nu_{1}\to \nu_{max}$ and also $\delta_{1} \to \delta$.  

There is finally a third possibility (see Fig. \ref{FF10a}) where  $a\, H/M_{P}$ has a pair of local maxima implying the presence of two inflationary stages separated by a decelerated phase. We shall denote by $N^{(1)}$ and $N^{(2)}$ the number of $e$-folds associated with the first and second inflationary stage. In analogy with Eq. (\ref{MULTI2}) the maximal number of $e$-folds accessible to large-scale observations can be determined by requiring that $H_{1}^{-1} (a_{0}/a_{i}) \simeq H_{0}^{-1}$ \cite{HOR}; however since 
in this case there are two inflationary stages this requirement maximizes 
the sum of the numbers of $e$-folds of the two phases: 
\begin{eqnarray}
N^{(1)}_{max} + N^{(2)}_{max} &=& 61.88 - \ln{\biggl(\frac{h_{0}}{0.7}\biggr)} +  \frac{\delta_{1} -1}{2 (\delta_{1} + 1)} \ln{\xi_{1}}  
\nonumber\\
&+& \frac{1}{4} \ln{\biggl(\frac{r_{T}}{0.06}\biggr)}
+ \frac{1}{4} \ln{\biggl(\frac{{\mathcal A}_{{\mathcal R}}}{2.41\times 10^{-9}}\biggr)} + 
\frac{1}{4} \ln{\biggl(\frac{h_{0}^2 \, \Omega_{R0}}{4.15 \times 10^{-5}}\biggr)}.
\label{MULTI2a}
\end{eqnarray} 
The principal inflationary stage (lasting for $N^{(1)}$ $e$-folds) ends at $a_{1}$ when $\nu_{max} \simeq a_{1} \, H_{1}$ while 
the secondary phase (lasting for $N^{(2)}$ $e$-folds) ends after $a_{r}$ and it is characterized by a second maximum. Since $a_{2} \, H_{2} < a_{r} \, H_{r} < a_{1} \, H_{1}$ the corresponding frequencies will also be in the same hierarchy. 
 The timeline of Fig. \ref{FF10a} is analog to the one already explored in Ref. \cite{liddle2} where the authors discussed a second burst of inflation  after a radiation phase where $\delta_{1} \to 1$.  The class of profiles illustrated in Fig. \ref{FF10a}  leads to a spectral energy density that always decreases as a function of the frequency, as we shall see in the last part of this section.  
 
The intermediate phases could be more than two but the most relevant constraints are always associated with the pair of earliest stages of expansion that develop a maximum. This conclusion follows by generalizing the previous results to $n$ expanding phases where:
\begin{equation}
\xi_{i} = \frac{H_{i+1}}{H_{i}}<1, \qquad\qquad \prod_{i= 1}^{n} \,\xi_{i} = \frac{H_{1}}{H_{r}}, \qquad \mathrm{where} \qquad 
i = 1, \,.\,.\,., n.
\label{NN1}
\end{equation}
According to Eq. (\ref{NN1}) $\xi_{i}$ just measures the duration of each intermediate 
stage of expansion.
With the same logic we have $n$ different typical frequencies 
and $\nu_{max}$ will be given by:
\begin{equation}
\nu_{1} = \prod_{i=1}^{n} \,\, \xi_{i}^{\frac{\delta_{i} -1}{2 (\delta_{i} +1)}}\,\, \overline{\nu}_{max}, \qquad \nu_{2} = \sqrt{\xi_{1}}\,\,\prod_{i=2}^{n} \,\, \xi_{i}^{\frac{\delta_{i} -1}{2 (\delta_{i} +1)}}\,\, \overline{\nu}_{max}, \qquad
\nu_{3} = \sqrt{\xi_{1}}\,\,\prod_{i=3}^{n} \,\, \xi_{i}^{\frac{\delta_{i} -1}{2 (\delta_{i} +1)}}\,\, \overline{\nu}_{max},
\,.\,.\,.\,.\,
\label{NN3}
\end{equation}
where the ellipses stand for the frequencies associated with the remaining ranges of expansion. As before we have that for all the intermediate stages the evolution 
is always decelerated:
\begin{equation}
\delta_{i} >0, \qquad\qquad \xi_{i} < 1, \qquad\qquad i = 1, \,.\,.\,., n.
\label{NN1a}
\end{equation}
The first frequency of the partition is always identified with $\nu_{1}$ (i.e. $\nu_{max} = \nu_{1}$) and the last one\footnote{This does not mean that $\nu_{r}$ is the lowest 
frequency {\em of the spectrum} but rather the lowest frequency of 
the frequency range affected by the intermediate modification of the expansion rate.} with $\nu_{r}$. As we shall see, the two successive phases examined before turn out to be sufficiently general for the present purposes since for an arbitrary number of intermediate phases the most relevant constraints always follow from the largest value of the spectral energy density in a given frequency range. 
From the profiles discussed here we excluded the possibility that the two successive 
phases are both expanding at a rate that is either faster or slower than radiation;
for instance we could have $\delta_{1} >1$ and $\delta_{2} >1$ or 
 $\delta_{1} <1$ and $\delta_{2} <1$. These cases have been omitted just for the 
 sake of conciseness since their discussion is very similar to the one already presented  in the previous section. 

\subsection{The shapes of the spectra and their phenomenological signatures}
When a single post-inflationary stage precedes the radiation epoch $h_{0}^2\,\Omega_{gw}(\nu,\tau_{0})$ consists of three separate branches that have been discussed in section \ref{sec3} and illustrated in Fig. \ref{FF6a} . Conversely the spectral energy density computed from the profiles of Figs. \ref{FF7a} and \ref{FF8a} is characterized by four distinct frequency domains. Besides the aHz region (i.e. $\nu_{p} < \nu < \nu_{eq}$) and part of the intermediate branch (for $ \nu_{eq} < \nu < \nu_{r}$), the slopes in the two supplementary ranges (i.e. $\nu_{r} < \nu < \nu_{2}$ and  $\nu_{2} < \nu < \nu_{max}$) depend on the values of $\delta_{1}$ and  $\delta_{2}$. With a unified notation the corresponding spectral slopes (denoted hereunder by  $m_{1}$ and $m_{2}$) are
\begin{equation}
m_{i} = \frac{32 - 4 r_{T}}{16 - r_{T}} - 2 \delta_{i}, \qquad \qquad r_{T} \ll 1, \qquad\qquad i = 1,\,\,2,
\label{MULTISP1}
\end{equation}
where $r_{T} \ll 1$ denotes, as usual, the tensor to scalar ratio. Equation (\ref{MULTISP1}) follows  
from the same considerations leading to Eqs. (\ref{SING4d})--(\ref{SING5}) with the proviso that the pair of spectral slopes now follow from the different evolutions of $a\, H$ at the time the corresponding wavelengths cross the Hubble radius after inflation. If $r_{T}\leq 0.06$ \cite{RT1,RT2,RT3} we have, in practice, 
\begin{equation}
m_{i} = 2 ( 1 - \delta_{i}) + {\mathcal O}(r_{T}), \qquad\qquad i = 1,\,\,2, 
\label{MULTISP1aa}
\end{equation}
According to Eqs. (\ref{MULTISP1})--(\ref{MULTISP1aa}) the spectral energy density decreases (i.e. $m_{i} < 0$) whenever $\delta_{i} > 1$ and the background expands faster than radiation; conversely $h_{0}^2\,\Omega_{gw}(\nu,\tau_{0})$ increases (i.e. $m_{i} >0$) if the expansion rate is slower than radiation and $\delta_{i} <1$. 

\begin{figure}[!ht]
\centering
\includegraphics[height=7cm]{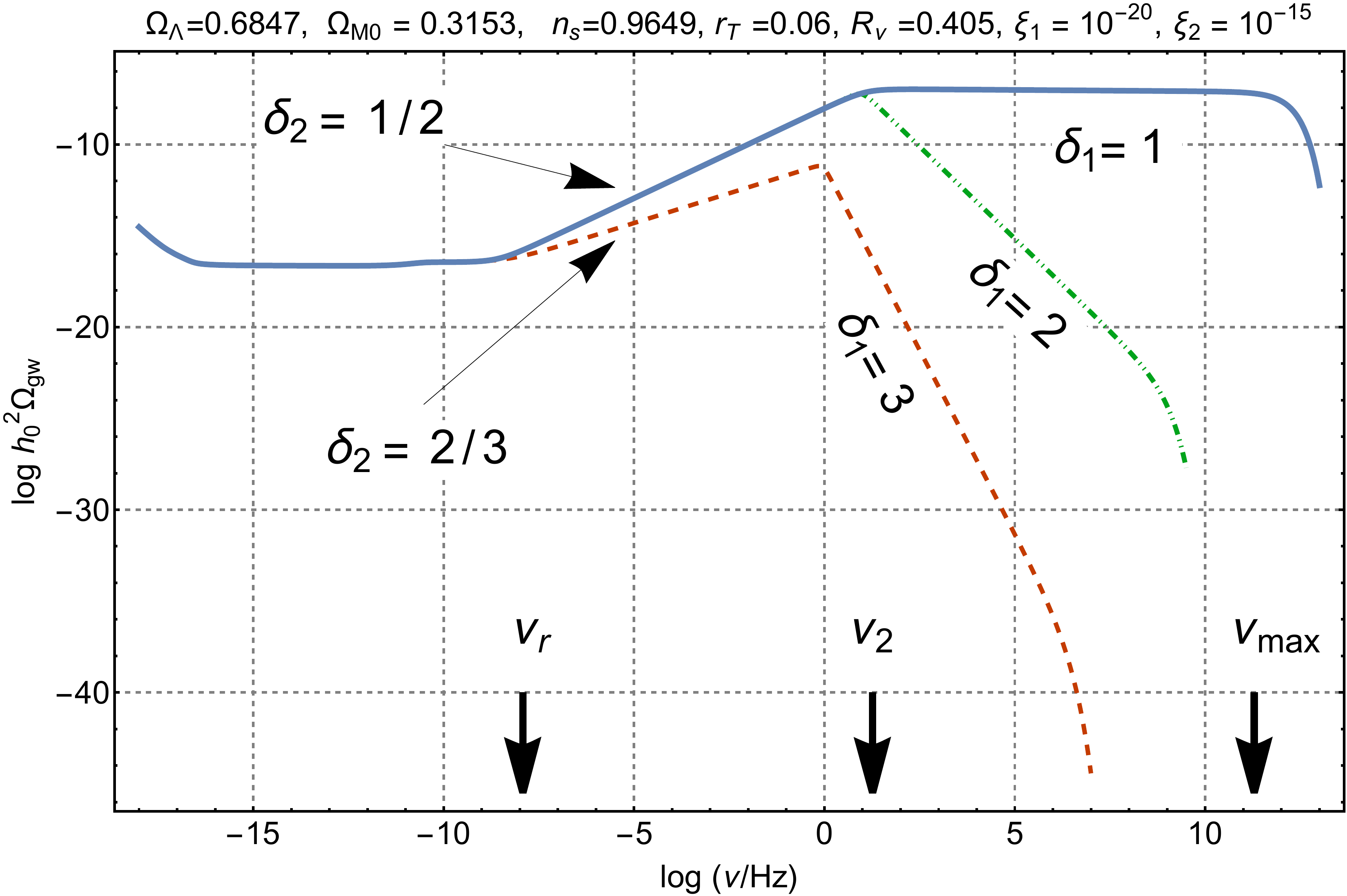}
\caption[a]{The spectral energy density is reported when $\delta_{2}<1$ and $\delta_{1}\geq 1$. The spectral energy develops  a maximum for $\nu\simeq \nu_{2}$.  For the parameters chosen here, the typical frequencies $\nu_{r}$, $\nu_{2}$ and $\nu_{max}$ (see Eqs. (\ref{MULTIF1})--(\ref{MULTIF2}) and discussion thereafter) have been explicitly illustrated. For consistency the late-time parameters correspond to the ones 
already discussed in Fig. \ref{FF6a}.}
\label{FF10b}      
\end{figure}

During the first post-inflationary stage illustrated in the left panel of Fig. \ref{FF7a} we have $\delta_{1} > 1$ while $\delta_{2} < 1$ in the second stage. From Eq. (\ref{MULTISP1}) we therefore deduce that the the spectral slope is negative between $\nu_{2}$ and $\nu_{1}\, =\,\nu_{max}$ (i.e. $m_{1} <0 $) while it is positive between $\nu_{r}$ and $\nu_{2}$ (i.e. $m_{2} >0$). Recalling the parametrization of Eq. (\ref{SING9b}) the high-frequency spectrum now consists of two branches\footnote{Since Eq. (\ref{MULTISP2}) applies in the case   $\delta_{1} > 1$ (i.e. $m_{1} <0$) it is practical write the spectral energy density in terms of $|m_{1}|$: in this way the potential confusions with the other cases (where $m_{1} >0$) are avoided.}:
\begin{eqnarray}
h_{0}^2\,\Omega(\nu,\tau_{0}) &=& \overline{{\mathcal N}}_{\rho}(r_{T},\nu) \biggl(\frac{\nu}{\nu_{r}}\biggr)^{m_{2}}, \qquad\qquad \nu_{r} < \nu < \nu_{2}, 
\nonumber\\
h_{0}^2\,\Omega(\nu,\tau_{0}) &=& \overline{{\mathcal N}}_{\rho}(r_{T},\nu) \biggl(\frac{\nu_{2}}{\nu_{r}}\biggr)^{m_{2}}\biggl(\frac{\nu}{\nu_{2}}\biggr)^{-|m_{1}|}, \qquad\qquad \nu_{2} < \nu < \nu_{max}.
\label{MULTISP2}
\end{eqnarray}
The spectral energy density given of Eq. (\ref{MULTISP2}) exhibits a maximum for $\nu= {\mathcal O}(\nu_{2}$) and when $\delta_{1} \to 1$ the maximum is replaced by a plateau since $h_{0}^2\,\Omega_{gw}(\nu, \tau_{0})$ flattens out (i.e. $m_{1} \to 0$  for $\nu > \nu_{2}$). Equation (\ref{MULTISP2}) 
is corroborated by the numerical result illustrated in Fig. \ref{FF10b} for different values of $\delta_{1}$ and $\delta_{2}$. The curve at the top (full line) corresponds to the choice $\delta_{1} =1$ and $\delta_{2} =1/2$: in this case the Universe first expands like radiation and then the rate becomes slower than radiation. At intermediate frequencies (i.e. for $\nu < \nu_{2}$) the high-frequency plateau  is replaced by an increasing branch $h_{0}^2 \, \Omega_{gw}(\nu,\tau_{0}) \propto (\nu/\nu_{2})^{m_{2}}$ (i.e. $m_{2} \to 1$ if $\delta_{2} = 1/2$ and $\delta_{1} \to 1$).
Always in Fig. \ref{FF10b}  with the dot-dashed line we report the case $\delta_{1}=2$ and $\delta_{2} =1/2$; as anticipated with this choice we expect a maximum for $\nu = {\mathcal O}(\nu_{2})$. Finally the dashed curve accounts for the case $\delta_{2}=2/3$ and $\delta_{1} =3$: also in this example, on the basis of Eq. (\ref{MULTISP2}) we expect a maximum for $\nu = {\mathcal O}(\nu_{2})$. 

Even if the values of $\xi_{1}$ and $\xi_{2}$ are the same for all the curves 
of Fig. \ref{FF10b}, the comparison between the dashed and dot-dashed 
lines suggests that the values of $\nu_{2}$ (corresponding to the location of the maximum) are slightly different. This happens since $\nu_{2}$ not only depends on $(\xi_{1},\,\xi_{2})$ but also on the specific value of $\delta_{2}$ as implied by Eq. (\ref{MULTI4}). More specifically, 
recalling that $(\xi_{1}, \,\xi_{2})$ parametrize the duration of the two successive phases (see Eq. (\ref{MULTI1})), in Fig. \ref{FF10b} we considered the illustrative choice $\xi_{1} = 10^{-20}$ and $\xi_{2} = 10^{-15}$. The values of $\xi_{1}$ and $\xi_{2}$ selected in Figs.  \ref{FF10b} and \ref{FF10c} imply $\xi_{1} \, \xi_{2}= 10^{-35}$; this value is not unrealistic since, according to  Eqs. (\ref{MULTI1}) and (\ref{NN1}),
 $\xi_{1} \, \xi_{2}$ coincides with $H_{1}/H_{r}$ that must be larger than $10^{-38}$ to guarantee that the synthesis of light nuclei takes place when the Universe is already dominated by radiation: 
\begin{equation}
\xi_{1} \, \xi_{2} = (H_{1}/H_{r}) > 10^{-38}.
\label{MULTISP8}
\end{equation}
From Eqs. (\ref{MULTI3})--(\ref{MULTI4}) (and for the typical values of the late-time parameters of Figs. \ref{FF10b}) the explicit value of $\nu_{r}$ is:
\begin{equation} 
\nu_{r} = 0.8 \, \biggl(\frac{\xi_{1}}{10^{-20}}\biggr)^{1/2}\, \biggl(\frac{\xi_{2}}{10^{-15}}\biggr)^{1/2}\mathrm{nHz}.
\label{MULTIF1}
\end{equation}
Since $\nu_{r}$ does not depend upon $\delta_{1}$ and $\delta_{2}$, Eq. (\ref{MULTIF1}) explains why $\nu_{r}$ is the same for the three curves of Fig. \ref{FF10b} while $\nu_{max}$ and $\nu_{2}$ (see Eqs. (\ref{MULTI3})--(\ref{MULTI4})) are slightly different because of their explicit dependence on the rates of the two phases:
\begin{equation}
\nu_{max} = 269.33 \,\,\xi_{1}^{ \frac{\delta_{1}-1}{2 (\delta_{1} + 1)}}\,\,
 \xi_{2}^{ \frac{\delta_{2}-1}{2 (\delta_{2} + 1)}} \,\,\,\mathrm{MHz},\qquad\qquad
\nu_{2} = 269.33 \,\,\sqrt{\xi_{1}}\,\,\,
 \xi_{2}^{ \frac{\delta_{2}-1}{2 (\delta_{2} + 1)}} \,\,\,\mathrm{MHz}.
\label{MULTIF2}
\end{equation}
In the case of the full curve at the top in Fig. \ref{FF10b} (i.e.  $\xi_{1} =10^{-20}$, $\xi_{2} = 10^{-15}$, $\delta_{1} =1$ and $\delta_{2}=1/2$) Eq. (\ref{MULTIF2}) implies $\nu_{max} = 85 \,\, \mathrm{GHz}$ and  $\nu_{2} = 8.5 \,\, \mathrm{Hz}$. Note finally that for $\nu > \nu_{1} \simeq \nu_{max}$ there is the usual exponential suppression that also occurs in the case of a single phase.

The values of $\xi_{1}$ and $\xi_{2}$ can be consistently 
chosen in a way that $\nu_{r}$ falls in the nHz domain without conflicting with the bound of Eq. (\ref{MULTISP8}). The local maxima of the spectral energy density (or the high-frequency plateau) could in principle explain the preliminary results of the  PTA (see Eq. (\ref{PTAb1}) and discussion thereafter). By looking at Fig. \ref{FF10b} we observe that $h_{0}^2\,\Omega_{gw}(\nu,\tau_{0})$ 
may be large as $10^{-8}$ in the nHz range. For this purpose, we should have, however, that either $\nu_{r}$ is much smaller than the nHz (which is impossible because of Eq. (\ref{MULTISP8})) or $m_{2} \gg 1$. When $0< \delta_{2} < 1$ the spectral 
slope can be, at most, of order $1$ (i.e. $m_{2} = {\mathcal O}(1)$) since the maximal value of $m_{2}$ is saturated\footnote{This conclusion follows, for instance, if the expansion rate is driven by a perfect 
fluid with barotropic index $w$: in this case $\delta_{2} = 2/(3 w +1)$ and  $\delta_{2} \leq 1/2$ whenever $w \leq 1$. } 
when $\delta_{2} \to 1/2$.  
 \begin{figure}[!ht]
\centering
\includegraphics[height=5.4cm]{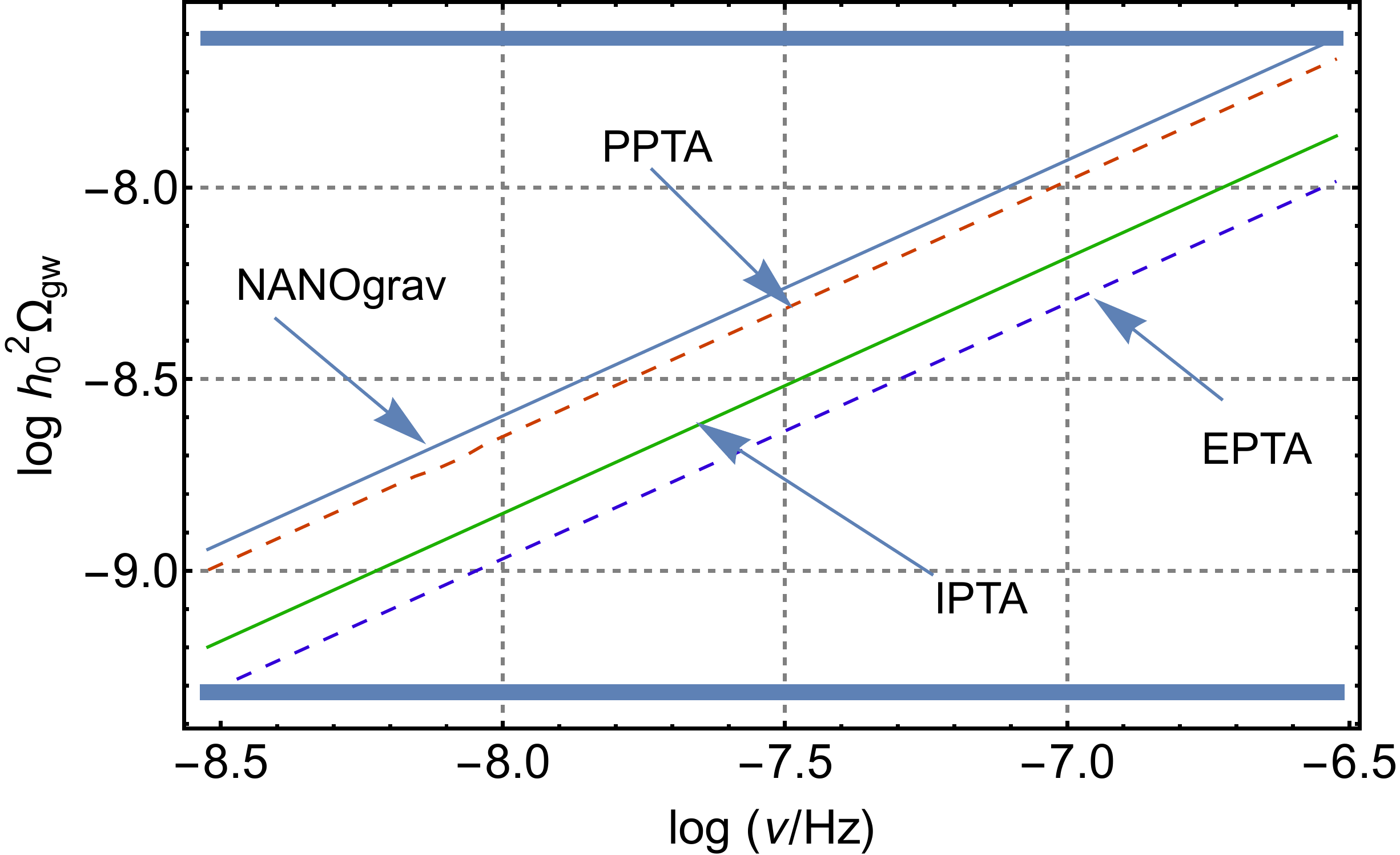}
\includegraphics[height=5.4cm]{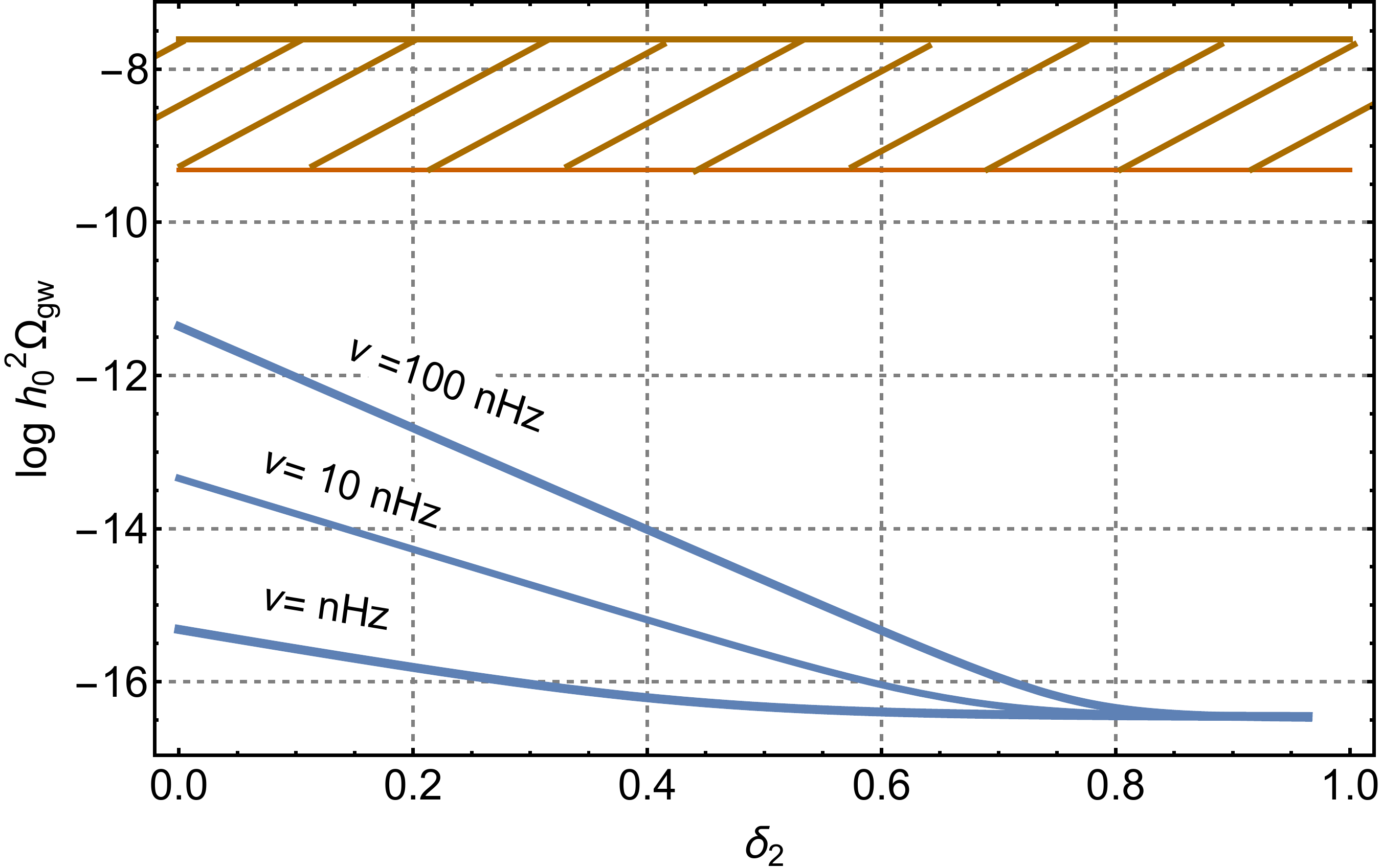}
\caption[a]{In the plot at the left we illustrate the claimed signal of the PTA collaborations (see Eqs. (\ref{PTAb1})--(\ref{NOTT8}) and discussions therein) in the case $\beta =-2/3$ which implies, according to Eq. (\ref{NOTT8}) that the spectral energy density scales as $\nu^{2/3}$. In the plot at the right the spectral energy density is computed by estimating $\xi_{1}$ and $\xi_{2}$ from Eqs. (\ref{EXX3})--(\ref{EXX4}). Since the dashed region in the right panel accounts for the potential signal of the PTA collaborations, the spectral energy density computed for different frequencies and as a function of $\delta_{2}$ always undershoots the PTA data.}
\label{FF10e}      
\end{figure}
The qualitative argument of the previous paragraph is consistent with a more rigorous estimate based on the bound of Eq. (\ref{CC2}).  
Since the maximum of the spectral energy density is reached for $\nu = {\mathcal O}(\nu_{2})$ we have that $h_{0}^2\,\Omega_{gw}(\nu_{2}, \tau_{0})$ must not exceed a maximal value ${\mathcal O}(10^{-5})$:
\begin{equation}
h_{0}^2\, \Omega(\nu,\tau_{0}) = \overline{{\mathcal N}}_{\rho}(r_{T},\nu) \biggl(\frac{\nu_{2}}{\nu_{r}}\biggr)^{\overline{n}_{2}} = h_{0}^2 \overline{\Omega}^{(max)}_{gw}< 10^{-5}.
\label{EXX2}
\end{equation}
For actual estimates, we can assume, for instance, $h_{0}^2 \overline{\Omega}^{(max)}_{gw}= 10^{-6}$
and  recalling from Eq. (\ref{MULTI4}) that $\nu_{2}/\nu_{r}= \xi_{2}^{-1/(\delta_{2}+1)}$, Eq. (\ref{EXX2}) implies:
\begin{equation}
\log{\xi_{2}} = \frac{(\delta_{2} +1)(16 - r_{T})}{32(\delta_{2} -1) + 2 r_{T}( 2 -\delta_{2})}\biggl[\log{h_{0}^2 \overline{\Omega}^{(max)}_{gw}} - \log{\overline{{\mathcal N}}_{\rho}}(r_{T},\nu)\biggr],
\label{EXX3}
\end{equation}
where it is understood, as verified in Eq. (\ref{SING9b}), that $\overline{{\mathcal N}}_{\rho}(r_{T},\nu)$ 
is practically constant in frequency since, by definition, it is the value of the spectral energy density for $\nu < \nu_{r}$.
From Eq. (\ref{MULTI1}) we must also require that $\xi_{1} \, \xi_{2} = H_{r}/H_{1} > 10^{-38}$, so that Eq. (\ref{EXX3}) finally demands
\begin{equation}
\log{\xi_{1}} = \log{(H_{r}/H_{1})} - \frac{(\delta_{2} +1)(16 - r_{T})}{32(\delta_{2} -1) + 2 r_{T}( 2 -\delta_{2})}\biggl[\log{h_{0}^2 \overline{\Omega}^{(max)}_{gw}} - \log{\overline{{\mathcal N}}_{\rho}(r_{T},\nu)}\biggr].
\label{EXX4}
\end{equation}
The spectral energy density in the frequency range of the PTA 
can then be estimated from Eqs. (\ref{EXX3})--(\ref{EXX4}) for different values 
of $H_{r}$ and $H_{1}$. It is not surprising that the largest spectral energy density in 
the nHz range follows by saturating the bound on $H_{1}$ and $H_{r}$ since, in this 
case, the frequency interval between $\nu_{r}$ and the nHz is wider and $h_{0}^2\,\Omega_{gw}(\nu,\tau_{0})$ 
can reach a larger value for a fixed value of $\delta_{2}$.
In Fig. \ref{FF10e} we illustrate the various evidences of the PTA collaborations 
and the spectral energy density that only depends on $\delta_{2}$; the dependence on $\xi_{1}$ and 
$\xi_{2}$ has been eliminated thanks to Eqs. (\ref{EXX3})--(\ref{EXX4}). To maximize the potential 
signal we imposed $h_{0}^2 \overline{\Omega}^{(max)}_{gw}= 10^{-6}$, $H_{1} = 10^{-6} M_{P}$ 
and $H_{r} = 10^{-40} \, M_{P}$ so that $\xi_{1} \xi_{2} = 10^{-34} > 10^{-38}$. Largest values of $H_{r}$ 
lead to comparatively smaller signals in the nHz band, smaller values of $H_{r}$ are dangerously 
close to the limit of Eq. (\ref{MULTISP8}).

 For different frequencies (and in spite of the value of $\delta_{2}$) the 
 spectral energy density is always smaller than the PTA region which is 
 represented by the shaded area in the right panel\footnote{As done before, in Fig. \ref{FF10e} we fixed $\delta_{1}\to 1$ since this is the situation 
 where the signal is larger for a wider interval of frequencies.} of Fig. \ref{FF10e}. 
  In the present case the bound (\ref{NOT2}) should be applied at high-frequencies and we have $ \zeta =- 2 \epsilon/ (1- \epsilon)$ with $\epsilon < 0.003$. To leading order in $\epsilon$ 
Eq. (\ref{NOT2}) implies 
\begin{equation}
\log{\overline{\Omega}}(\epsilon) < -\,8.236 -\, 0.335\, \epsilon -0.393 \epsilon^2,
\end{equation}
and $\overline{\Omega}(\zeta)$ becomes in fact $\overline{\Omega}(\epsilon)$.
We could now assume, by fiat, that the spectral slope of our spectrum coincides 
with a chirp amplitude scaling as $\beta = - 2/3$ (see Eq. (\ref{NOTT1}) and discussion thereafter); this choice of $\beta$, however, does not
 maximise the growth of the spectral energy density\footnote{ Indeed the 
 PTA collaborations \cite{CCPP1,CCPP2,CCPP3,NANO1} specifically 
 consider the case $\beta = -2/3$ as a potential signal  implying that the slope of the 
 spectral energy density is actually $+2/3$, as it follows from Eq. (\ref{NOTT8}).}
since 
\begin{equation}
 \beta= -2/3,\qquad \Rightarrow\qquad \delta_{2} = 2/3, \qquad \Rightarrow \qquad m_{2} = 2/3 + {\mathcal O}(r_{T}).
 \end{equation}
It turns out that the growth of the spectral energy density is instead maximized for $\delta_{2} = 1/2$ implying $m_{2} = 1 + {\mathcal O}(r_{T})$ ( see, in this respect, the full and dot-dashed curves in Fig. \ref{FF10b}).
 
When $h_{0}^2\,\Omega_{gw}(\nu,\tau_{0})$ 
reaches a local maximum (as in Fig. \ref{FF10b}) we should also worry that 
the bounds of Eqs. (\ref{NOT1})--(\ref{NOT2}) coming from the audio band are satisfied for typical 
frequencies $\nu_{KLV}$ ranging between $30$ and $100$ Hz.
\begin{figure}[!ht]
\centering
\includegraphics[height=7.4cm]{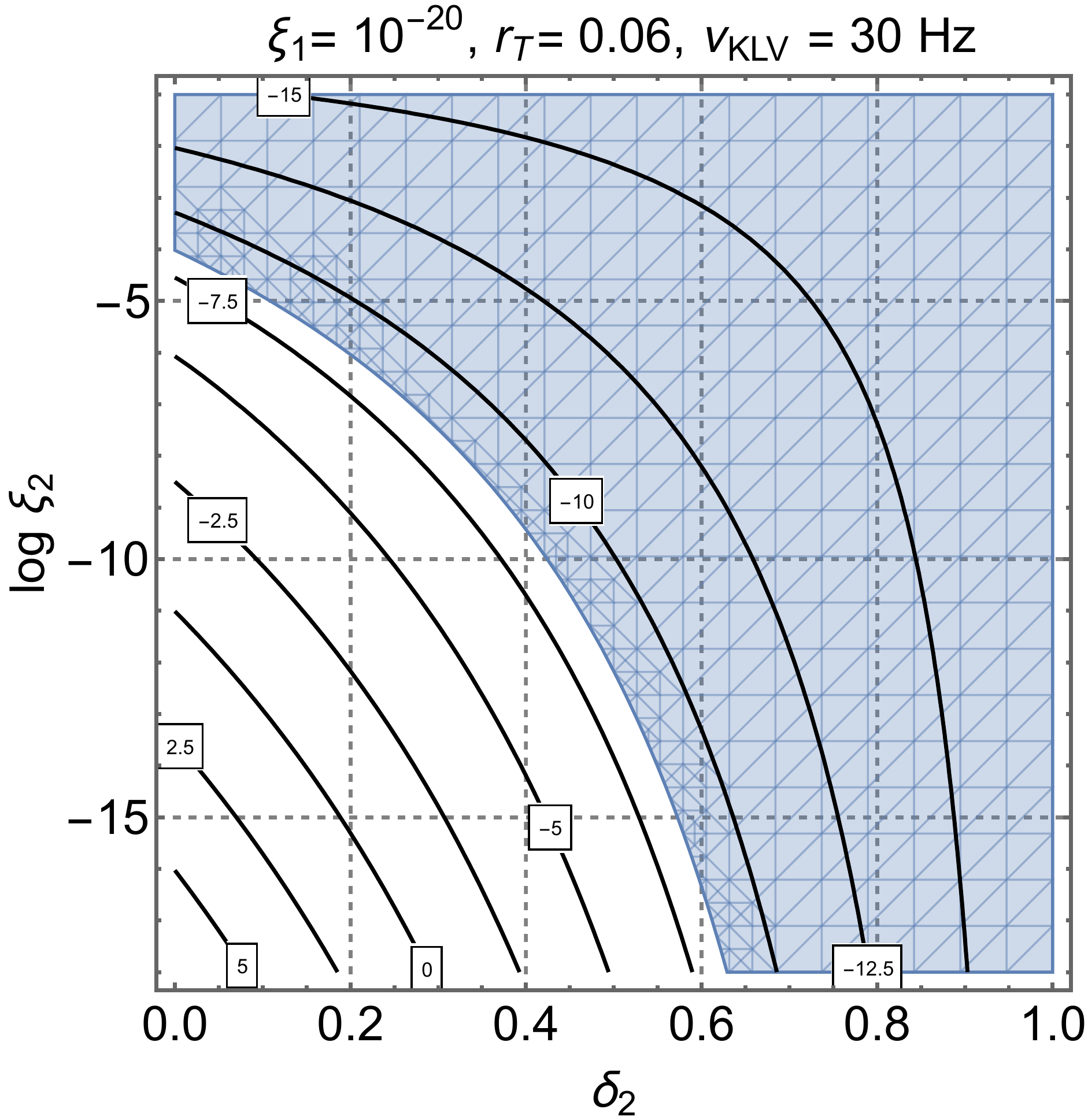}
\includegraphics[height=7.4cm]{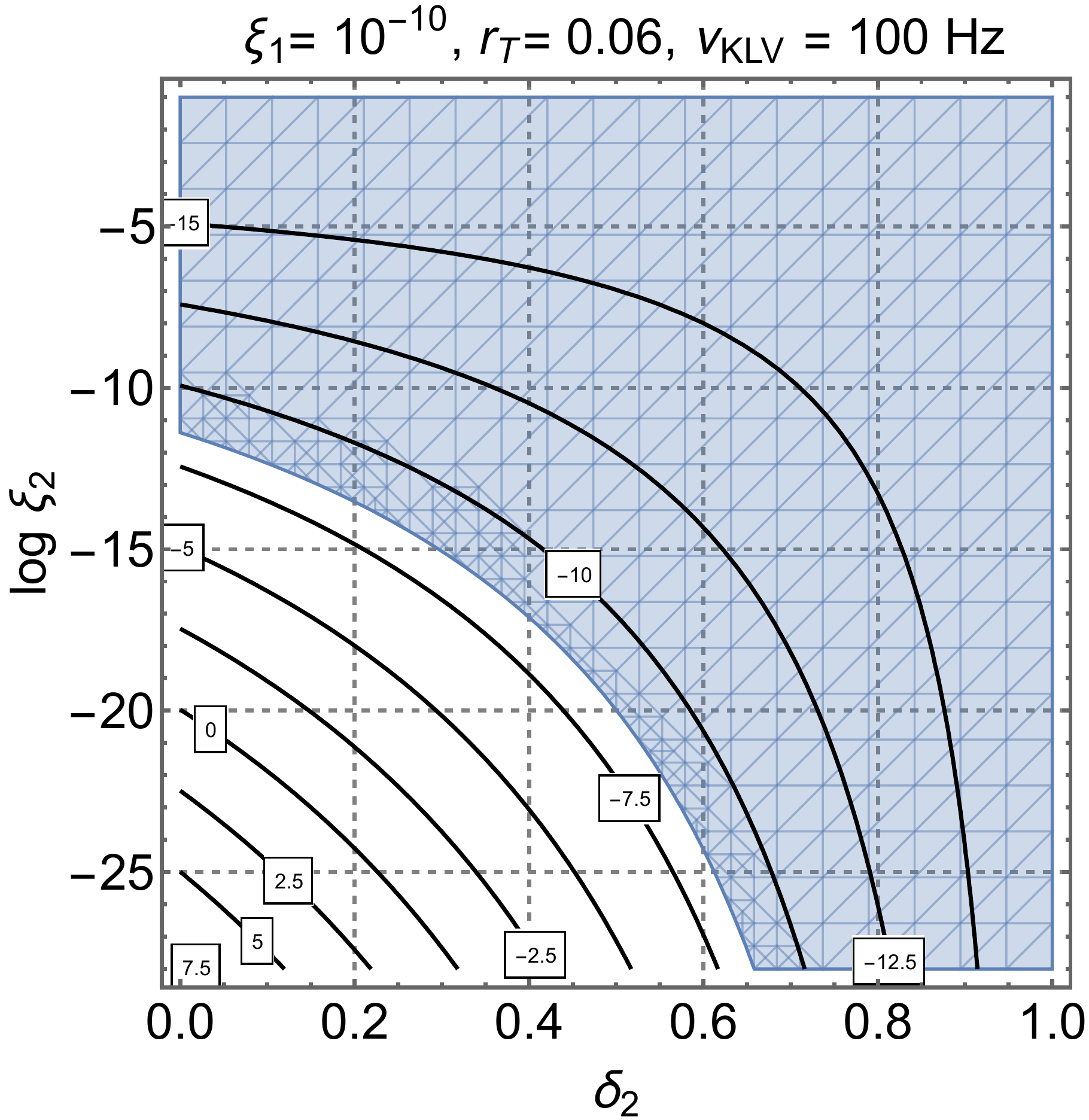}
\caption[a]{The Kagra-Ligo-Virgo bound is applied to the case where 
the post-inflationary stage contains two successive phases where $\delta_{1} \geq 1$ and $0< \delta_{2} \leq 1$. The spectral energy density associated with this dynamical profile is illustrated in Fig. \ref{FF10b}. The labels 
on the different curves in both plots correspond to the value of the common logarithm of $h_{0}^2 \, \Omega_{gw}(\nu,\tau_{0})$ which is constant along each of the contours. }
\label{FF10f}      
\end{figure}
This point is addressed in Fig. \ref{FF10f} where the most constraining situation has been assumed by requiring that $\delta_{1} =1$ and $0< \delta_{2} \leq 1$. Two illustrative values of $\xi_{1}$ have been selected and, for each of these 
values, the range of $\xi_{2}$ follows by requiring $\log{\xi_{1} \xi_{2}} \geq -38$. 
The shaded area of both plots represents the allowed region. The excluded corner of the parameter 
space corresponds, as expected, to the region where $\xi_{2}<10^{-10}$ and $\delta_{2}< 1/2$.  The choices $\xi_{2} \ll 1$ and $\delta_{2} \ll 1$  imply that $h_{0}^2\,\Omega_{gw}(\nu,\tau_{0})$ increases for a larger interval of frequency and may potentially jeopardizes the KLV bounds.

\subsection{Complementary considerations}
\begin{figure}[!ht]
\centering
\includegraphics[height=7cm]{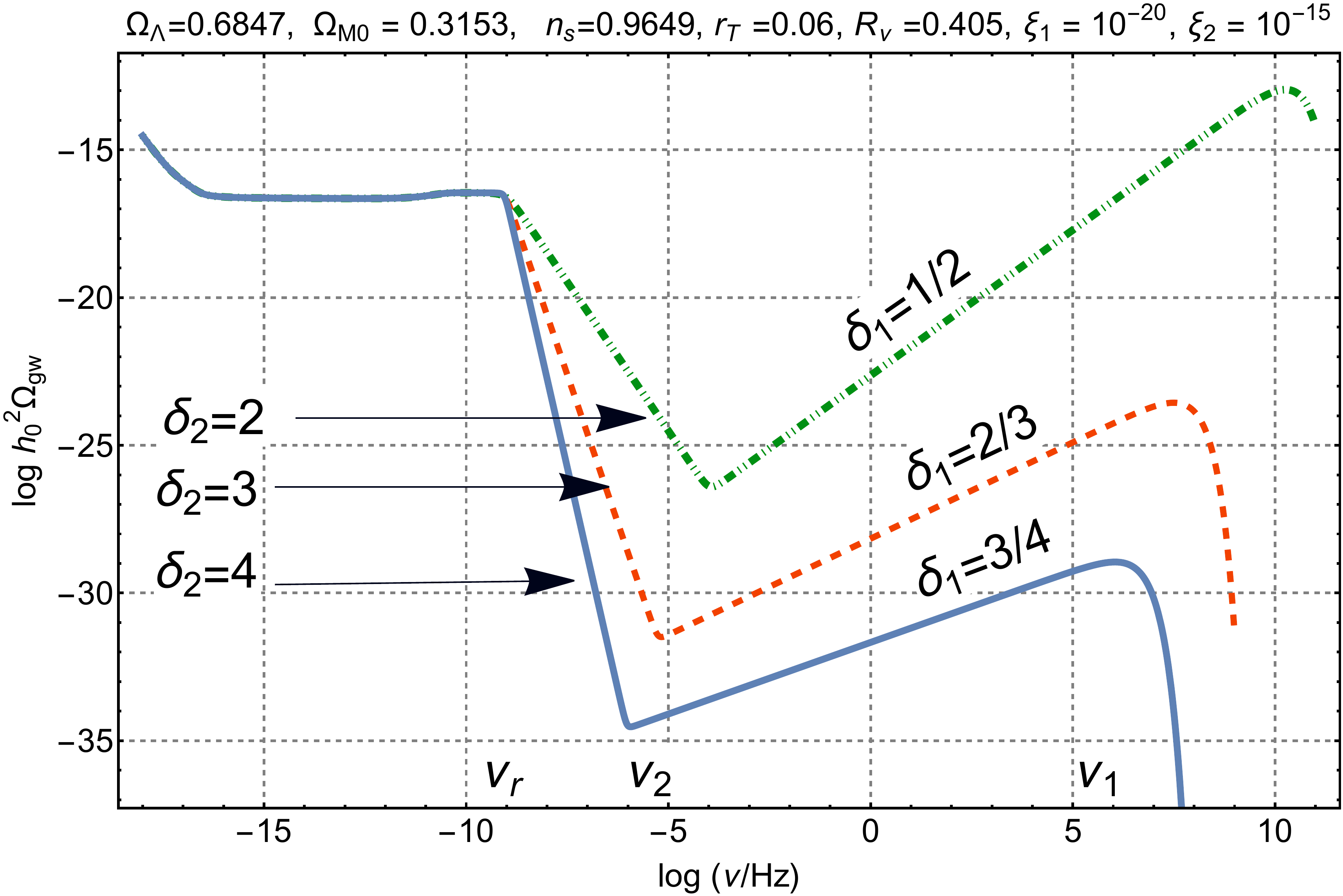}
\caption[a]{The spectral energy density is illustrated in the situation where $\delta_{2}>1$ and $\delta_{1} \leq 1$.
 In this case the first post-inflationary phase is slower than radiation 
while the second stage expands at a rate which is faster than radiation. 
The spectral energy density is standard for $\nu< \nu_{r}$, it decreases at intermediate frequencies (i.e. for $\nu_{r} < \nu < \nu_{2}$) and it increases again in the high-frequency 
branch (i.e. for $\nu_{2} < \nu < \nu_{max}$). However since the increase always occurs {\em after} the absolute minimum it is clear that the maximal $h_{0}^2\,\Omega_{gw}(\nu,\tau_{0})$ always undershoots the profile obtained in the case of a single phase with $\delta<1$ (see Fig. \ref{FF6a}) for the same choice of the other late-time parameters.}
\label{FF10c}      
\end{figure}

In the case of Fig. \ref{FF10b} the transfer function for the spectral energy density
is computed exactly with the same technique applied in the case of the 
radiation-matter transition (see Eq. (\ref{SING7}) and discussion thereafter). 
There is however a difference  since ${\mathcal T}_{high}^2(\nu,\nu_{r},\nu_{2})$ now depends on two frequency scales instead of one:
\begin{equation}
{\mathcal T}_{high}^2(\nu,\nu_{r},\nu_{2}) = \frac{\sqrt{1 + a_{1} (\nu/\nu_{r})^{m_{2}} + a_{2} (\nu/\nu_{r})^{2 m_{2}} }}{\sqrt{1 + b_{1} (\nu/\nu_{2})^{m_{2} +|m_{1}|} + b_{2} (\nu/\nu_{2})^{2(m_{2} +|m_{1}|)} }},
\label{MULTISP3}
\end{equation}
where $a_{i}$ and $b_{i}$ (with $i = 1,\, 2$) are numerical coefficients of order $1$ that depend on the specific choice of $\delta_{1}$ and $\delta_{2}$ and cannot be written in general terms. Since  Eq. (\ref{MULTISP3}) depends on two 
different scales,  there are three relevant limits of ${\mathcal T}_{high}^2(\nu,\nu_{r},\nu_{2})$ that must be considered:
\begin{eqnarray}
{\mathcal T}_{high}^2(\nu,\nu_{r},\nu_{2}) &\to& \sqrt{\frac{a_{2}}{b_{2}}}\,\,\biggl(\frac{\nu_{2}}{\nu_{r}}\biggr)^{m_{2}} \,\,\biggl(\frac{\nu}{\nu_{2}}\biggr)^{- |m_{1}|}, \qquad\qquad \nu \gg \nu_{2},
\nonumber\\
{\mathcal T}_{high}^2(\nu,\nu_{r},\nu_{2}) &\to& \sqrt{a_{2}} \biggl(\frac{\nu}{\nu_{r}}\biggr)^{m_{2}},\qquad\qquad \nu_{r}< \nu< \nu_{2},
\nonumber\\
{\mathcal T}_{high}^2(\nu,\nu_{r},\nu_{2}) &\to&  1, \qquad\qquad \nu< \nu_{r}.
\label{MULTISP4}
\end{eqnarray}
We remind that the parametrization of Eqs. (\ref{MULTISP3})--(\ref{MULTISP4}) is also applicable for $m_{1} \to 0$ when the first post-inflationary stage is dominated by radiation (i.e.  $\delta_{1} \to 1$). In the complementary situation (i.e. $\delta_{1} < 1$ and $\delta_{2} > 1$) we have that $m_{1} >0 $ while $m_{2} < 0$ the transfer function is given by:
\begin{equation}
{\mathcal T}_{high}^2(\nu,\nu_{r},\nu_{2}) = \frac{\sqrt{1 + c_{1} (\nu/\nu_{2})^{|m_{2}| + m_{1}} + c_{2} (\nu/\nu_{2})^{2(|m_{2}| + m_{1})} }}{\sqrt{1 + d_{1} (\nu/\nu_{r})^{|m_{2}|} 
+ d_{2} (\nu/\nu_{r})^{2 |m_{2}|} }},
\label{MULTISP6}
\end{equation}
and  $c_{i}$ and $d_{i}$ (with $i = 1,\, 2$) are numerical coefficients of order $1$
depending on the numerical values of $\delta_{1}$ and $\delta_{2}$.
As in Eq. (\ref{MULTISP3}) we introduced the absolute value of $m_{2}$ since, this time, 
$m_{2} <0$ and $\delta_{2} > 1$. The transfer function of Eq. (\ref{MULTISP6}) has 
again three different limits:
\begin{eqnarray}
{\mathcal T}_{high}^2(\nu,\nu_{r},\nu_{2}) &\to& \sqrt{\frac{c_{2}}{d_{2}}}\,\,\biggl(\frac{\nu_{r}}{\nu_{2}}\biggr)^{|m_{2}| + m_{1}} \,\,\biggl(\frac{\nu}{\nu_{r}}\biggr)^{m_{1}}, \qquad\qquad \nu \gg \nu_{2},
\nonumber\\
{\mathcal T}_{high}^2(\nu,\nu_{r},\nu_{2}) &\to& \frac{1}{\sqrt{d_{2}}} \biggl(\frac{\nu}{\nu_{r}}\biggr)^{-|m_{2}|},\qquad\qquad \nu_{r}< \nu< \nu_{2},
\nonumber\\
{\mathcal T}_{high}^2(\nu,\nu_{r},\nu_{2}) &\to&  1, \qquad\qquad \nu< \nu_{r}.
\label{MULTISP7}
\end{eqnarray}
Equations (\ref{MULTISP6})--(\ref{MULTISP7}) are also applicable when $m_{1} \to 0$, i.e. when 
the first phase coincides with radiation and $\delta_{1} \to  1$. In the situation described by Eq. (\ref{MULTISP7}) 
the high-frequency spectrum consists of two branches with a minimum in $\nu_{2}$:
\begin{eqnarray}
h_{0}^2\,\Omega(\nu,\tau_{0}) &=& \overline{{\mathcal N}}_{\rho}(r_{T},\nu)\,\,\biggl(\frac{\nu}{\nu_{r}}\biggr)^{-|m_{2}|}, \qquad\qquad \nu_{r} < \nu < \nu_{2}, 
\nonumber\\
h_{0}^2\,\Omega(\nu,\tau_{0}) &=& \overline{{\mathcal N}}_{\rho}(r_{T},\nu)\,\,\biggl(\frac{\nu_{2}}{\nu_{r}}\biggr)^{-|m_{2}|}\biggl(\frac{\nu}{\nu_{2}}\biggr)^{m_{1}}, \qquad\qquad \nu_{2} < \nu < \nu_{max},
\label{MULTISP5}
\end{eqnarray}
while for $\nu > \nu_{max}$ there is the usual exponential suppression.
The spectral energy density corresponding to Eqs. (\ref{MULTISP7})--(\ref{MULTISP5}) does not lead to a prominent signal at intermediate 
or high frequencies. In this case the full spectral energy density is illustrated in  Fig. \ref{FF10c} when $\delta_{1} <1$ and $\delta_{2}> 1$.  For comparison the values of $\xi_{1}$ and $\xi_{2}$ coincide 
with the ones already employed  in Fig. \ref{FF10b}.  
Above $\nu_{r}$ the spectral energy density first 
decreases and then it increases between $\nu_{2}$ and $\nu_{1}$. Since $\nu_{r}$ does not depend on $\delta_{1}$ and $\delta_{2}$ we have that its 
value coincides with Eq. (\ref{MULTIF1}) and it is therefore the same for Figs. \ref{FF10b} and \ref{FF10c}. The KLV bound (see Eqs. (\ref{NOT1})--(\ref{NOT2}) and discussion therein) is always satisfied even when the spectral energy density decreases as a function of the frequency; the most relevant limit is at low frequencies and it is associated with $r_{T}$. As before Eqs. (\ref{MULTIF1})--(\ref{MULTIF2}) can be used to determine the explicit values of the typical frequencies\footnote{For instance if $\delta_{2} = 2$ and $\delta_{1} =1/2$
(see Fig. \ref{FF10c}) we have $\nu_{max} =1.8\,\, \mathrm{GHz}$, $\nu_{2} = 85 \,\, \mu\mathrm{Hz}$ and $\nu_{r}= 0.8\,\, \mathrm{nHz}$.  Once $\xi_{1}$ and $\xi_{2}$ are fixed $\nu_{2}$ falls always within a similar range while $\nu_{max}$ may take quite different values depending on $\delta_{1}$.}.
While the results of Fig. \ref{FF10c} are interesting in their own right they also suggest that the most relevant constraints on $h_{0}^2 \, \Omega_{gw}(\nu,\tau_{0})$ always fall in the low-frequency region and are associated with the limits on $r_{T}$. Since the spectral energy density  exhibits a minimum for $\nu \simeq \nu_{2}$,  the bounds at intermediate and high-frequencies are even less relevant than in the conventional situation. 

It is finally appropriate to consider, for the sake of completeness the possible occurrence of an intermediate inflationary stage whose dynamical profile has been already described in Fig. \ref{FF10a}. The first inflationary stage is standard while the second epoch of accelerated expansion takes place at a lower curvature scale; three different ranges naturally appear in the problem.  If $a\, H$ evolves monotonically (for instance like in Figs. 
\ref{FF7a} and \ref{FF8a}) for $a_{1}< a_{2}< a_{r}$ 
we also have $H_{1} \, a_{1}> H_{2} \,a_{2} > H_{r}\, a_{r}$ and this 
hierarchy ultimately implies that $\nu_{1}> \nu_{2} > \nu_{r}$.
On the contrary, if the evolution of $a\, H$ is non-monotonic the hierarchy 
of the frequencies does not reflect the hierarchy of the scale factors. 
Consider, for instance, the situation illustrated in Fig. \ref{FF10a} where 
$a_{1}< a_{2}< a_{r}$ but $H_{r}\, a_{r}$ (corresponding to the second peak) 
is larger than $H_{2}\, a_{2}$ which coincides with the intermediate minimum; in this situation $ \nu_{2} < \nu_{r}$.
\begin{figure}[!ht]
\centering
\includegraphics[height=5.5cm]{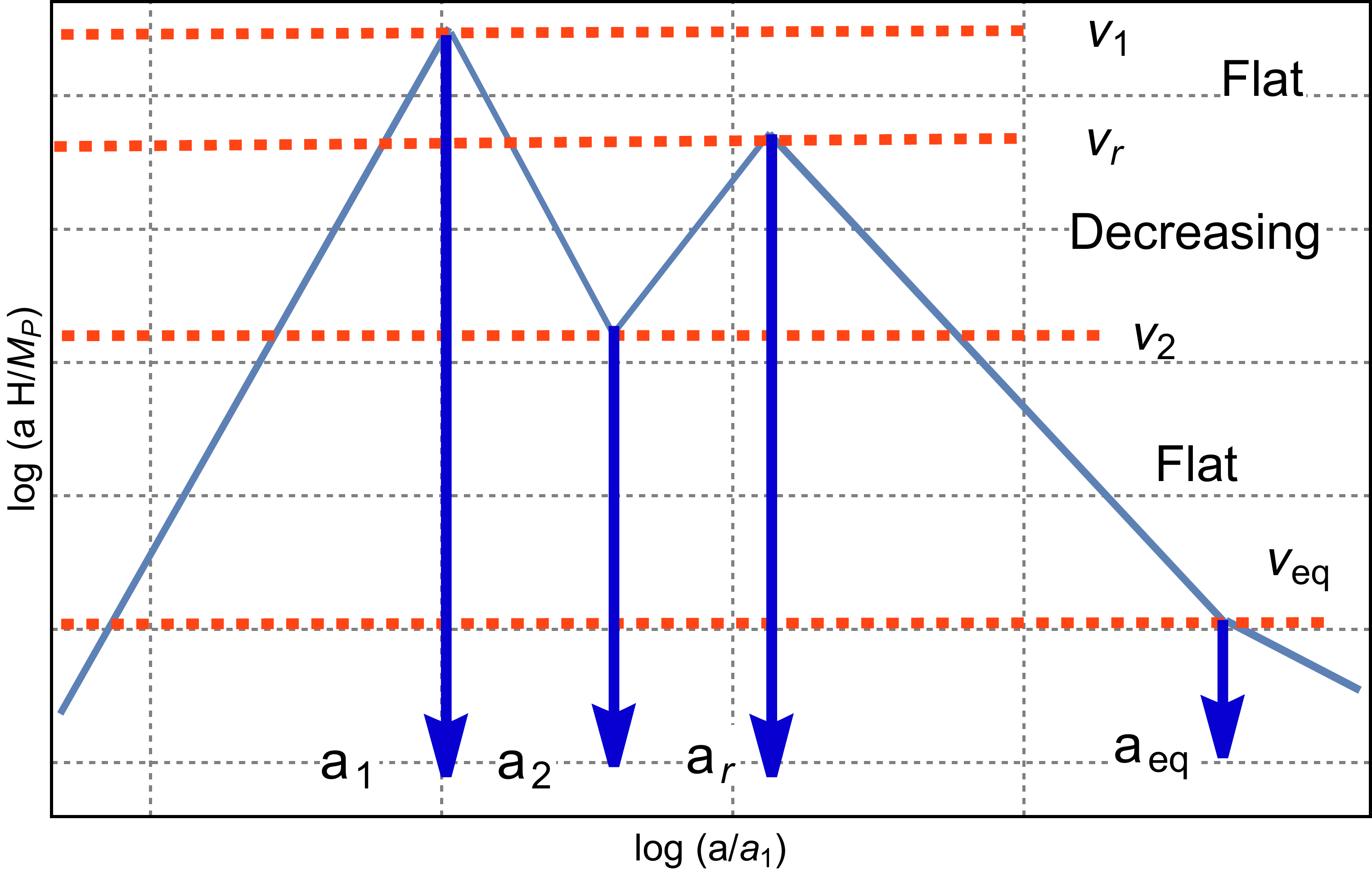}
\includegraphics[height=5.5cm]{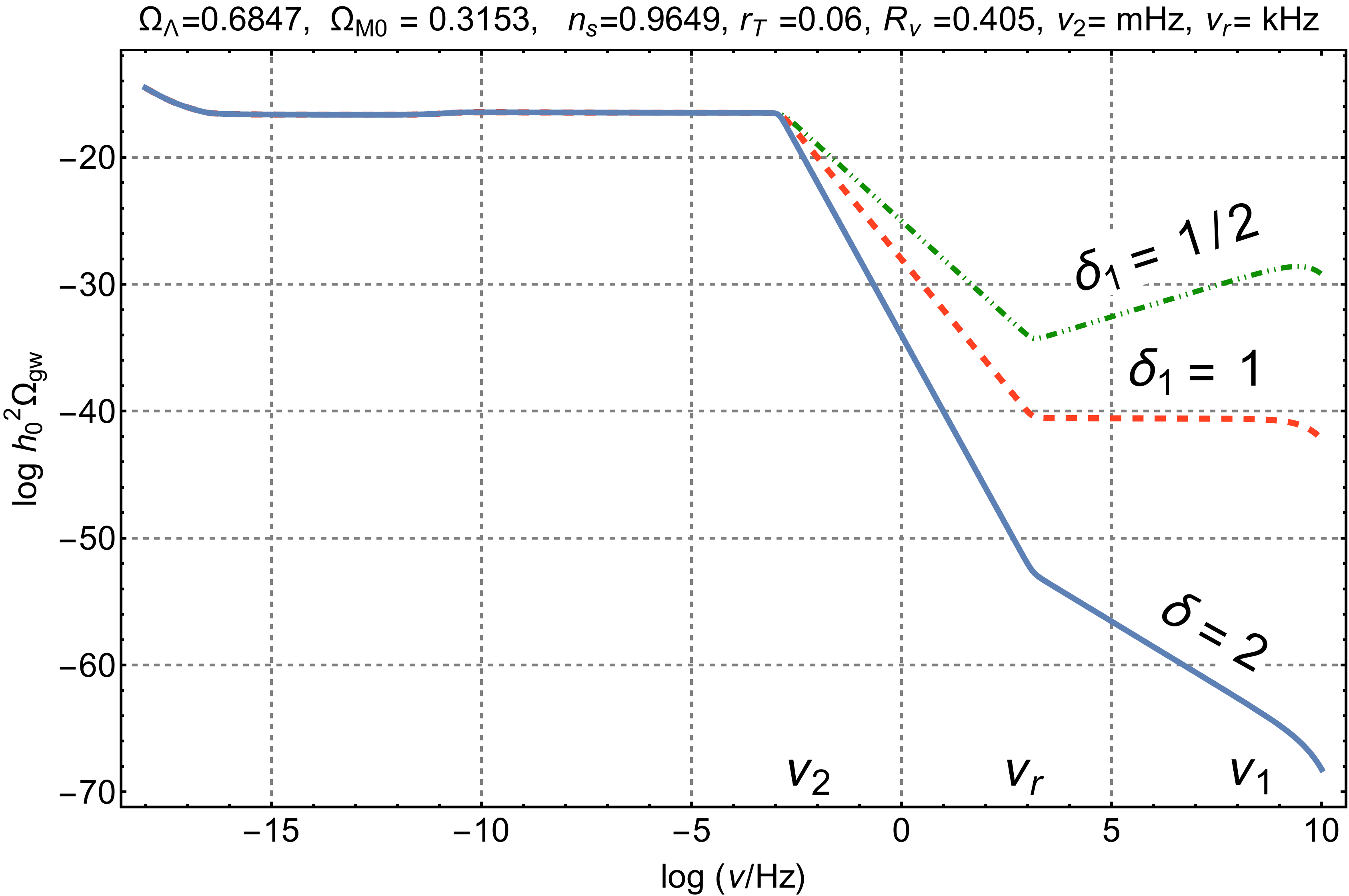}
\caption[a]{In the plot at the left we schematically illustrate $a\, H/M_{P}$ and the different
frequency ranges. In the right panel the corresponding spectral energy density is reported. While 
$\delta_{1}$ changes, the value of $\delta_{2}$ is now fixed, in both plots, by the occurrence of a second 
inflationary phase implying $\delta_{2} \to -1$. For the illustrative choice of the 
parameters of the plot at the right we have 
$\nu_{r} = \mathrm{kHz}$ while $\nu_{2} = \mathrm{mHz}$.}
\label{FF10d}      
\end{figure}
Given that the evolution of $a\, H$ is non-monotonic the different ranges of frequency have been specifically illustrated in the left panel of Fig. \ref{FF10d}. From top to bottom 
we first see the two dashed lines corresponding to $\nu_{1}=\nu_{max}$ and $\nu_{r}$: in this range 
the corresponding wavelengths exit the Hubble radius during inflation and reenter during a decelerated 
stage. Between $\nu_{r}$ and $\nu_{1}$ the spectral slope coincides with 
the one already determined in Eq. (\ref{MULTISP1}). In the right plot of Fig. \ref{FF10d}  three 
different slopes are illustrated (i.e. $\delta_{1}= 1/2$, $\delta_{1} =1$ and $\delta_{1} =2$); the case 
$\delta_{1} = 1$ has been already discussed in Ref. \cite{liddle2} and it implies a flat slope at high-frequencies.

By always following the left panel of Fig. \ref{FF10d} the region $\nu_{2}< \nu< \nu_{r}$ between the two dashed lines 
corresponds to the wavelengths that left the Hubble radius during inflation, reentered for a while during 
the decelerated phase and then exited again during the second stage of inflation. 
With the same analysis leading to Eq. (\ref{SING4d})
 the spectral index  can be written as
\begin{equation}
s_{T} = 4 - 2 (\mu_{1} + \lambda_{1}) -2 (\mu_{2} + \lambda_{2}),
\label{ESTIM}
\end{equation}
where $(\mu_{1}, \, \mu_{2})$ are the Bessel indices associated with the decelerated 
stages of evolution (between $a_1$ and $a_2$ and also after $a_{r}$);  $(\lambda_{1}, \, \lambda_{2})$
are instead the Bessel indices associated with the two inflationary stages (for $a< a_{1}$ and for $a_{2} < a< a_{r}$).  Neglecting the slow-roll corrections we have that during the two successive inflationary stages $\lambda_{1} \simeq \lambda_{2} \simeq 3/2$. During the intermediate decelerated stage we have $\mu_{1} = \delta_{1} -1/2$ and if we assume that after $a_{r}$ the background is dominated by radiation we gate $\mu_{2} =1/2$. 
Putting all together we have that $s_{T} = - 2 (\delta_{1} +1)$ and this result explains why,  
in the right plot of Fig. \ref{FF10d}, the spectral energy density is steeply decreasing in the range\footnote{More specifically, when  $\delta_{1}= 1/2$, $\delta_{1} =1$ and $\delta_{1} =2$ we have, respectively, $s_{T} = -3$, $s_{T} = -4$ and $s_{T} = -6$.
} $\nu_{2}< \nu< \nu_{r}$. 
\subsection{Preliminary summation}
All in all, if there are two successive phases of expansion the main results can be summarized as follows:
\begin{itemize}
\item{} the profiles leading to the largest signal (and hence to the most stringent constraints) 
are the ones illustrated in the left panels of Figs. \ref{FF7a} and \ref{FF8a}; these are also the most promising if the potential signals in the nHz range are viewed as a genuine effect due to relic gravitons;
\item{} however, in spite of the possible presence of a local maximum the combined 
constraints lead to a signal that always undershoots the PTA measurements; 
\item{} in all the remaining cases
the only relevant limits are the ones 
in the aHz region and they are associated, as in the conventional case, with 
the smallness of the tensor to scalar ratio $r_{T}$. 
\end{itemize}
The above results can be generalized to the case of $n$ successive stages 
of expansion. Recalling  Eqs. (\ref{NN1})--(\ref{NN3}) it is possible to show  
that the spectral energy density is 
always decreasing except than in the case of a local maximum. For the $jth$ 
stage of expansion there is a local maximum in the spectral energy density
 if $\delta_{j} <1 $ and $\delta_{j +1} > 1$. At the local maximum occurring for 
$\nu_{j}$ all the previous considerations can be applied. We can also imagine a situation where there is a
maximum for $\nu_{j}$,  a minimum for $\nu_{j+1}$, then again a maximum for $\nu_{j+2}$ and so on. 
From the explicit analytic estimates of the spectral energy density given in Eq. (\ref{SING4d}) 
we can evaluate $\Omega_{gw}(\nu,\tau_{0})$ for the two successive maxima and then take the ratio 
of the obtained results 
\begin{equation}
\frac{\Omega_{gw}(\nu_{j}, \tau_{0})}{\Omega_{gw}(\nu_{j+2}, \tau_{0})} = \biggl(\frac{H_{j}}{H_{j +2}}\biggr)^2 \, 
\biggl(\frac{a_{j}^2 \, H_{j}}{a_{j+2}^2 \, H_{j + 2}}\biggr)^2 > 1,
\label{FIRSTjj}
\end {equation}
where the inequality is a simple consequence of the fact that, in the absence of secondary 
inflationary phases, $a H/M_{P}$ is always decreasing during the post-inflationary stage of expansion. 
From Eq. (\ref{FIRSTjj}) we therefore conclude:
\begin{equation}
\Omega_{gw}(\nu_{j}, \tau_{0}) > \Omega_{gw}(\nu_{j+2}, \tau_{0}), \qquad \mathrm{with} \qquad 
 \nu_{j} > \nu_{j+2}.
\label{SECONDjj}
\end{equation} 
Therefore, according to Eqs. (\ref{FIRSTjj})--(\ref{SECONDjj}),  the most relevant constraints are always associated with the pair of earliest stages of expansion that develop a maximum in the spectral 
energy density. For this maximum all the considerations reported in the present section can be repeated 
and they are qualitatively unaltered.  If the intermediate phases are sufficiently short the succession 
of the maxima approximately reproduces the spectrum of a single phase with 
$\xi= H_{r}/H_{1}$. When the post-inflationary stage consists 
of multiple phases expanding, respectively, faster and slower than radiation 
the limits imposed by the BBN considerations and by the KLV bounds are satisfied but the corresponding signal in the nHz band is always smaller than the PTA observations. Future measurements below and around the Hz might be extremely relevant for 
direct constraints on the expansion rate just before the BBN stage and around the 
electroweak time. In spite of the obvious 
problems of the seismic noises (that are customarily addressed by considering 
space-borne detectors) we regard as particularly 
interesting, in this respect,  the case for the atomic gravitational wave interferometric sensors described 
in Refs.  \cite{SENS2,SENS3,SENS4}.

\renewcommand{\theequation}{5.\arabic{equation}}
\setcounter{equation}{0}
\section{Refractive index and related spectra}
\label{sec5}

\subsection{The effective evolution the expansion rate} 
The results of the two previous sections will now be complemented  with the physical situation described in Fig. \ref{FF2} where the evolution of the expansion rate is modified during inflation. As we shall see, in this case, the role of $n_{T}$ (the high-frequency slope) and $\overline{n}_{T}$ (the intermediate-frequency slope) 
are exchanged: while in section \ref{sec3} $n_{T}$ was typically positive 
and $\overline{n}_{T} \ll 1$, in the present section we shall basically  have 
the opposite, namely $\overline{n}_{T}>0$ and $n_{T} \ll 1$.
This is in fact what happens when the refractive index of the relic gravitons is dynamical during the early stages of the inflationary expansion. The basic observation is that while they propagate in curved backgrounds, the gravitational waves may acquire an effective index of refraction \cite{CC1a,CC1b}.  It was later observed \cite{CC2} that even if the geometry undergoes a stage of conventional accelerated expansion the intermediate slope 
of the spectral energy density increases depending on the evolution of the refractive index.
\begin{figure}[!ht]
\centering
\includegraphics[height=5.6cm]{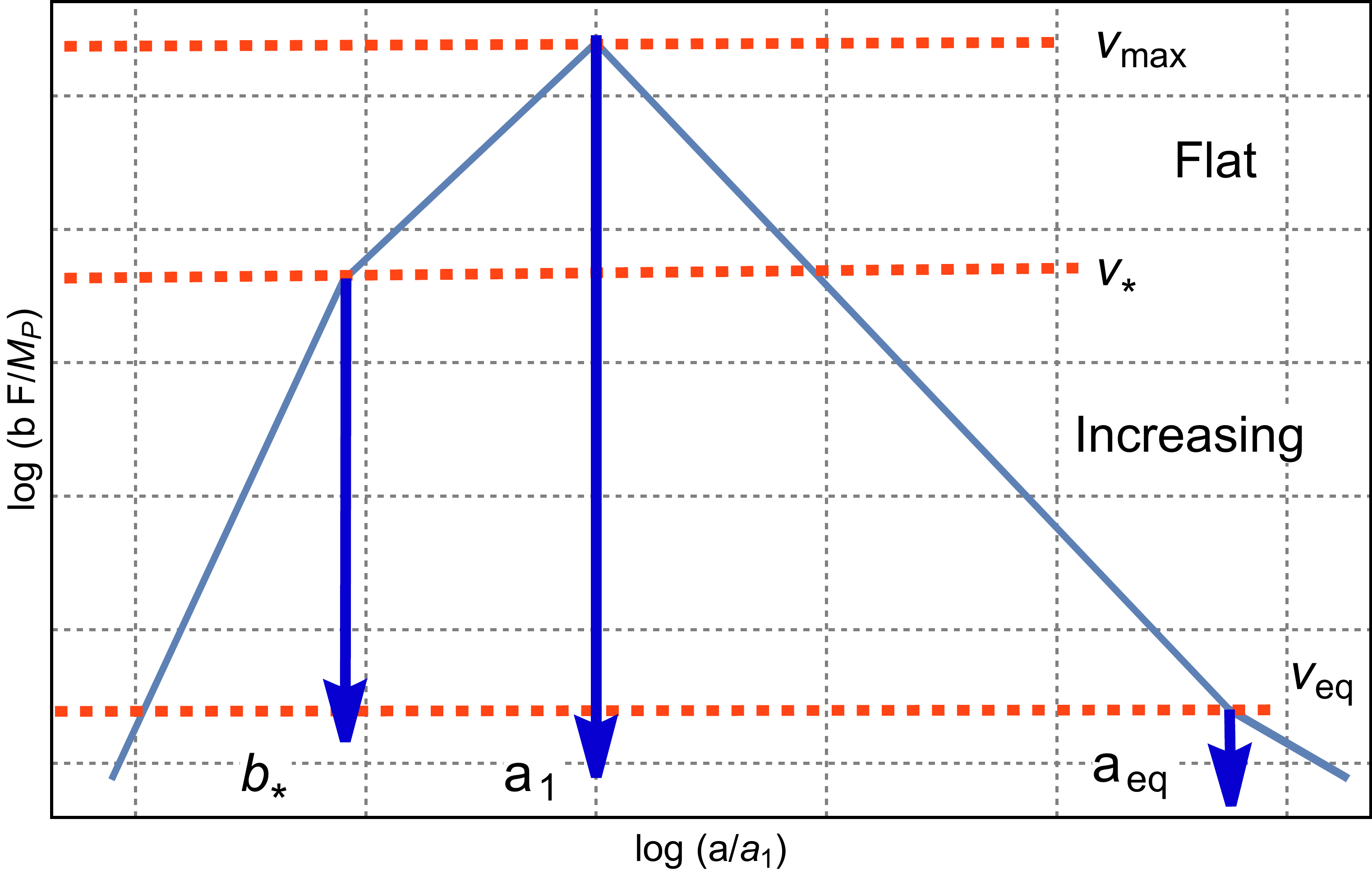}
\includegraphics[height=5.6cm]{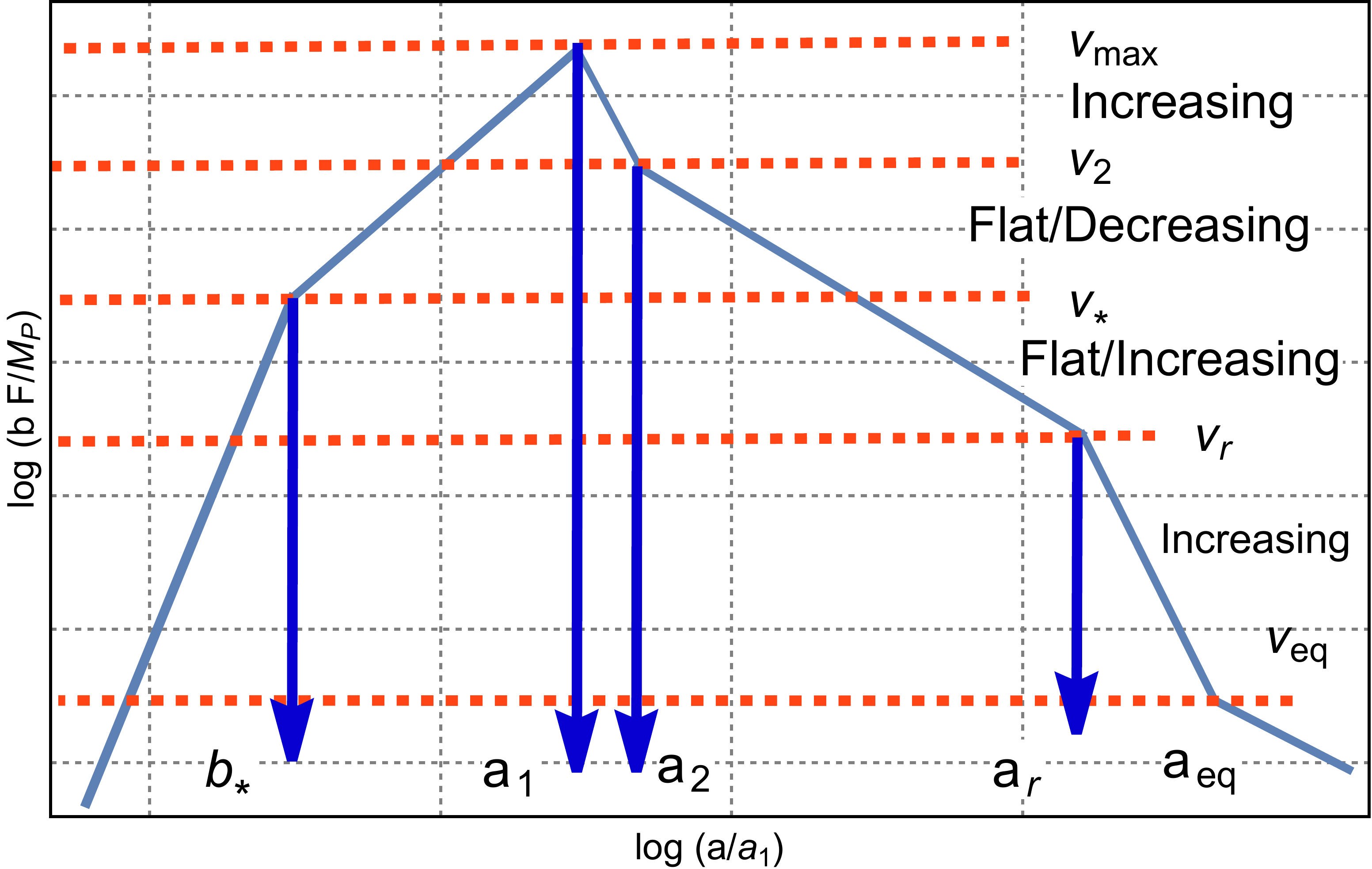}
\caption[a]{In the plot at the left we report a profile for the effective expansion 
rate where the refractive phase takes place during inflation and then the radiation background dominates after inflation. In the right plot the dominance of radiation is delayed. We not that on the vertical axis we report $F = \dot{b}/b$ where 
the overdot now denotes a derivation with respect to the $\eta$-time that differs from the conformal time during the refractive stage but coincides 
with $\tau$ time after inflation.}
\label{FF10g}      
\end{figure}
A concrete realisation of the evolution outlined in the profiles of Fig. \ref{FF2} is illustrated in Fig. \ref{FF10g} where the refractive index evolves in the initial stages of inflation 
and then the post-inflationary evolution is either dominated by radiation (as in the left plot of Fig. \ref{FF10g}) or complemented by some other intermediate stage\footnote{All the considerations of the previous sections could 
be repeated in the presence of a dynamical refractive index. For the sake of conciseness 
the focus will now be on the two options illustrated in Fig. \ref{FF10g} where
the post-inflationary stage is either dominated by radiation or it expands 
initially at a rate that is slower than radiation between $a_{1}$ and $a_{2}$.}. 

The logic behind the suggestion of Fig. \ref{FF10g} 
is that the effective action of single-field inflationary models involves all the different terms that include four derivatives and are suppressed by the negative powers of a large mass scale \cite{ONEW}. There are non-generic models of inflation where the higher-order corrections assume a specific form since the inflaton has some particular symmetry or because the rate of inflaton roll remains
constant (and possibly larger than $1$). This happens in certain fast-roll scenarios \cite{NON1} (see also, for instance, \cite{NON2,NON3}) but other examples involve the higher-order curvature corrections  given in terms of the Gauss-Bonnet combination and weighted (in four space-time dimensions) by some inflaton dependent-coupling \cite{NON4,NON5,NON6}. In \cite{CC2} (see also \cite{CC5,CC3}) it has been argued that in all these situations the effective action of the relic gravitons is modified and ultimately assumes the following general form:
\begin{equation}
S_{g} = \frac{1}{8 \ell_{P}^2} \int d^{4} x \biggl[ A(\tau)\, \,\partial_{\tau} h_{i j} \, \partial_{\tau} h^{i j} -  B(\tau)\, \partial_{k} h_{i j} \partial^{k} h^{i j} \biggr].
\label{ONE}
\end{equation}
If parity breaking terms are included in the effective action \cite{TWO}, the relic graviton background may be polarized but this possibility has been already discussed in a related context \cite{FOUR} and will not be 
specifically analyzed here.  While both terms $A(\tau)$ and $B(\tau)$ depend on the conformal time coordinate $\tau$ we can always factor $A(\tau)$ and introduce an effective refractive index $n(\tau)$ associated with the interactions with the background geometry \cite{CC1a,CC1b,CC2}:
 \begin{equation}
S_{g} = \frac{1}{8 \ell_{P}^2} \int d^{4} x \,\,A(\tau) \,\,\biggl[ \partial_{\tau} h_{i j} \, \partial_{\tau} h^{i j} - \frac{\partial_{k} h_{i j} \partial^{k} h^{i j}}{n^2(\tau)}  \biggr],\qquad\qquad n^2(\tau) = \frac{A(\tau)}{B(\tau)}.
\label{TWO}
\end{equation}
After Eq. (\ref{TWO}) has been proposed in Ref. \cite{CC2} apparently different parametrizations have been later introduced and the difference between these strategies consists in modifying the first term (rather than the second) inside the squared bracket of Eq. (\ref{TWO}):
 \begin{equation}
S_{g} = \frac{1}{8 \ell_{P}^2} \int d^{4} x \,\,B(\tau) \,\,\biggl[ n^2(\tau) \partial_{\tau} h_{i j} \, \partial_{\tau} h^{i j} - \partial_{k} h_{i j} \partial^{k} h^{i j} \biggr].
\label{TWOa}
\end{equation}
This choice is immaterial since the two parametrizations of the effect are related by a rescaling of the four-dimensional metric through a conformal factor that involves the refractive index itself \cite{CC5,CC3}. In spite 
of the preferred parametrization,  Eqs. (\ref{TWO})--(\ref{TWOa}) can always be rephrased in terms of a new time coordinate conventionally referred to as the $\eta$-time. So, from Eq. (\ref{TWO}) we can deduce:
 \begin{equation}
S_{g} = \frac{1}{8 \ell_{P}^2} \int \,\,d^{4} x \,\,b^2(\eta)\,\, \biggl[ \partial_{\eta} h_{i j} \,\, \partial_{\eta} h^{i j} - \partial_{k} h_{i j} \,\,\partial^{k} h^{i j} \biggr],
 \qquad b(\eta) = \frac{a(\eta)}{\sqrt{n(\eta)}},
\label{THREE}
\end{equation}
where $b(\eta) = a(\eta)/\sqrt{n(\eta)}$ and the scale factor is assumed everywhere continuous with its first derivative; the $\eta$-time parametrzation is defined by $n(\eta) \, d\eta \,=\,d\tau$. The profile of the effective 
expansion rate given Fig. \ref{FF10g} where we introduced $F = \dot{b}/b$ where the overdot denotes, in this context,  a derivation with respect to the $\eta$ coordinate introduced in Eq. (\ref{THREE}), and not a derivation with respect to the cosmic time as usually implied\footnote{After Eq. (\ref{tauHa}) we introduced the standard definition of the slow-roll parameter and denoted with an overdot the derivation 
with respect to the cosmic time coordinate $t$. In this section, however, the overdot 
will {\em only} denote a derivation with respect to the $\eta$-time. With this caveat we believe that no confusion is possible. }. Equation (\ref{THREE}) generalizes the standard Ford-Parker action \cite{AC2,AC3} to the case of a dynamical refractive index. 

The evolution of the refractive index is specified unambiguously by assigning $n(a)$.  Even though the phase velocity of the relic gravitons is not required to be sub-luminal we consider here the situation where $n(a) \geq 1$. The situation described in Fig. \ref{FF10g} is reproduced when $n(a)$ changes appreciably during inflation and it goes to $1$ in the standard decelerated stage of expansion\footnote{In Eq. (\ref{NEX}) $a_{i}$ and $a_{1}$ mark, respectively, the beginning and the end of the inflationary epoch; $a_{*}$ defines the boundary of the refractive stage and $N_{*}$ is the corresponding number of $e$-folds.}:
\begin{equation}
n(a) = n_{\ast} \frac{ (a/a_{\ast})^{\alpha} \,\,e^{- \gamma (a/a_{1})}}{(a/a_{*})^{\alpha} + 1} + 1, \qquad\qquad
n_{\ast} = n_{i} (a_{\ast}/a_{i})^{\alpha} = n_{i} e^{\alpha \, N_{\ast}}.
\label{NEX}
\end{equation}
Equation (\ref{NEX}) defines, in practice, three successive physical regimes \cite{PPNN}. For $a \gg a_{1}$ the refractive index goes to $1$ and the standard situation is recovered
depending on the value of $\gamma\geq 1$ which controls the sharpness of the transition.
When $a_{*} < a < a_{1}$ the refractive index is practically constant but still larger than $1$, i.e. $n(a)\simeq n_{\ast} > 1$.
Finally for $a< a_{\ast}$ we have the truly refractive stage where $n (a) \simeq n_{\ast} (a/a_{\ast})^{\alpha}$.  

 In the plot at the left of Fig. \ref{FF10g} the refractive stage precedes $b_{*}$ and after inflation the radiation background dominates down to the equality time. In this case at intermediate frequencies (between $\nu_{eq}$ and $\nu_{*}$) the 
spectral energy density increases while it is flat (or slightly decreasing) for $\nu > \nu_{*}$. 
The frequency dependence of $h_{0}^2\,\Omega_{gw}(\nu,\tau_{0})$ can be more complicated 
if the post-inflationary evolution is not immediately dominated by radiation. For instance 
in the plot at the right in Fig. \ref{FF10g} the post-inflationary evolution is first slower than radiation.
In this case it is possible to have an increasing slope in the high-frequency region. 

\subsection{Typical frequencies and spectral energy density}
As in the case of Fig. \ref{FF10d}, the pivotal frequencies of the 
spectrum can be deduced by looking at the dashed lines 
reported in Fig. \ref{FF10g}. The range $\nu_{eq} < \nu < \nu_{\ast}$ (where $\nu_{\ast} = k_{\ast}/(2 \pi)$ and $k_{\ast} = 1/\eta_{\ast}$) correspond to the wavelengths leaving the Hubble radius during the refractive stage and reentering when the background is dominated by radiation. The frequencies $\nu_{\ast} < \nu < \nu_{max}$ are instead associated with the wavelengths crossing the Hubble radius during the inflationary 
phase and reentering either in the decelerated phase (i.e. either during radiation 
or during some decelerated epoch not necessarily coinciding with radiation).

Since the wavelengths crossing the Hubble
when $n(a)\to 1$ are not affected by the evolution of the 
refractive index the maximal frequency does not change in comparison 
with the previous cases and it is ${\mathcal O}(200)$ MHz. In particular,
is we assume that the post-inflationary phase is dominated by radiation (as in the 
left profile in Fig. \ref{FF10g}) the maximal frequency $\overline{\nu}_{max}$ has the same expression of Eq. (\ref{SING3a}). In terms of the maximal frequency it is 
possible to estimate the value of $\nu_{\ast}$ that defines, as we shall see in a moment, the knee of the spectrum:
\begin{equation}
\nu_{\ast} = \biggl(1 + \frac{\alpha}{1 - \epsilon}\biggr) e^{N_{\ast} (\alpha+1) - N_{t}} \, \, \nu_{max},\qquad\qquad \nu_{max} = \overline{\nu}_{max},
\label{nuast}
\end{equation}
where $N_{t}$ is the total number of $e$-folds while, as already mentioned, $N_{\ast} = \ln{(a_{\ast}/a_{i})}$ has been already introduced in Eq. (\ref{NEX}). 
 In Eq. (\ref{nuast}) as well as in the forthcoming discussion we shall always be assuming that $n_{i} \to 1$; different choices are possible (provided $n_{i} \geq 1$) but their effect does not modify the conclusions since the value of $n_{i}$ can always be traded for a shorter refractive phase. The relevant point to appreciate here is that $\nu_{\ast}$ controls the typical frequency of the knee of the spectrum and it does depend on $N_{*}$, $N_{t}$ and $\alpha$. 

In the $\eta$-time the Hamiltonian associated with Eq. (\ref{THREE}) is simpler 
than in the conformal time coordinate $\tau$ and it is given by:
\begin{equation}
H_{g}(\eta) = \int d^{3} x \biggl[ \frac{8 \ell_{P}^2}{b^2} \pi_{i\,j} \, \pi^{\,i \, j} + \frac{b^2}{8 \ell_{P}^2} \partial_{k} h_{i\,j} \, 
\partial^{k} h^{\, i \, j} \biggr], \qquad\qquad \pi_{i\, j} = \frac{b^2}{8 \ell_{P}^2} \partial_{\eta} h_{i\,j}.
\label{THREEa}
\end{equation}
There is an obvious similarity between Eqs. (\ref{SEC1four}) and (\ref{THREEa}) but we remind that $\eta$ and $\tau$ only coincide for after the end of inflation; the same is true for $b(\eta)$ and $a(\tau)$.  
After promoting the classical fields and their conjugate momenta  to the status of quantum operators, from Eq. (\ref{THREEa}) the governing equations for $\widehat{h}_{i\,j}$ and 
 $\widehat{\pi}_{i\,j}$ have the same content of the ones already discussed in Eq. (\ref{HPI}) with the difference that now the time variable is $\eta$. In particular the evolution of the mode functions is now\footnote{Again we note that there is a formal similarity 
between Eq. (\ref{SEC1seven}) and Eq. (\ref{FIVE}) with the caveat that 
$\eta \to \tau$ and $b(\eta) \to a(\tau)$ only after the end of inflation.}:
\begin{equation}
\dot{G}_{k,\,\lambda} = - k^2 \,b^2 \, F_{k,\,\lambda}, \qquad \dot{F}_{k, \, \lambda} = \frac{G_{k,\, \lambda}}{b^2},
\label{FIVE}
\end{equation}
where we recall, as already mentioned after Eq. (\ref{THREE}) and in the caption of Fig. \ref{FF10g} that 
the overdot now denotes a derivation with respect to $\eta$ (i.e. $\dot{F}_{k, \, \lambda} = 
\partial_{\eta} \,F_{k, \, \lambda}$).  

The mode functions are normalized during the refractive phase where, according to Eq. (\ref{NEX}), the $\eta$-time and the conformal time coordinates are related as $(- \eta/\eta_{\ast}) = 
(- \tau/\tau_{\ast})^{\alpha/(1- \epsilon) +1}$ and $\eta_{\ast} = \tau_{\ast} (1 - \epsilon)/[n_{\ast} 
(1 + \alpha -\epsilon)]$; both relations follow from the definition of the 
$\eta$-time (i.e. $n(\eta) d\eta =d\tau$) and also from 
$n(a) = n_{*} (a/a_{\ast})^{\alpha}$, as implied by Eq. (\ref{NEX}) when $a< a_{\ast}$. In the $\eta$-time the explicit expression of $b(\eta)$ is: 
\begin{equation}
b(\eta) = b_{\ast} \biggl(-\frac{\eta}{\eta_{\ast}}\biggr)^{-\sigma}, \qquad 
\sigma= \frac{2- \alpha}{2( 1 + \alpha - \epsilon)}, \qquad \mathrm{for}\qquad \eta < - \eta_{\ast},
\label{EIGHT1}
\end{equation}
where $b_{\ast} = a_{\ast}/\sqrt{n_{\ast}}$. In the refractive regime the solution of Eq. 
(\ref{FIVE}) is therefore given by:
\begin{equation}
F_{k}(\eta) = \frac{{\mathcal N}}{\sqrt{ 2 k} \, b(\eta)} \, \sqrt{- k \eta} \, H_{\mu}^{(1)}(- k\, \eta), \qquad \qquad G_{k}(\eta) = - {\mathcal N} \, b(\eta)\,\sqrt{\frac{k}{2}}  \,\sqrt{- k\eta} \, H_{\mu-1}^{(1)}(- k \eta), 
\label{EIGHT2}
\end{equation}
where $\mu = \sigma +1/2$ is the Bessel index \cite{abr1,abr2}. Note also that ${\mathcal N}$ is complex but because of the Wronskian normalization 
condition (i.e. $F_{k} G_{k}^{\ast} - F_{k}^{\ast} G_{k} =\,i$)  that preserves the canonical commutation relations, 
 $|{\mathcal N}| = \sqrt{\pi/2}$ and the modulus of ${\mathcal N}$ is fixed. Equation (\ref{EIGHT2}) is actually the analog of Eq. (\ref{INCOND1}) that has been used in order to set the initial conditions in the absence of a refractive index. 
 As in Eq. (\ref{INCOND1}) $H_{\mu}^{(1)}(-k\eta)$ is the Hankel function of first kind \cite{abr1,abr2} with the crucial difference that $\eta$ only coincides with $\tau$ when $n(a) \to 1$, which is not the case during the refractive stage. From Eq. (\ref{EIGHT2}) we can compute the tensor power spectrum and eventually determine the low-frequency normalization of the spectral energy density:
\begin{eqnarray}
 P_{T}(k, \eta) &=& 
\biggl(\frac{H_{1}}{M_{P}}\biggr)^2\,\,{\mathcal C}(n_{T}, N_{*}, N_{t}, \epsilon) \,\, \biggl(\frac{k}{a_{1} H_{1}}\biggr)^{\overline{n}_{T}},
 \label{EIGHT3}\\
 \qquad {\mathcal C}(n_{T}, N_{*}, N_{t}, \epsilon) &=& \frac{ 2^{6 - \overline{n}_{T}}}{\pi^2} \biggl| 1 + \frac{\alpha}{1 -\epsilon}\biggr|^{2 - \overline{n}_{T}}\, \Gamma^2\biggl(\frac{3 - \overline{n}_{T}}{2}\biggr)e^{\alpha\, N_{\ast}( 3 - \overline{n}_{T}) - \overline{n}_{T} (N_{\ast} - N_{t})}.
 \label{EIGHT3a}
 \end{eqnarray}
Equation (\ref{EIGHT3}) corresponds 
to the limit $| k \eta| \ll 1$ where $P_{T}(k, \eta)$ becomes actually constant in time.
 In this regime the tensor to scalar ratio becomes:
\begin{equation}
r_{T}(\nu) = \frac{\epsilon}{\pi} \,  {\mathcal C}(n_{T}, N_{*}, N_{t}, \epsilon) \biggl(\frac{\nu}{\nu_{max}}\biggr)^{\overline{n}_{T}}. 
\label{EIGHT5}
\end{equation}
In the limit $\alpha\to 0$ we have that ${\mathcal C}(n_{T}, N_{*}, N_{t}, \epsilon)\to 16\pi$ so that $r_{T}\to 16 \epsilon$ and the standard consistency 
condition is recovered. In the same limit the tensor spectral index goes to $ - 2\epsilon$: 
\begin{equation}
\overline{n}_{T} = \frac{3 \alpha - 2 \epsilon}{1+ \alpha - \epsilon} = \frac{3 \alpha}{1 + \alpha} + 
 \frac{\epsilon (\alpha - 2)}{(1 + \alpha)^2} + {\mathcal O}(\epsilon^2).
\label{EIGHT4}
\end{equation}
Equation (\ref{EIGHT4}) defines the tensor spectral index {\em in the intermediate 
frequency range}  where $\alpha$ is  larger than $\epsilon$ and this 
is why the exact result can always be expanded in the limit $\epsilon \ll 1$. 
 The slope of the tensor 
power spectrum evaluated for the wavelengths larger than the Hubble radius 
coincides, after reentry, with the slope of spectral energy density before the knee:
\begin{equation}
h_{0}^2 \,\,\Omega_{\mathrm{gw}}(\nu,\tau_{0}) = {\mathcal N}_{\rho} \,\,r_{T}(\nu_{p})\,\, {\mathcal T}_{low}^2(\nu, \nu_{eq})  \,\,\biggl(\frac{\nu}{\nu_{p}}\biggr)^{\overline{n}_{T}}, \qquad \nu < \nu_{\ast},
\label{EIGHT6}
\end{equation}
Above $\nu_{\ast}$ the spectral energy density is instead quasi-flat, as anticipated in Fig. \ref{FF10g} 
and it is approximately given by:
\begin{equation}
h_{0}^2\,\Omega_{\mathrm{gw}}(\nu,\tau_{0}) = {\mathcal N}_{\rho} \,\,r_{T}(\nu_{p}) \,\, {\mathcal T}_{low}^2(\nu, \nu_{eq})\,\,\biggl(\frac{\nu_{\ast}}{\nu_{p}}\biggr)^{\overline{n}_{T}}\,\,\biggl(\frac{\nu}{\nu_{\ast}}\biggr)^{n_{T}}\,  \qquad \nu_{\ast} < \nu < \nu_{max},
\label{EIGHT8}
\end{equation}
where $n_{T} \simeq - 2 \, \epsilon$ is the high-frequency spectral index.
where, as before, the contribution of ${\mathcal T}_{low}^2(\nu, \nu_{eq})$ is, in practice, frequency-independent for 
$\nu > \nu_{\ast}$. To avoid possible confusions it should be clear that we are here using 
exactly the same notations employed in the previous section where 
$\overline{n}_{T}$ denotes the spectral index at intermediate frequencies 
which is typically blue in this discussion (but is red in the conventional situation) while $n_{T}$ is the spectral index at high-frequency (which is now flat or even red but was blue in some examples of the previous section).
The results of Eqs. (\ref{EIGHT6})--(\ref{EIGHT8}) agree with the previous estimates of Ref. \cite{CC2} and are a natural candidate for exploring a potential signal in the nHz range \cite{PPNN}. This expectation will now be scrutinized
in connection with the perspective of this analysis.

\subsection{The PTA data and the other phenomenological constraints}
\begin{figure}[!ht]
\centering
\includegraphics[height=7cm]{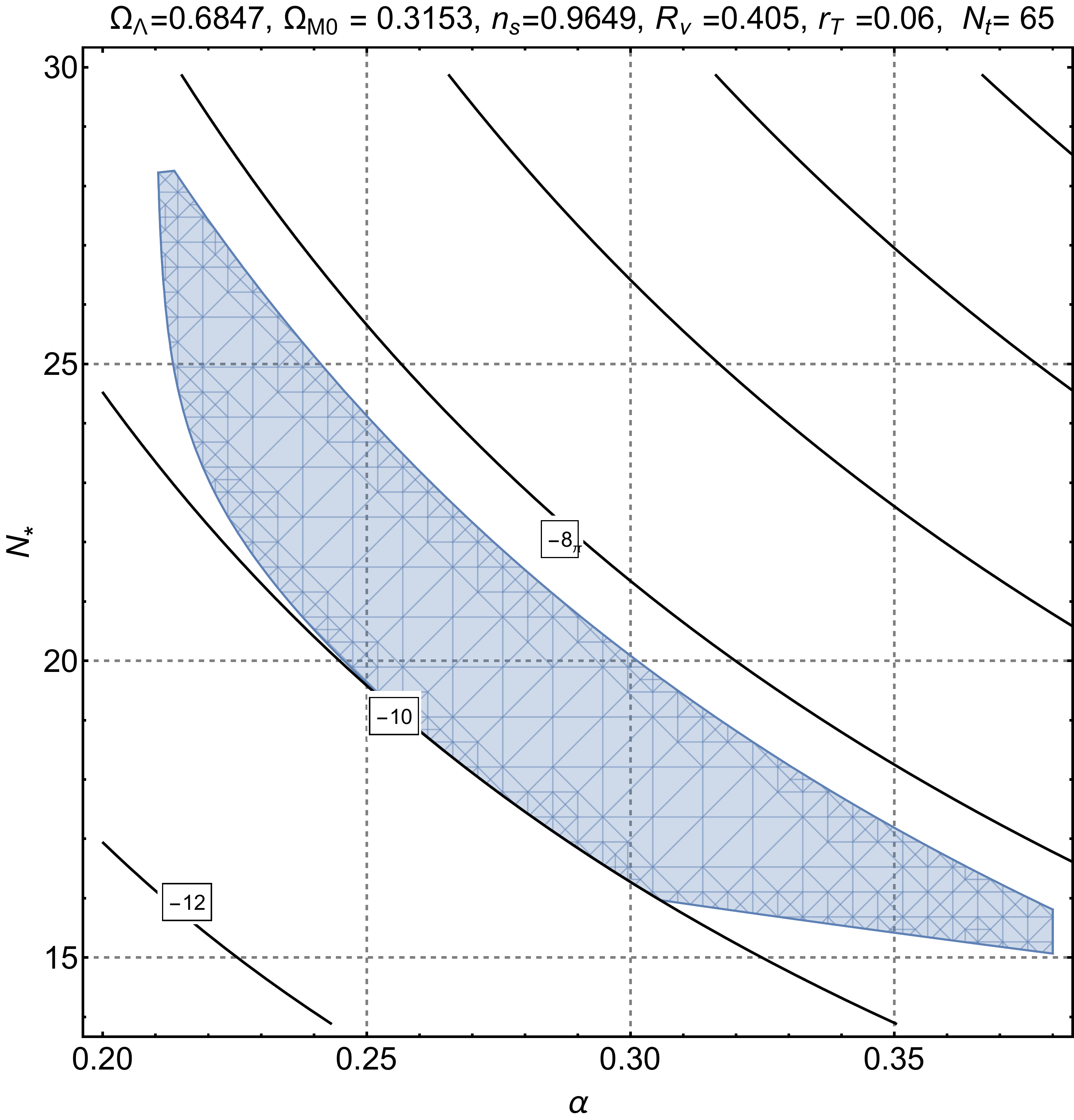}
\includegraphics[height=7cm]{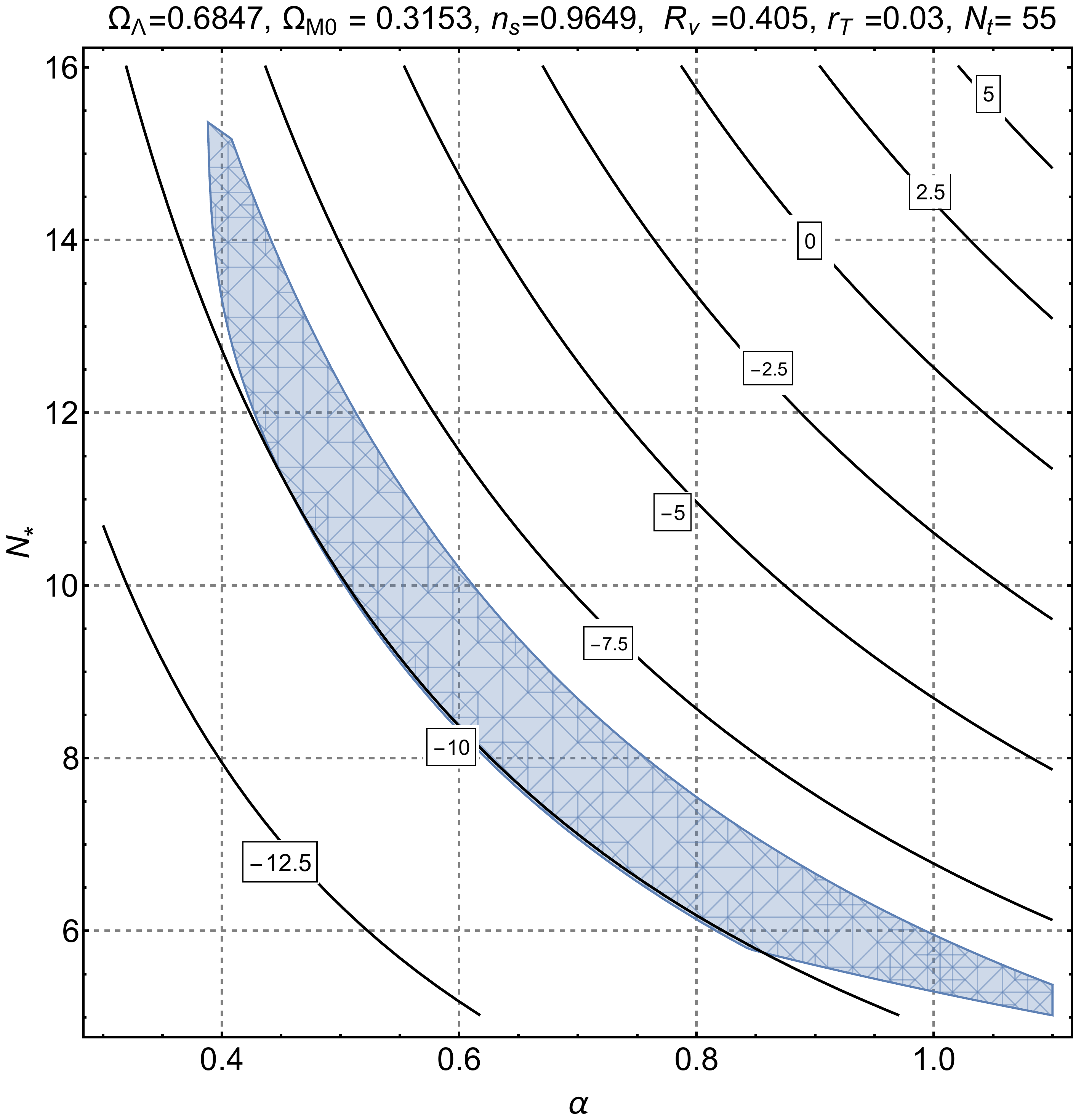}
\caption[a]{The shaded areas in both plots illustrate the allowed regions 
of the parameter space where the spectral energy density is 
compatible with the PTA measurements (see Eq. (\ref{PTAb1}) and also the 
left plot in Fig. \ref{FF10e}). The different contours 
appearing in the plots correspond to the values of $h_{0}^2 \, \Omega_{gw}(\nu_{KLV}, \tau_{0})$ 
where $\nu_{KLV}$ approximately denotes most sensitive frequency domain  
of the wide-band interferometers (see Eqs. (\ref{CONS2}) and (\ref{NOT1})--(\ref{NOT2})). Within the allowed regions the BBN constraints of Eqs. (\ref{CC2})--(\ref{CC3}) are always satisfied.}
\label{FIG2}      
\end{figure}

The evolution of refractive index of the gravitons 
during the early stages of the inflationary evolution leads to a blue (i.e. slightly increasing) slope
of the spectral energy density at intermediate frequencies above the fHz. 
The specific values of the slopes are determined by the competition of the slow-roll parameter and of the rate of variation of the refractive index. The general idea explored here suggests that increasing frequency spectra can also be obtained in the framework of conventional inflationary 
scenarios. 
The pivotal parameters that determine the spectrum are primarily $\alpha$, $N_{\ast}$ and $N_{t}$. When $N_{*}$ and $N_{t}$ are of the same order the transition to normalcy occurs at the end of inflation
but in this case it is impossible to get a large signal in the nHz range without jeopardizing the big-bang nucleosynthesis constraint of Eqs. (\ref{CC2})--(\ref{CC3}).  If we ought to address the PTA measurements (see Eq. (\ref{PTAb1}) 
and discussion thereafter) we must require  $N_{*} <N_{t}$ since, in this case, the transition to normalcy takes place before the onset of the radiation-dominated epoch (i.e. when the background is still inflating deep inside the quasi-de Sitter stage of expansion). In Fig. \ref{FIG2} we therefore illustrate the allowed region that also leads to a large 
signal in the nHz range. In the plot at the left we choose $N_{t} =65$ and set $r_{T}=0.06$ \cite{RT1,RT2,RT3}, as already discussed in the previous sections. In the exclusion plot 
at the right, for the sake of illustration, we consider a shorter inflationary stage
(i.e. $N_{t} =55$)  and $r_{T} =0.03$.
The shaded areas appearing in both plots of Fig. \ref{FIG2} are constructed by requiring that the spectral energy density is sufficiently large at intermediate frequencies as required by the considerations 
related to Eq. (\ref{PTAb1}). If the parameters fall in the shaded 
area of Fig. \ref{FIG2}  the spectral energy density in critical units
is within the limits provided by the PTA collaborations. For larger frequencies 
in the audio band we instead enforce the KLV bound discussed in Eqs. (\ref{CONS2}) and 
(\ref{NOT1})--(\ref{NOT2}). In the simplest situation where the post-inflationary evolution 
is dominated by radiation these two classes of constraints imply the BBN bounds of Eqs. (\ref{CC2})--(\ref{CC3}). However if the post-inflationary evolution expands at a rate that is slower 
than radiation (as suggested in the right plot of Fig. \ref{FF10g}) the BBN limit must be separately 
imposed. The exclusion plots of Fig. \ref{FIG2} can be presented in slightly different 
manners which are, however, equivalent. For instance we could fix $\alpha$ and study the allowed region in the plane $(N_{\ast}, \, N_{t})$.  An amusing value of $\alpha$ (which is incidentally falling 
within the allowed region of Fig. \ref{FIG2}) is $\alpha = 2/7$. In this case, according to Eq. 
(\ref{EIGHT4}) the spectral index at intermediate frequencies is given by $n_{T} \simeq 2/3 $, up to 
 slow-roll corrections ${\mathcal O}(\epsilon)$ which are negligible since $\epsilon < 10^{-3}$. The PTA results are often reported in terms of a chirp amplitude scaling as $\nu^{-2/3}$ for a typical reference frequency ${\mathcal O}(\mathrm{yr}^{-1})$. In the language of Eqs. 
(\ref{NOTT1})--(\ref{NOTT8})  the value of $\alpha$ corresponds to $\beta= -2/3$ is obtained by setting $n_{T} \simeq 2/3$ in Eq. (\ref{EIGHT4}). 
Consequently we have that $\alpha = (2 + 4 \epsilon)/7$ 
which can be approximated as $\alpha = 2/7 + {\mathcal O}(\epsilon)$ 
since $\epsilon < 10^{-3}$. From this kind of analysis we can infer, for instance, that for $\alpha\simeq 2/7$  the maximal signal of the model occurs when $N_{t} = {\mathcal O}(60)$ and $N_{\ast} = {\mathcal O}(20)$.

The results of Fig. \ref{FIG2} suggest that a large signal in the nHz range is obtained when the variation of the refractive index occurs sufficiently early during the inflationary stage and anyway not beyond the first $20$ $e$-folds. In units of the inflationary Hubble rate, the rate of variation of the refractive index must roughly fall in the range $0.2 < \alpha < 0.5$.   When the parameters are selected within the shaded region the corresponding spectral energy density falls within the PTA box 
and the frequency of the knee is in the nHz range while the limits coming from wide-band interferometers are satisfied.  
\begin{figure}[!ht]
\centering
\includegraphics[height=5.8cm]{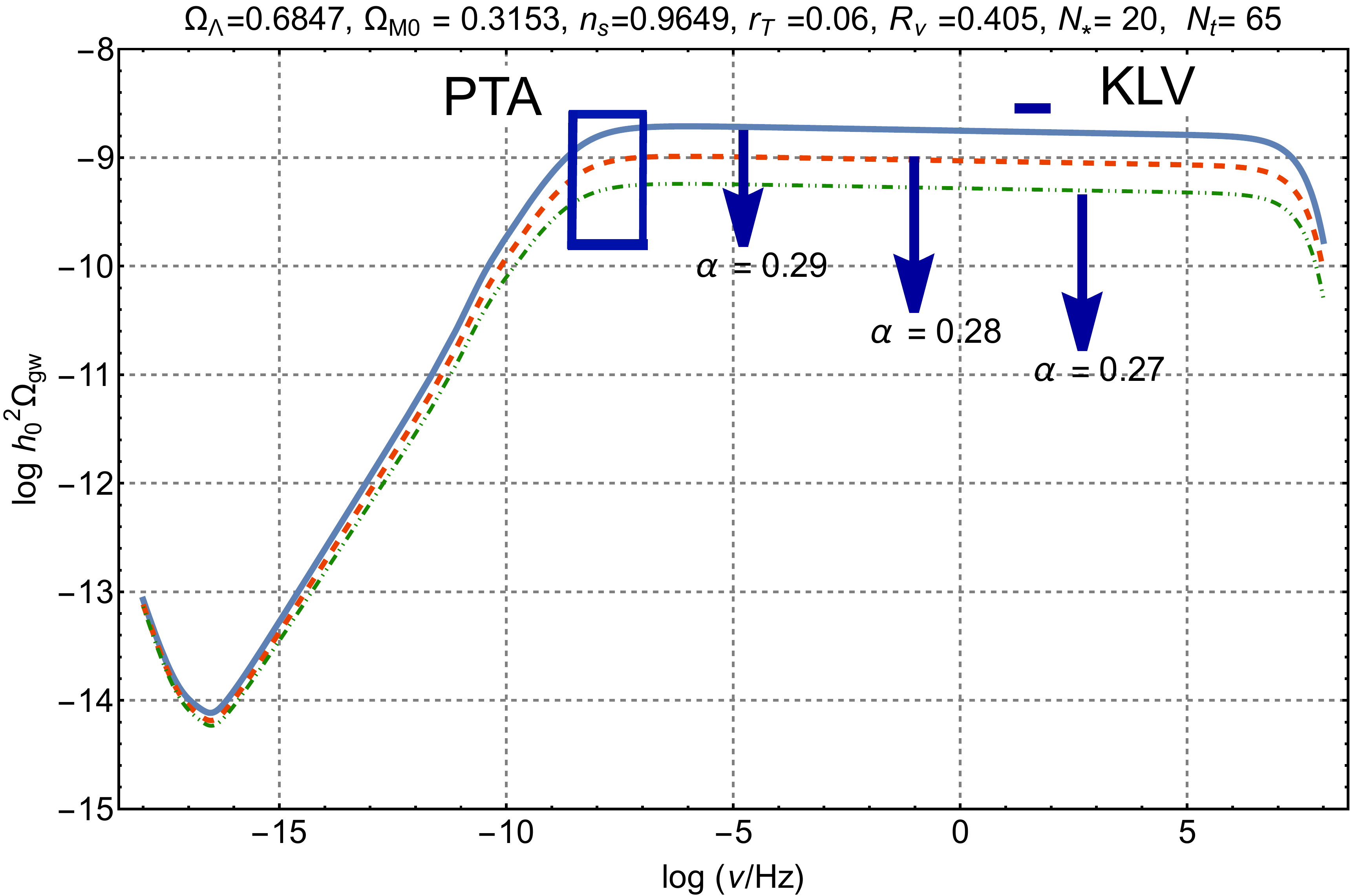}
\includegraphics[height=5.8cm]{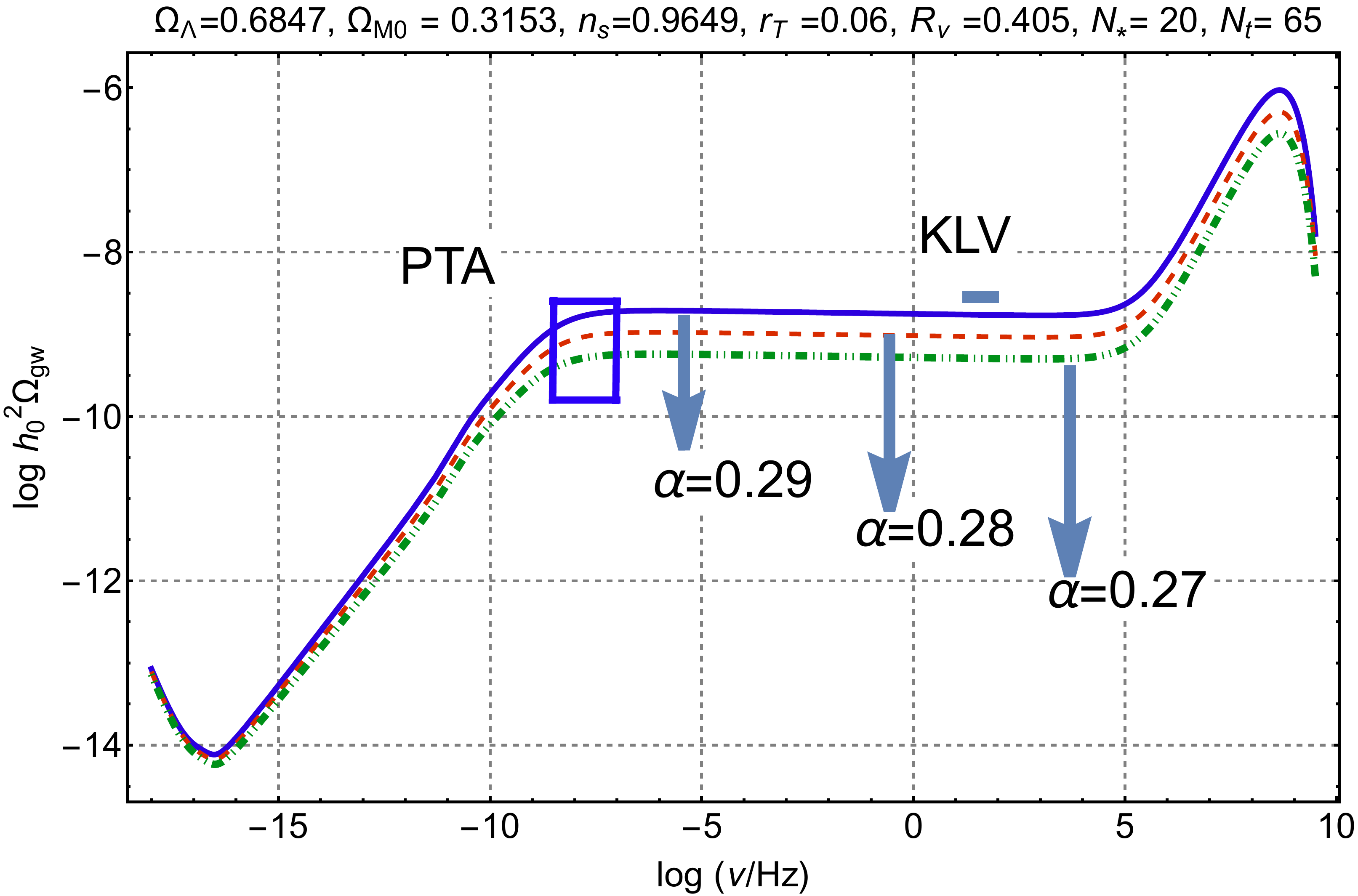}
\caption[a]{We illustrate the spectral energy density for a set of parameters selected 
within the shaded region appearing in the exclusion plot of Fig. \ref{FIG2}.
In the plot at the left we consider the situation where the post-inflationary evolution is dominated 
by radiation, as illustrated in the left cartoon of Fig. \ref{FF10g}. In the right plot a the post-inflationary evolution includes instead a first stage expanding at a rate slower than radiation (see also the right profile of Fig. \ref{FF10g}).}
\label{FIG3}      
\end{figure}
The spectral energy density of the relic gravitons in critical units is plotted 
in Fig. \ref{FIG3} for three particular values of $\alpha$ drawn from the allowed 
region of Fig. \ref{FIG2}. The PTA box corresponding to Eq. (\ref{PTAb1}) 
is also approximately illustrated. The left plot of Fig. \ref{FIG3} is 
computed by assuming that the post-inflationary 
evolution is dominated by radiation while in the right panel the post-inflationary epoch includes an initial stage expanding at a rate that is slower than radiation. This is why we can observe a final spike after the high-frequency plateau, as already 
suggested in the past in a related context \cite{FOUR}. The ultra-high-frequency spectral slope appearing in Fig. \ref{FIG3} is given by $m_{T} = 4 - 2/(1-\epsilon) - 2 \delta$ 
where $\delta$ has the same meaning discussed in section \ref{sec3}. The right plot of Fig. \ref{FIG3} illustrates, in particular, the case $\delta =1/2$.  In this case the onset of the radiation-dominated phase is delayed by the presence of a stiff phase; 
a spike appears in the GHz region and the signal is comparatively more constrained. This potential signal might be interesting 
for electromagnetic detectors of gravitational radiation which have been proposed and partially developed in the past \cite{brag1,cav1} (see also \cite{ten1,ten2,ten3,ten4,ten5}). 

\newpage

\renewcommand{\theequation}{6.\arabic{equation}}
\setcounter{equation}{0}
\section{Concluding remarks}
\label{sec6}
The spectral slopes of the cosmic gravitons do not only depend upon the early (e.g. inflationary) evolution of the space-time curvature and this is one of the major obstacles to inferring 
the expansion rate of the Universe from the energy density of the relic gravitons.
 For instance, the quasi-flatness of the spectral energy density for frequencies larger than the nHz is a simultaneous consequence of an early stage of accelerated expansion combined with the post-inflationary dominance of radiation down to the scale of matter-radiation equality. The recent measurements of the pulsar timing arrays and the improved limits on relic gravitons in the audio band call for a more thorough analysis of the connection between the expansion rate and the slopes of the spectral energy density. Provided the post-inflationary evolution is modified before the onset of big-bang nucleosynthesis, the standard form of the spectral energy density remains practically unaltered below the nHz but it is modified for higher frequencies.  

Assuming the Universe expands faster than radiation prior to nucleosynthesis, the only relevant constraints always stem from the aHz region. Conversely if the post-inflationary expansion rate is slower than radiation the largest signal is above the MHz and must be 
constrained by high-frequency measurements.  In this respect the most stringent bound in the audio band are associated with Kagra-Ligo-Virgo limits. When multiple post-inflationary phases evolve at different rates the local maxima of the spectral energy density are always smaller than the signals potentially attributed to cosmic gravitons in the nHz band.
If the spatial correlations associated with cosmic gravitons are not observed the pulsar timing arrays might anyway set a unique and relevant upper limit on the expansion rate before big-bang nucleosynthesis. In spite of the specific profile of the expansion rate, the detection of the tensor modes in the aHz domain is necessary if we want to disentangle the different possibilities. A value of the tensor to scalar ratio just below the current limits would only leave room for an enhancement of the spectral energy density in the high-frequency region where, probably, the only hope would be the extensive use of electromagnetic detectors. 

Even thought the post-inflationary expansion rate is modified, the slope of the spectral energy density between $100$ aHz and the nHz would always seem quasi-flat. However even this conclusion is challenged if the refractive index of the gravitons is dynamical during the early stages of the inflationary expansion. The effective evolution of the refractive index leads in fact to a blue slope at intermediate frequencies while above a fraction of the $\mu$Hz  the spectral energy density is still quasi-flat. If the dynamics of the refractive index is combined with the modifications of the post-inflationary expansion rate a further bump may arise in the MHz domain and the signatures of the refractive index could explain the claimed evidences of the pulsar timing arrays.  

All in all it is plausible to analyze the spectral energy density of the relic gravitons with the aim of inferring the early expansion history of the Universe which would be otherwise unaccessible. For this purpose it is however essential to develop a model-independent approach whose basic features have been described in the present investigation. If combined with the latest limits at intermediate and high-frequencies, the scheme proposed in this paper already pins down the physical timelines that are potentially compatible with a detectable signal in the nHz domain. An improvement of the Kagra-Ligo-Virgo limits by two orders of magnitude combined with a more definite origin of the signal observed by the pulsar timing arrays could severely constrain  multiple post-inflationary phases and even exclude certain classes of profiles.

\section*{Acknowledgements} 
The author wishes to thank T. Basaglia, A. Gentil-Beccot, S. Rohr and J. Vigen of the CERN Scientific Information Service for their valuable help.

\end{document}